\documentclass[prb,twocolumn,showpacs,amsmath,amssymb]{revtex4}
\usepackage{graphicx}       % Include figure files
\usepackage{dcolumn}        % Align table columns on decimal point
\usepackage{xcolor}
\usepackage{ulem}
\newcommand{\beq}{\begin{equation}}
\newcommand{\eeq}{\end{equation}}
\newcommand{\beqa}{\begin{eqnarray}}
\newcommand{\eeqa}{\end{eqnarray}}

\newcommand{\qvec}{{\bf q}}

\begin{document}
\title{The charge density fluctuations and the Shrinking Fermi Liquid scenario for strange metallicity in cuprates}

% Authors, for the paper (add full first names)
\author{Sergio Caprara$^{1}$*, Sauri Bhattacharyya$^1$, Carlo Di Castro$^{1}$, Giovanni Mirarchi$^{2}$, G\"otz Seibold$^{3}$,  and Marco Grilli$^{1,}$*
}

%\author{S. Caprara$^1$, C. Di Castro$^1$, M. Grilli$^1$, G. Mirarchi$^1$, G. Seibold$^2$}
%\newcommand{\orcidauthorA}{0000-0001-8041-3232} % Add \orcidA{} behind the author's name
%\newcommand{\orcidauthorB}{0000-0003-1089-1444} % Add \orcidA{} behind the author's name
%\newcommand{\orcidauthorC}{0000-0003-2124-6993} % Add \orcidA{} behind the author's name
%\newcommand{\orcidauthorD}{0000-0002-1772-1563} % Add \orcidB{} behind the author's name
%\newcommand{\orcidauthorE}{0000-0001-5607-7996} % Add \orcidB{} behind the author's name

\affiliation{$^1$CNR-INFM-SMC, and Dipartimento di Fisica, Sapienza Universit\`a di 
Roma, P.$^{le}$ Aldo Moro 5, 00185 Roma, Italy}
\affiliation{$^2$ Institute for Theoretical Physics and Astrophysics, 
Computational Quantum Materials, 
Julius-Maximilians-Universit\"at, 
97074 W\"urzburg,
Germany}
\affiliation{$^3$ Institut f\"ur Physik, BTU Cottbus-Senftenberg - PBox 101344, D-03013 Cottbus, Germany}
\affiliation{* corresponding authors, email: sergio.caprara@roma1.infn.it, marco.grilli@roma1.infn.it}

\begin{abstract}
{We interpret the strange metal properties of slightly overdoped cuprates in terms
of the recently proposed Shrinking Fermi Liquid theory. This proposal is based on the
pervading presence in the cuprate phase diagram of charge density fluctuations (CDF),
which have recently been identified and characterized in Resonant X-Ray Scattering experiments.
These charge fluctuations are abundant and have a rather low energy due to the proximity of the well-known charge density wave
quantum critical point hidden under the superconducting dome of cuprates, but have a short range and non-critical
character with a finite energy $M/\gamma$ of order $10 $ meV as measured above T$_c$. Here $M\sim \xi^{-2}$ is determined
by the rather short density-density correlation length $\xi$, while $\gamma$ encodes
the Landau damping ruling the lifetime of the charge fluctuations. 
Besides these low energy charge density fluctuations cuprates also display phonons of rather
high energy and, importantly, a broad continuum of particle-hole excitations, mostly due to
spin paramagnons arising from the strongly correlated character of these systems. With these
experimentally characterized ingredients we show that above $T_c$ the strange metal properties in transport
are well described in terms of fermionic Landau quasiparticles scattering with CDF and phonons.
The optical properties can instead be interpreted by the combined effect of low energy CDF
determining the temperature dependence, and of the paramagnon continuum determining a linear in frequency scattering rate.
Remarkably, the combined effect of these simple ingredients also induces $\omega/T$ scaling properties
for frequencies larger than $M/\gamma$. When superconductivity is suppressed by strong magnetic fields 
the strange metal properties extend down to a few Kelvin. By simply assuming that the CDF dissipation
parameter $\gamma$ grows logarithmically by lowering $T$, we account for all anomalous transport and thermodynamic
properties of cuprates (specific heat, Seebeck, heat transport, resistivity, and magnetoresistance) thereby providing a consistent
scenario for the strange metal phase of cuprates.
}
\end{abstract}
\date{\today}
%\pacs{}
\maketitle

\section{Introduction}
The occurrence of a `Strange Metal' state in high temperature 
superconducting cuprates, (see e.g. Refs.\,
\onlinecite{martin90,greene98,taillefer09,legros-2019,
greene23,hussey-2022,zhao22,chung24,yuan24,hartnoll-2022,paschen24,varma02,
science21,phillips22,fratini23} and  references therein), the T-linear
resistivity down to very low temperatures  being its most distinctive feature,
is a long-standing
theoretical and experimental challenge. The fact that similar strange 
metal properties also occur in many 
other systems like heavy fermions \cite{tomita15,hartnoll20,yuan20,paschen22,stewart01,vonloehneysen-2007,kenzie13}, pnictide 
superconductors \cite{fernandes22,hayes-2016}, and twisted bilayer graphene 
\cite{ball20,randeria21}, raises the issue of a new generic state of matter requiring
a revision of the paradigmatic Landau Fermi liquid (FL) theory valid for ordinary 
metals.  We will focus on the hole doped high temperature superconductors.
At the moment the main theoretical trends to explain the strange metal properties of cuprates
involve the coupling of the itinerant fermions with low-energy nearly 
local dynamical boson degrees of freedom, which provide
a source of strong scattering among the itinerant fermions, thereby disrupting the Landau FL state.
Recently, it has been 
proposed to represent these low-energy excitations by a so-called 
Sachdev-Ye-Kitaev model\cite{patel-2018,sachdev23_1,sachdev23_2} or 
by two-level systems\cite{schmalian-2024}. While in this framework disorder plays a crucial role 
to render these excitations local,
%These boson excitations might arise from some small-momentum order 
%parameter fluctuations (e.g., nematic) near a quantum critical 
%point (QCP). Disorder is then responsible for turning these long-ranged 
%excitations into local excitations.
the local character of the bosonic scatterer could also arise from the
topologic nature of excitations in a dissipative effective XY model
describing nearly critical loop currents\cite{varma-2020,varma-2015,varma-2025}.

The local character of the excitations and their abundance at arbitrarily low 
energies naturally allows the above models in some parameter regions 
to display marginal Fermi liquid (MFL)
properties. MFL theory, is a well known phenomenological scheme  proposed 
by Varma and coworkers \cite{varmaMFL,varma89} in the early days of cuprates
where  the interaction mediators are assumed to be momentum independent and 
%have a spectral density of the type
%\begin{eqnarray}
%\label{MFLmediator}
%\text{Im}\,D_\mathrm{MFL}(\omega,T) &\simeq &-N^*\frac{\omega}{T}\,\,\,\, |\omega|<T \nonumber \\
%\text{Im}\,D_\mathrm{MFL}(\omega,T) &\simeq &-N^* \text{sgn}({\omega}) \,\,\,\, |\omega|>T
%\end{eqnarray}
%($\hbar=1$ and the Boltzmann constant $k_B=1$ henceforth, while $N^*$ is
%the bare fermionic density of states)
to induce a fermionic retarded (R) self-energy of the form 
\begin{equation}
\label{MFLapp}
\text{Im}\,\Sigma^R_\mathrm{MFL}(\omega,T) \simeq -\lambda_{MFL} 
\max(|\omega|,T)
\end{equation}
where $\lambda_{MFL}$ is a dimensionless coupling. Notice that
 temperature is the only low energy scale.

MFL was very successful and has become a benchmark to 
phenomenologically describe many of the anomalies in
cuprates, but it does not identify  the microscopic 
mechanisms leading to the interaction mediators. 
A step forward in this sense was provided by the recent 
discovery \cite{arpaia-2019,arpaia-2023} 
of nearly local charge density fluctuations (CDF) in the form of `aborted', 
very short ranged, charge density waves (CDW). This has triggered the idea that CDF 
are the sought nearly local low-energy degrees of freedom. 
These fluctuations with almost the same modulation vector of the CDW, have a low-enough energy to 
leave T as a relevant energy scale thereby accounting well for 
transport and spectroscopic anomalies\cite{seibold-2021,SFLoptics-2024}.
At the same time they have a finite energy scale that, as we will see, controls
many features of the strange metal behavior. Indeed, in slightly
overdoped samples, their energy  $\omega_{CDF}\sim 5-10$ meV\cite{arpaia-2023} and 
at temperatures below this energy scale
they no longer scatter effectively and FL properties are recovered.
Therefore, in order to account for the strange metal behavior
observed down to substantially lower temperatures when superconductivity is 
suppressed (usually by strong magnetic fields) a mechanism has to be assumed 
to lower the CDF energy scale. This is precisely the main outcome of the 
so-called Shrinking Fermi Liquid scenario \cite{caprara-2022,grilli-2023,SFLoptics-2024,mirarchi-2024}.
This will be summarized in section I.B.3, after the 
evidences and properties of the CDF will be reported in the next subsection.

\subsection{Charge density waves and charge density fluctuations}\label{ssec:cdf}
\begin{figure}[thb]
\includegraphics[width=8cm,clip=true]{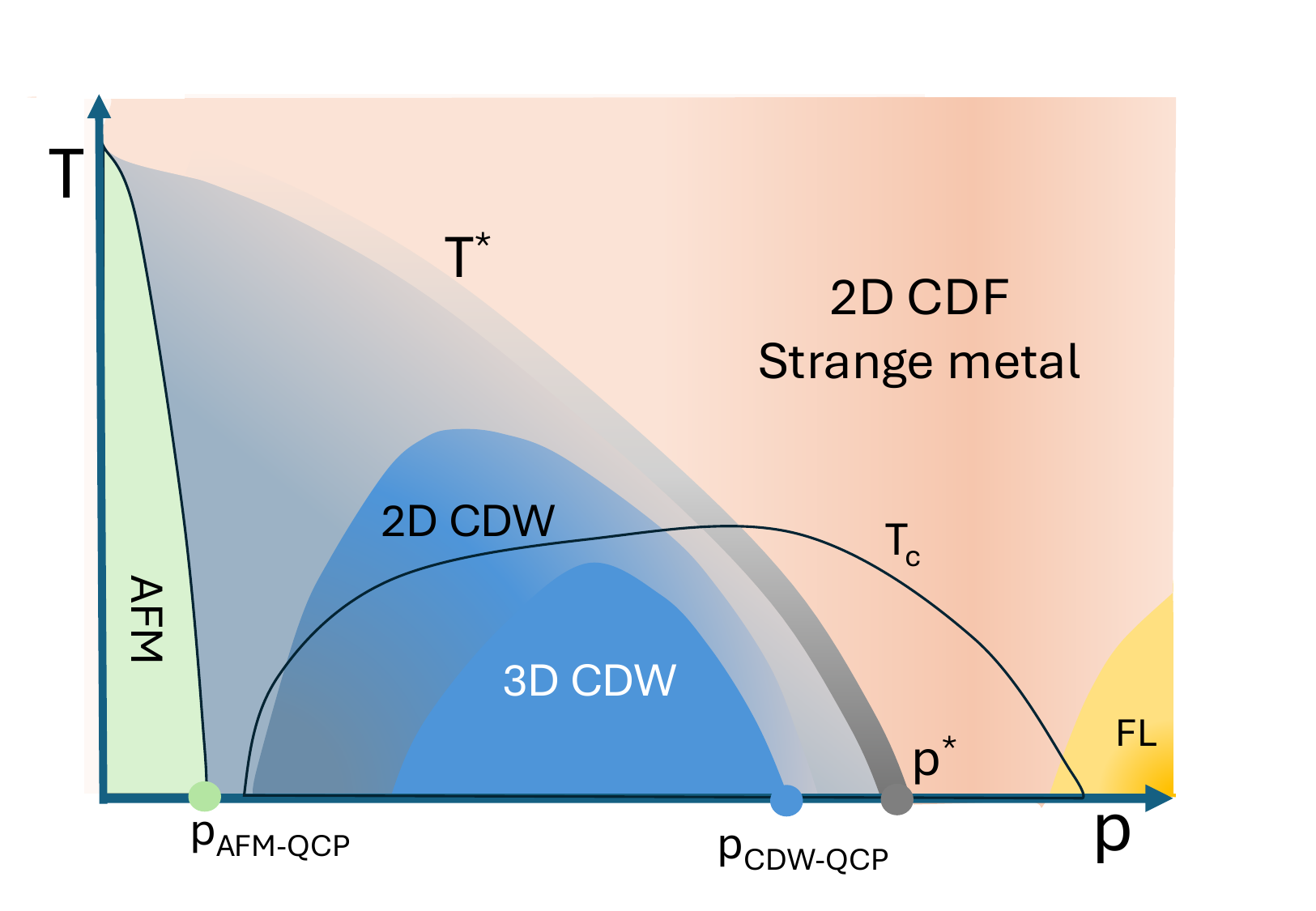}
\caption{Schematic phase diagram of the cuprates.
$p$ is the doping of the CuO$_2$ planes. $T^*$ is the pseudogap crossover line
ending at $T=0$ at the doping $p^*$}
\label{phase-diag}                                                   
\end{figure}

This work is the completion of a long story: the search of a mediator of scattering among 
the Fermi quasiparticles suitable to provide a d-wave paring with high
critical temperature $T_c$ in the particle-particle channel and strong enough 
to disrupt the Fermi liquid behavior for the normal phase in the particle-hole channel. 
It was then natural to search for strong critical fluctuations near some Quantum Critical Point (QCP).  
Many researchers emphasized the presence of the low doping antiferromagnetic QCP  at doping $p=p_{AFM-QCP}$ 
(see Fig. \ref{phase-diag}) and the relative critical spin fluctuations
(see, e.g., Refs. \onlinecite{millis-1990,abanov-2003} and references therein). 
We instead investigated the optimal and 
high doping regions of the phase diagram for two reasons: the optimal doping peak of 
$T_c(p)$  (at the doping $p_{opt}$ at which the superconducting critical temperature is maximum)
is far from the low doping $p_{AFM-QCP}$; the strange metal behavior is shifted towards high doping 
in the V shaped region of the T-p phase diagram around the doping $p^*$ (see Fig. \ref{phase-diag}).
It was found that coming from the high doping  region the strongly correlated FL, 
in the absence of superconductivity, could easily be unstable for CDW formation at a hidden QCP ($p_{CDW-QCP}$)
near optimal doping, see Fig. \ref{phase-diag}  
\cite{castellani-1995,reviewQCP1,reviewQCP2,seibold-2000,andergassen-2001,caprara-2017}.  
Evidences of charge (and sometimes of intertwined spin) ordering soon started to
emerge\cite{tranquada-1995} and kept accumulating along the years (see, e.g.
 Refs. \onlinecite{kivelson-review,keimer-2015} and references therein), but the 
occurrence of a CDW-QCP was eventually
confirmed by quantum oscillation and transport experiments under strong magnetic fields \cite{taillefer-2007-1,badoux-2016}.
 The critical  CDW  fluctuations are good mediators 
of d-wave pairing for gap and pseudogap formation\cite{perali-1996,andergassen-2001}, 
but are strongly anisotropic and peaked around
the critical wavevectors. This feature makes them unsuited to disrupt the
FL and cannot explain all the experimental challenges posed by the strange metal\cite{hlubina}. 

The last decade was characterized by an impressive progress 
in Resonant Inelastic X-ray Scattering (RIXS), which not only unambiguously confirmed 
\cite{ghiringhelli-2012,chang-2012,dasilvaneto-2014,comin-2016} the
strong tendency to form CDW in underdoped and optimally doped cuprates, but also
 stimulated the search for the related
critical fluctuations. Indeed, more recently, RIXS experiments observed short-range dynamical CDF, pervading 
most of the cuprates phase diagram \cite{arpaia-2019,arpaia-2023,arpaia-2021}. 
These fluctuations have nearly the same periodicity of the CDW, but, while 
they coexist in the underdoped region below the doping-dependent pseudogap crossover temperature 
$T^*(p)$, CDF alone are found up to the highest temperatures for $T>T^*$ and deep in the 
overdoped region $p\gtrsim p^*$ ($p^*$ is the doping at which the pseudogap crossover 
temperature vanishes $T^*(p^*)=0$). This is the reddish region reported in the schematic phase diagram of
Fig.\ref{phase-diag}.
In Ref. \onlinecite{andergassen-2001} the crossover line marking the
onset of the pseudogap was ruled by the proximity to the charge-ordering quantum criticality 
and the two doping points $p^*$ and $p_{CDW}$ were related to one another, the second being shifted 
to lower doping due to the CDW flucuations. 
%it was also proposed that dynamical CDWs
 %may mediate strong pairing
%and have a momentum structure that makes them
%favorable for d-wave superconductivity\cite{perali-1996,andergassen-2001}.

In comparison to CDW, the CDF have a substantially shorter coherence length 
and a higher (but still rather small) characteristic energy  
$\omega_{CDF}\sim 8-25$\,meV, which remains finite even when extrapolated 
to zero temperature\cite{arpaia-2019} (the CDW have instead a larger coherence length
and a somewhat smaller energy $\omega_{CDW}\sim 0-3$\,meV). 
When superconductivity is (at least partly) 
suppressed by strong magnetic fields the CDW sets in as a true static 
phase\cite{taillefer-2007-1,badoux-2016}.

Roughly 
speaking, owing to this distinct dynamical behavior, one can think of the CDF 
as `aborted' CDW, which are nonetheless more robust both in temperature and 
doping, so that they can be detected over much broader regions of the phase diagram 
and essentially in all classes of cuprates\cite{arpaia-2019,arpaia-2021,arpaia-2023}. Remarkably, 
the abundant CDF are the only excitations that are detected at relatively low 
energy at temperatures above $T_c$ in the optimal and overdoped regions, where 
optical phonons and spin excitations are at much higher energy (if any) and where 
the strange metal behavior is present (see Fig \ref{phase-diag}).
Moreover, while the rather long-ranged CDW taking place in the underdoped region
of cuprates are rather peaked in momentum space, CDF, on the contrary, have a very mild
momentum dependence and not only can be detected along the (1,0,0) and (0,1,0)
directions, but the broad peak in their momentum response extends and overlap
substantially also along (1,1,0) so that their effects look rather isotropic in the
RIXS response function\cite{arpaia-2023}. 

\begin{figure}[thb]
    \centering
    \includegraphics[width=1.\linewidth]{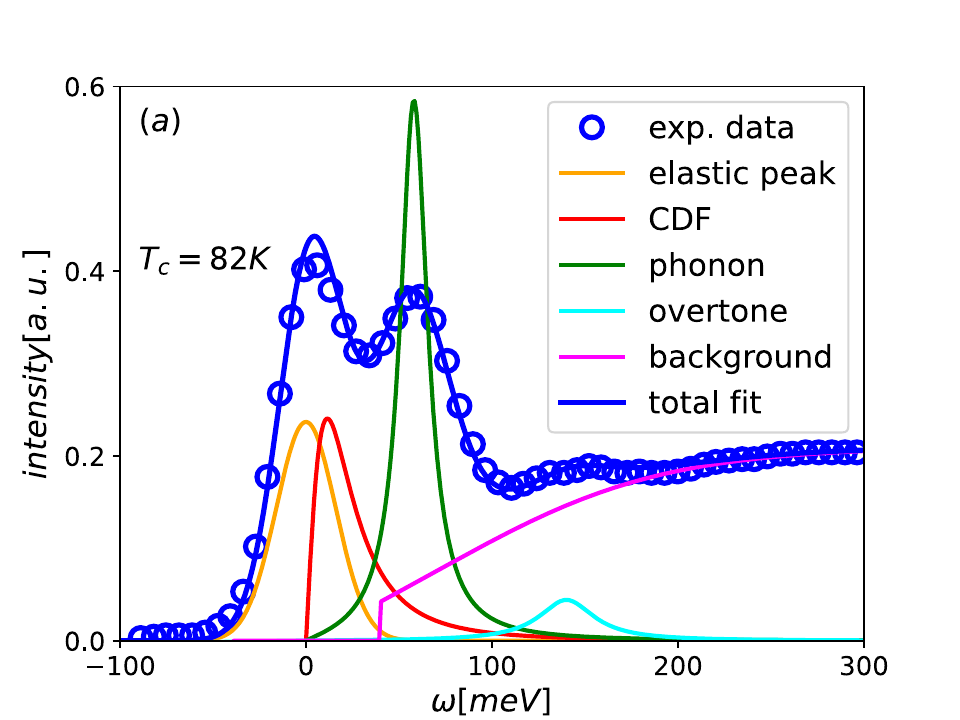}
    \includegraphics[width=1.\linewidth]{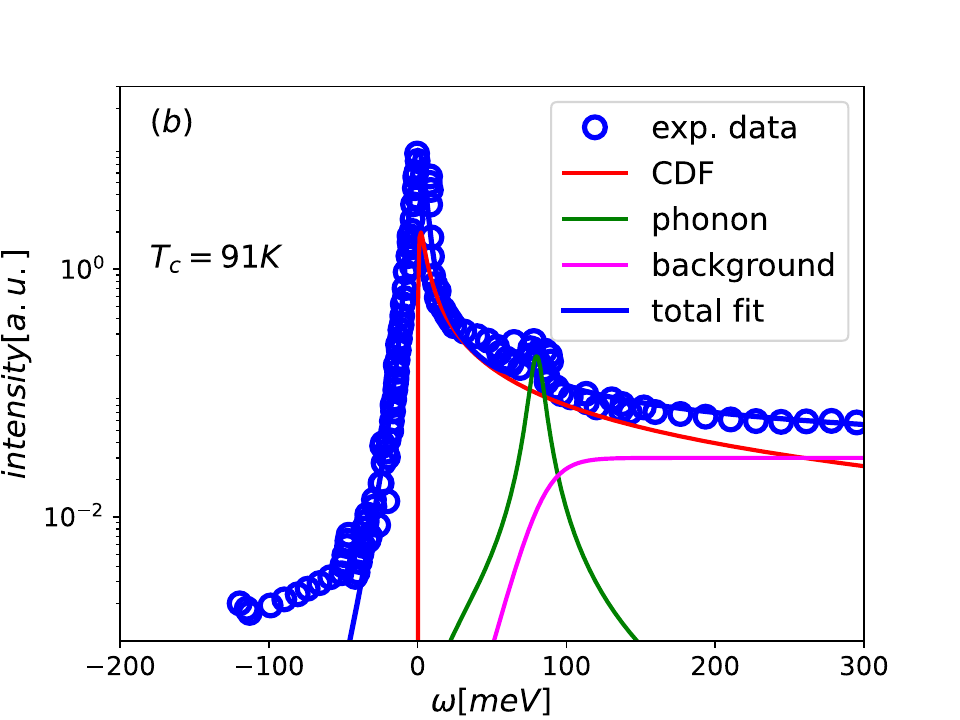}
    \caption{High resolution RIXS spectrum  (a) and EELS spectrum (b) for Bi2212 with $T_c$ indicated in the panels. 
    RIXS data are from Ref. \onlinecite{arpaia-2023} whereas EELS data have been taken from Ref. \onlinecite{abbamonte-2025}.
    The experimental spectra (blue circles) are decomposed into the individual excitations (elastic peak, CDF, phonons, background) as described in the text. }
    \label{fig-rixs}
\end{figure}
Fig.\ref{fig-rixs}(a) reports 
the RIXS response at $\qvec=\qvec_c$ obtained from Ref. \onlinecite{arpaia-2023} 
together with the CDF lineshape as a red line (the spectrum is broader since 
it is obtained by a convolution with an experimental resolution Gaussian of width
$20-30$ meV). The magenta and the green curves represent the phonon (half-breathing mode)
and higher energy excitations: the paramagnons and the incoherent particle-hole continuum.

%In Fig.\ref{fig-rixs} we report the structure factors of low-intermediate energy
%excitations that all might in principle determine transport and spectroscopic properties
%of cuprates. Despite some differences, that we will comment below, both RIXS and EELS
%show the presence of low-energy rather local excitations that were identified as 
%CDF in Refs. \cite{arpaia-2019,arpaia-2023}. Above the characteristic energy
%of CDF (5-10 meV in RIXS, 3-10 meV in EELS) the presence of phonons up to 50-70 meV
%is also detected. Finally at energies of 100 meV and above there are paramagnon and 
%incoherent particle-hole excitations.

Electron Energy Loss spectrocopy (EELS) also made impressive progresses in terms of
momentum and energy resolution in the last years. Then they have nicely confirmed 
the presence of electronic excitations at energy as low as $3$ meV having a quite
isotropic character\cite{abbamonte-2025} (see Fig.\ref{fig-rixs}(b)).
It is quite tempting to identify these 
excitations with the CDF, which on the surface layer probed by EELS acquire a much
more static character (and indeed short-ranged charge ordering is also 
detected by STM\cite{how03,hana4,versh04}).
%{\textcolor{red}{ ELIMINATE THIS PARAGRAPH? TOO CONJECTURAL?
%While there is a qualitative agreement between RIXS and EELS
%spectra [cf. Fig. \ref{fig-rixs}(a) and (b)], 
%the completely different intensities arise from the different nature of 
%the probes and of the probed regions: photon in - photon out  bulk sampling for RIXS and
%electron in - electron out surface sampling for EELS. It is therefore not so surprising that 
%quantitative difference arise in the cross sections of the various excitations and of
%their energies. }}

In summary, CDF are abundant low-energy excitations and are 
a natural candidate to explain the strange metal properties, 
most prominently the linear-in-temperature scattering, 
in the metallic state of cuprates \cite{seibold-2021,
caprara-2022,grilli-2023,mirarchi-2024}. 
This triggered the Shrinking Fermi liquid (SFL) 
scenario that we summarize in the next section.

\subsection{The Shrinking Fermi Liquid scenario: An updated summary}
\label{sec:sfl}

\subsubsection{CDF as the low-energy scattering mechanism}
The main idea is that the CDF fulfill two rather general sufficient conditions 
to obtain linear-in-temperature resistivity:
(a) they are sufficiently short ranged (i.e., nearly local in space) to provide a broad
range of momenta in the scattering processes of quasiparticles. This first condition is
needed in order to have the quasiparticles scattered all 
over the Fermi surface, thereby bypassing the objection of 
Ref.\,\onlinecite{hlubina} related to the violation of Fermi liquid properties, which applies whenever the scattering is peaked 
at finite momenta as is the case for 
CDW or spin wave critical fluctuations. In this latter case the Fermi surface is partitioned in strongly 
scattered (hot) regions and weakly scattered (cold) regions, which dominate 
transport keeping overall Fermi liquid features; (b) the CDF have sufficiently low 
energy to obey a semiclassical statistical distribution down to $T_c$.
CDF have the above properties with $\omega_{CDF} \lesssim T_c$. 
Roughly speaking, owing to condition (a), the scattering is almost 
uniform over the entire Fermi surface and the scattering
integral in the Boltzmann equation can be approximated with only one
effective scattering time for each single scattering event. 
Its inverse is determined by the effective characteristic frequency 
of the scatterers i.e.  $\omega_{CDF}$.
The full scattering rate
must be proportional to the characteristic frequency of the single event 
multiplied by the population of the scatterers, given by the Bose function (
$\tau^{-1}\sim b(\omega,T) \omega_{CDF}$).
Owing to condition  (b), the Bose distribution of the collective scatterers is well 
approximated by $T/\omega$, resulting in a linear-in-temperature scattering 
rate. Since the scattering is nearly isotropic this also induces a linear-in-temperature resistivity.

Our starting point is that the observed (and experimentally characterized) CDF, 
near the modulation wave vector ${\bf Q}_c$,
have a propagator of the typical form for fluctuations around a Gaussian 
QCP, ~\cite{enss-2007} in the presence of Landau damping $\sim \gamma$ 
\begin{equation}\label{eq:cdf} 
    D_{CDF}({\bf q},\omega)=\frac{1}{M+\nu|\qvec-{\bf Q}_c|^2
    -\frac{\omega^2}{\overline{\Omega}}-\mathrm i\gamma \omega}\,.
\end{equation}
Here, $M=\nu \xi^{-2}$ with $\xi$ being the charge-fluctuation correlation length and $\nu$ is
the stiffness of the CDF.
$\overline{\Omega}$ is a crossover scale above which CDF acquire a 
more propagating character. This energy scale also acts as an ultraviolet cutoff. 
Previous work \cite{seibold-2021} has shown that CDF, having
a characteristic energy $\omega_{CDF}=M/\gamma$ of the order of 5-10 meV 
as measured by high resolution RIXS experiments, typically smaller than $T$
when $T>T_c$, can well account for the linear-in-temperature inelastic 
quasiparticle scattering down to temperatures of order $T_c$. 

\subsubsection{What happens at high frequencies?}
The RIXS and EELS spectra in Fig. \ref{fig-rixs} not only report the large peaks 
at low energy ($\sim 10$ meV) due to CDF and (at a bit larger energies $\sim 50-70$ meV)
due to phonons, but also a broad continuum usually attributed to overdamped 
paramagnon and incoherent particle-hole excitations. The general
form of these spectra have a generic resemblance with the interaction
mediator proposed by Varma and coworkers,  
with the important difference that in the MFL scheme the low energy cutoff of the continuum scales with temperature.
%The main features of this mediator are the $1/T$ slope at low frequencies and the constant character at high frequencies, 
This spectral density of the scattering mechanism results in the $\omega$-linear
behavior of Im$\Sigma(\omega)\sim\omega$ at $\omega>T$. 
It is clear that the CDF alone would not be able to produce this linearity
in frequency: Once the external frequency exceeds the characteristic energies of
the CDF, Im$\Sigma(\omega) $ would saturate at a constant value,
see also Sec. II.C and Fig. \ref{opt-scatt-rate}.
%\textcolor{red}{I have eliminated the previous Fig. 3 which is the light version of Fig. 12 in Sec. 2d}
%This is clearly visible from the dotted line in FIg.\ref{optics_split_M10}.
%\begin{figure}[thb]
%\includegraphics[width=8cm,clip=true]{optics_split_M10.pdf}
%\caption{HERE WE SHOULD REPLACE THIS FIGURE WITH A M ORE SCHEMATIC ONE AT %ONLY ONE TEMPERATURE.
%THE MORE COMPLETE ANALYSIS IS CARRIED OUT IN SECT. II.D.4}
%\label{optics_split_M10}                                                   
%\end{figure}  
Instead, the
actual spectra of cuprates do have the high energy nearly constant part
(although its origin is unrelated to the low-energy CDF), which
naturally accounts for the $\omega$-linear behavior of Im$\Sigma(\omega)$
and of the optical scattering rate (see also Sect. II.C).
%Therefore, while the MFL scenario clumps into a single phenomenological form,
%Eq. (\ref{MFLmediator}), 
%the temperature and frequency features
%responsible for T-linear resistivity and $\omega$-linear scattering rates
Our scenario starts from the experimentally observed and rationalized 
RIXS spectra and attributes the two behaviors to different physical
mechanisms: the T-linear behavior arises from the nearly classical 
character of the statistical weight of the CDF [$b(\omega,T)\sim T/\omega$, 
see condition (b) in the previous subsection] when the temperature
is larger than the characteristic energy of the CDF, $T\gtrsim \omega_{CDF}=M/\gamma$.
On the other hand, the $\omega$-linear behavior is due to the 
nearly constant weight of the overdamped paramagnons for frequencies 
above $30-50$ meV. 
%Fig. \ref{optics_split_M10} reports the total 
%Im$\Sigma(\omega)=$Im$\Sigma_{cdf}(\omega)+$Im$\Sigma_{pm}(\omega)$
%as well as the separate contribution due to CDF (dotted lines) and paramagnons(dashed lines)
%at three different temperatures. Clearly the scattering rate due to CDF saturates
%at frequencies above $M/\gamma$ ($M=10$ meV and $\gamma=1$ for all
%temperatures in the figure), while the damped spin excitations produce
%a linear in frequency scattering rate allover the range of their broad spectrum.
Then the total overall Im$\Sigma(\omega,T)$ due to both CDF and paramagnon scattering
(phonons in this rough argument can easily be included in the paramagnon contribution)
can finally be mimicked by a single 
analytic form like,
\begin{equation}
\label{SFLapp}
\text{Im}\,\Sigma_R(\omega,T)\simeq -
 (\lambda_T T+\lambda_\omega \omega)
\end{equation}
where $\lambda_{T,\omega}$ are dimensionless couplings between the quasiparticles and
the low ($\sim 5-10$ meV) and high energy ($\gtrsim 30$ meV) scattering excitations. 
%with $N^*$ being the
%electron density of states at the Fermi level and $g_{eff}$ the coupling 
%between the fermionic quasiparticles and the bosonic 
%fluctuations (or some average coupling with the CDF and the paramagnons). 
%Clearly this form is conceptually wrong since it merges in a single 
%analytic expression the scattering rate due to two distinct mechanisms,
%namely the low-energy CDF and the particle-hole continuum. Nevertheless, 
%Empirically this expression 
%rather well mimics the actual merged scattering rate from both mechanisms.
%By analyzing at different temperatures the total scattering due to both CDF and 
%particle-hole continuum, we also found (see Fig. \ref{opt-scatt-rate}) that 
%$\tilde{M}\approx M/\gamma$. Moreover, Eq.(\ref{SFLapp})
%displays an approximated scaling character whenever $\tilde{M}$
%is negligible with respect to $T$ and/or $\omega$. In particular, for 
%$\tilde{M}\sim M/\gamma \to 0$ it clearly reduces to the MFL form of Eq. (\ref{MFLapp}).
Owing to the close resemblance of this equation with the MFL form of Eq.(\ref{MFLapp}),
it is not surprising that the joined effect of CDF and particle-hole continuum might have been 
phenomenologically interpreted in terms of a single MFL behavior.
%However, it is worth emphasizing again, that this form is just a way to synthesize
%the separate contribution of different scattering mechanisms due to the 
%paramagnon spin excitation (with a rather large low-energy cutoff when
%the system is at or above optimal doping) and to CDF having energies
%ranging from $25-30 $ meV in heavily underdoped cuprates to $10 $ meV or less 
%above optimal doping.~\cite{arpaia-2023}
This issue and its relevance to interpret optical spectra
will be discussed in Sect. II.C.

\subsubsection{Why should a FL be `shrinking'?}\label{sec:sfl:shrink}
The possibility to attribute the occurrence of linear in T resistivity to  well 
identified and characterized low-energy excitations, namely the CDF,  is the main success of the SFL theory. 
However, this only occurs as long as $T>\omega_{CDF}=M/\gamma$,
with $\omega_{CDF}$ being experimentally found to be of order 40-60 K at least.
Therefore the major limitation of the theory is that, by lowering the temperature, 
below the CDF characteristic energy $T<\omega_{CDF}$ this scattering 
mechanism becomes ineffective and 
FL properties should be recovered. This would typically occur below $40-60$\,K, 
at least.
This is at odds with the observation of linear-in-temperature 
resistivity down to much lower temperature, when superconductivity 
is suppressed by strong magnetic fields \cite{legros-2019}. The condition (a) 
above prevents $\omega_{CDF}$ from decreasing as $M\sim\xi^{-2}$ decreases, 
which would otherwise induce the standard critical slowing down due to a
criticality with infinite correlation length and a strongly peaked momentum 
dependence of the fluctuation propagator. 

An alternative was recently proposed \cite{caprara-2022} to decrease the FL 
scale $M/\gamma$: assuming a temperature dependent dissipation parameter like, e.g.,
$\gamma \sim \log(1/T)$, the FL scale decreases {\it without} producing an 
increase of the spatial correlations (i.e., a reduction of $M$). 
The specific logarithmic form of $\gamma(T)$ was phenomenologically inferred from 
the logarithmic increase of the low-temperature specific heat coefficient 
$C_V/T$ under strong magnetic field \cite{michon-2019} (see Sect. III.A). In this case, the 
boson contribution to $C_V$ acquires a logarithmic contribution
$C_V/T\sim \gamma(T) \log (1/M)$. While the standard QCPs may display a logarithmic 
temperature dependence from a vanishing mass of the fluctuations 
$M\sim \xi^{-2}\sim T$, here the system is assumed to be close, but not too 
much, to the CDW-QCP in order to have abundant fluctuations at finite (actually, 
rather small) correlation length and mass $M$, as experimentally observed. Then, the observed 
logarithmic behavior stems from dissipation of the fluctuations $\gamma(T)$ and 
is neither due to the standard critical behavior of the fluctuations nor to 
the diverging fermion quasiparticle mass $m^*$, as found instead in the MFL theory. 

This scenario has been extended along two directions. In 
Ref.\,\onlinecite{grilli-2023}, a simple model has been introduced where the low-energy CDF 
not only can decay in particle-hole pairs, but also into slow diffusive modes 
that always occur even in rather clean Drude metals whenever $T$ is lower than 
the elastic scattering rate $1/\tau_0$. In two dimensions, this additional 
decay channel was shown to give rise to a logarithmic increase of the damping 
term $\gamma$. Of course, this is just one out of several other possibilities 
leading to an increasing damping, and further general effective mechanisms 
like the formation of a CDF nearly glassy state \cite{ketmaier-2026} are 
still under investigation. 

Notice also that an experimental access to $\gamma(T)$ is rather hard because
an increase of the Landau damping can only occur when the superconducting gap is
suppressed, opening the way to the low-energy decay of the collective modes
into particle-hole pairs. Unfortunately, this requires such strong magnetic fields 
that remain inaccessible to RIXS experiments and only an indirect 
inference of $\gamma(T)$ from transport and specific heat 
experiments under fields of several tens of Tesla is presently available.

From Eq. \ref{SFLapp} the frequency and temperature dependence of 
the quasiparticle self-energy turns out to be almost indistinguishable 
in the MFL and SFL scenarios whenever the FL scale $M/\gamma$ is smaller than
$\omega$ and/or $T$. It is then clear that
this similarity improves at low temperature if  superconductivity  is 
suppressed by very strong magnetic fields because $\gamma=\gamma(T)$ increases
when $T$ is lowered, correspondingly reducing the FL scale. 
This obviously makes the experimental distinction between these two scenarios
quite challenging (see the discussion in Section \ref{sec:discuss}). Despite 
the strong similarities between our SFL scenario and the MFL scenario, 
some crucial differences are present. First of all, while it is well known that 
the quasiparticle mass $m^*$ diverges logarithmically in $T$ within the MFL 
scheme, in the SFL the quasiparticle mass is logarithmically increased by the scattering with 
CDF, but it eventually saturates at low temperatures
and always stays finite. Another major difference is that in MFL 
the dynamical quantities like electron self-energy and conductivity display 
an exact $\omega/T$ scaling, while such property is broken by any finite
$M/\gamma$ in the SFL case. Of course when $\omega,T\gg M/\gamma$
an approximate scaling is also realized in the SFL case.
%Thus, although the lack of scaling in SFL is not surprising, because the 
%linear-in-temperature 
%behavior of the scattering rate in SFL merely arises from the 
%semiclassical statistical distribution function of the CDF while the 
%frequency dependence is related to the presence of a broad continuum of
%overdamped paramagnon spin excitaions,
%as a matter of fact, an approximate scaling sometimes still seems to take place for 
%$\omega,T \gg M/\gamma$.

In the following sections we benchmark the SFL with the presently available experimental landscape.
In particular, we will first compare the transport and spectroscopic results
with the theoretical calculations at temperature above the superconducting critical 
temperatures (Sect. II). In this case we will see that no specific assumptions have to be done
on the (irrelevant) temperature dependence of the Landau damping parameter $\gamma$.
This will provide a first complete scenario pointing out that CDF have by themselves 
low enough energy and weak momentum dependence to secure a possible explanation
of strange metal behavior. Afterwords, in Sect. III we will address the issue of the low energy
transport and thermodynamic anomalies below T$_c$, which require additional assumptions 
on the temperature dependence of $\gamma(T)$. With a common assumption of a logarithmic 
dependence $\gamma(T)\sim \log(T_0/T)$ we will show that the linear T dependence down to a few Kelvin
 as well as the Seebeck and specific heat anomalies are consistently explained.
A discussion and concluding remarks are finally reported in Sect. IV.

\section{The SFL scenario above $T_c$: 
Matching RIXS, resistivity, optical conductivity, and Raman scattering in cuprates}\label{sec:2}
The complex material structure of cuprates has often prevented to carry out
a complete set of experiments on the same set of samples. The
 experimentalists face many technical issues ranging from sample size, surface deterioration,
doping homogeneity and control, and so on together with the intrinsic limitations
of each technique (bulk or surface probe, dynamical or static probe, ...).
This is why often a patchwork analysis has to be carried out over different
sample sets, different material classes, and different technical tools.
Nevertheless we believe that a common coherent picture should emerge by comparing different
experiments, different techniques, and different material classes. This is precisely the aim
of the present section, where the scenario provided by the SFL theory is tested
with a cascade procedure linking RIXS experiments to transport (resistivity) and 
optical conductivity. We thus aim at showing that SFL theory is able to account in a
coherent manner for various experiments in different material samples, namely 
slightly overdoped Bi$_2$Sr$_2$CaCu$_2$O$_{8+\delta}$ (BSCCO), 
YBa$_2$Cu$_3$O$_{7-\delta}$ (YBCO) or 
(Y$_{1-x}$Ca$_x$)Ba$_2$Cu$_3$O$_{7-\delta}$ (YCBCO), and La$_{2-x}$Sr$_x$CuO$_4$ (LSCO). 
Specifically, in this section, for
each of these systems we start from RIXS experiments to identify and characterize  
the excitations (CDF, phonons, paramagnons/particle-hole pairs) responsible for the
scattering of the quasiparticles. This information will then be used to calculate
the quasiparticle scattering rate to fit both the resistivity and the frequency-dependent 
scattering rate as found from optical conductivity experiments
and also determining the Raman response. 

To set up the technical framework we now describe the calculation of the resistivity and
optical scattering rate. 
As far as resistivity is concerned, we assume that the parameters of the excitations 
(CDF, phonons, paramagnons,...) are temperature independent. Actually a weak temperature
dependence of the CDF mass $M$ [see Eq. (\ref{eq:cdf})] is experimentally found\cite{arpaia-2023}, but this
dependence is rather weak (of order 20-30 percent at most from room temperature down to below $T_c$)
and most importantly it extrapolates to a finite value at $T=0$. 
Moreover, we explicitly checked that keeping this additional temperature dependence does not hinder a 
satisfactory fitting and it does not spoil the T-linear behavior of $\rho(T)$. 
Owing to this substantial irrelevance, for the sake of
simplicity we neglect these weak dependencies.
For the phonons, instead we use a Lorentzian form centered at the energy found fitting
the RIXS spectra. The elastic contribution in RIXS spectra  (orange lines) does not enter the calculations of the
$T$ and $\omega$ dependent scattering rates, because it arises from mechanisms (like, e.g. 
surface diffuse scattering of the X-rays) that are not pertinent to the electron dynamics.
Elastic scattering rates $1/\tau_0$ are, however, considered in the fitting of the resistivity
and of $1/\tau(\omega)$ as they naturally arise from impurity elastic scattering.
Finally, and most importantly, we assume that the CDF peak is so broad in momentum space
that the momentum dependence of the CDF spectral density can be neglected (this was
explicitly shown in Ref. \onlinecite{seibold-2021}). As usually done, the momentum dependence of phonons (ph),
paramagnons (PM), is also assumed to be weak and neglected.
Thus, taking the spectral density of the CDF and all other excitations we fit RIXS spectra as
%\begin{eqnarray}
%    &&ImD(\omega)=\tilde\lambda_{CDF}ImD_{CDF}(\omega) \nonumber \\
%    &&+\tilde\lambda_{PH}ImD_{PH}(\omega)+\tilde\lambda_{PM}ImD_{PM}(\omega) 
%\end{eqnarray}
\begin{equation}
    D''(\omega)=W_{CDF}D''_{CDF}(\omega)
    +W_{PH}D''_{PH}(\omega)+W_{PM}D''_{PM}(\omega) \,.
\end{equation}
Here the parameters $W_\alpha$ ($\alpha=CDF,\,PH,\,PM$) 
are adjusted to match the relative weight of the various excitation
in the RIXS spectra and also include the corresponding RIXS matrix elements. 
Note that the matrix elements are rather local in RIXS and therefore only weakly momentum dependent. 
Further on, in the spirit of a 'fast collision approximation' \cite{devereaux_2011} we neglect any energy dependence. 
Since the fitted data are in arbitrary units, the $W_\alpha$ coefficient are redundant and the
relevant physical content is in the relative weight of the excitations $W_{PH}/W_{CDF}$ and $W_{PM}/W_{CDF}$.

In principle, the various excitations couple to the fermionic quasiparticles
with different couplings $g_\alpha$.
Therefore, when the quasiparticle self-energy is calculated, the relative `abundance' 
of excitations represented by the parameters  $W_\alpha$  is modified
into the following combination of factors $g_\alpha^2N^*W_\alpha $, which has the dimension of energy
and where the quasiparticle density of states $N^*$ is assumed to be constant.
However,  lacking detailed information on the $g_\alpha$'s  and to minimize the number of fitting parameters,
we assume all couplings to be the same $g_\alpha=g$ so that the combination becomes
$(g^2N^*)W_\alpha\equiv \tilde{\lambda} W_\alpha$. In this way the relative 
weight of the excitations in the quasiparticle self-energy is the same as in the
RIXS spectrum and the intensity of the coupling
between these excitations and the quasiparticles is 
encoded in the common  $\tilde{\lambda} =g^2N^*$ dimensional factor. 
%To be even more explicit, 
%the common effective coupling between excitations and quasiparticles
%can be represented by a single dimensionless fitting parameter $\lambda$
%[$g^2N^*=\lambda(g_0^2N^*)$] 
%that we will explicitly append in front of the self-energy for clarity.
We thus obtain the imaginary part of the quasiparticle self-energy
%\begin{eqnarray}\label{sigma2}
%  \lambda \Sigma''(\omega,T)&=&-\lambda \int d\omega' \left\lbrace \lambda_{CDF}ImD_{CDF}(\omega'-\omega) \right.\nonumber \\
%  &+&\lambda_{PH}ImD_{PH}(\omega'-\omega) \\
%    &+&\left.\lambda_{PM}ImD_{PM}(\omega'-\omega) \right\rbrace
%  \left\lbrack f(\omega')+b(\omega'-\omega)\right\rbrack \nonumber
%\end{eqnarray}
\begin{eqnarray}\label{sigma2}
 &&\Sigma''(\omega,T)\equiv \tilde{\lambda} \tilde\Sigma''(\omega,T)  \\
 &=&-\tilde{\lambda} \int d\omega' \left\lbrace D''_{CDF}(\omega'-\omega) +\frac{W_{PH}}{W_{CDF}}D''_{PH}(\omega'-\omega) \right. \nonumber\\
    &+&\left. \frac{W_{PM}}{W_{CDF}}D''_{PM}(\omega'-\omega) \right\rbrace
  \left\lbrack f(\omega')+b(\omega'-\omega)\right\rbrack \nonumber \,.
\end{eqnarray}
Here, $ImD_{CDF}(\omega)$ is obtained from Eq. (\ref{eq:cdf}) for ${\bf q}={\bf Q}_c$, 
while $ImD_{PH}(\omega)$ for the phonon and its overtone is
represented by a Lorentzian with frequency $\omega_{PH}$ (or $\sim 2\omega_{PH}$ for the overtone) and damping 
$\gamma_{PH}$. This has the following form- 
\begin{equation}
    D''_{PH} = \frac{\gamma_{PH}}{(\omega-\omega_{PH})^2 + \gamma_{PH}^2} - \frac{\gamma_{PH}}{(\omega+\omega_{PH})^2 + \gamma_{PH}^2} .
\end{equation}
For the paramagnon background we consider
\begin{eqnarray}
D''_{PM}(\omega)=\frac{1}{2{J_{PM}}}{C}(\omega)\left\lbrack \mbox{tanh}
(\frac{\omega-\omega_{PM}}{\gamma_{PM}}) \right.\nonumber \\
\left.+\mbox{tanh}(\frac{\omega+\omega_{PM}}{\gamma_{PM}})\right\rbrack
\end{eqnarray}
with a low-energy cutoff phenomenologically given by 
\begin{equation}
C(\omega)=1+\frac{1}{2}\left\lbrack \mbox{tanh}
(\frac{\omega-\Omega_{c}}{\Gamma_{c}})-\mbox{tanh}(\frac{\omega+\Omega_{c}}{\Gamma_{c}})\right\rbrack \,.
\end{equation}
Here $J_{PM}$ is a characteristic energy scale related to these excitations, which we choose to be $100$ meV.

In Eq. (\ref{sigma2}) $f(\omega)$ and $b(\omega)$ denote Fermi and Bose function. 
%and the couplings $\lambda_{CDF},\, \lambda_{e-ph},\, \lambda_{PM}$ 
% weight the relative contributions from CDF, phonons (plus overtone) and paramagnon 
% background.
Within this scheme, the FL quasiparticles, which are the starting point of the following considerations, encode all the
 high-energy electron-electron processes leading to the formation of the Landau 
 FL with a concomitant mass enhancement $m\equiv m_{QP}=m_{el}/z_{HE}$ related to the 
 strongly correlated character of the cuprates ($z_{HE}$ here is the quasiparticle residuum 
 due to the high energy (HE) processes).
% Thus, our  the mass enhancement related to the stro
% $N^*\propto m_{FL}^*=m/z_{FL}$ therefore encodes the mass enhancement related to the 
% strongly correlated character of the cuprates, whereas $m^*=m^*_{FL}/z$ arises from the
These quasiparticles will then undergo further renormalization via low-energy processes 
mostly due to the CDF, but also phonons and paramagnons. This is precisely what we are 
interested in in order to account for the strange metal properties.

\subsection{Optical scattering rate and resistivity}
The optical scattering rate is extracted from the 
optical conductivity for which we outline details of the calculation 
in appendix \ref{app:opt} within the so-called Allen approximation.~\cite{allen} 
Within this scheme, exploiting the momentum independence of the scattering mechanisms, the vertex corrections are
discarded and the frequency dependent mass and scattering rate are computed. 
In particular, it can be shown that at sufficiently low frequencies
the optical scattering rate is related to the electron self-energy
via 
\[1/\tau(\omega,T)=2 \tilde{\lambda} \tilde\Sigma''(\omega,T)\]
and we adopt this relation in the following for our fitting procedure.

%Since it is found
%that this optical scattering rate and the frequency dependent quasiparticle scattering rate
%of Eq. (\ref{sigma2}) are very similar (within the Allen approach they actually coincide
%in the small frequency limit), for the sake of simplicity and to keep as low as possible
%the fitting parameters, we here report only the results of $\Sigma''(\omega,T)$.
%Notice finally, that it is straightforward to compare $\Sigma''(\omega,T)$ with
%$1/\tau(\omega)$: only the dimensionless $\lambda_\omega =g^2N^*/M$ %coupling 
%and the elastic scattering rate $1/\tau_0$ have to be used as fitting parameters.

Besides the inelastic part, the experimentally determined scattering 
rate contains an elastic contribution $1/\tau_0$ which depends on
disorder. Following Matthiessen rule we thus write
\begin{equation}\label{eq:ac}
\frac{1}{\tau_{exp}(T,\omega)}=\frac{1}{\tau_0}+\tilde{\lambda}_\omega \tilde\Sigma''(\omega,T)
\end{equation}
where $\Sigma''(\omega,T)$ is determined from the RIXS data while
$1/\tau_0$ and $\tilde{\lambda}_\omega$ are  fitting parameters.

Adopting Drude theory $\rho=\frac{m}{ne^2\tau}$ the experimental DC conductivity can be analogously fitted by
\begin{equation}\label{eq:dc}
\rho_{exp}(T)=\rho_0 + \tilde{\lambda}_T \tilde\Sigma''(\omega=0,T)
\end{equation}
with $\rho_0=m/(ne^2\tau_0)$ and $\tilde{\lambda}_T=\tilde{\lambda}_\omega m/(ne^2)$.
Thus, the fitting parameters are in principle related by
\begin{equation}\label{eq:consist}
\frac{\tilde{\lambda}_T}{\tilde{\lambda}_\omega}=\tau_0 \rho_0 =\frac{m}{ne^2}\,.
\end{equation}
However, since available data for DC and AC conductivity are usually
for different samples we point out that Eq. (\ref{eq:consist})
is not strictly obeyed in the following.

%To fit the resistivity we calculate the
%DC scattering rate $\frac{1}{\tau}(T)=\Sigma''(\omega=0,T)$. 
%In this case other parameters should be considered
%because $\rho(T)=\frac{m}{ne^2}\frac{1}{\tau}(T)$, with possible %additional temperature 
%dependencies in $m$ and $n$. Here we adopt the simplest possible 
%scheme and  we neglect these additional dependencies. The fitting of %then proceeds by identifying $\rho(T)=\rho_0+\lambda_T \Sigma''(\omega=0,T)$
%and using $\rho_0$ and $\lambda_T$ as fitting parameters.
%We then check that the value of $\lambda_T$ and $\lambda_\omega$ are consistent 
%once reasonable values for $m$ and $n=1+p$ are considered (together with
%the known interlayer distance $d$ of each cuprate material).

%[all the above should be discussed and checked....????
%Should we also discuss the issue of $m*=m/z$, $1/\tau=z \Sigma''$ ????
%add a note???? ]

\subsubsection{BSCCO}
We consider here lightly overdoped BSCCO. One such sample with doping $p = 0.19 \approx p^*$ 
and $T_c=82 K$ was investigated in Ref. \onlinecite{arpaia-2023} by RIXS. In particular [cf.
Fig. 1C of Ref. \onlinecite{arpaia-2023}] we report here 
in Fig. \ref{fig-rixs}(a) the experimental intensity of the RIXS 
spectrum at ${\bf q}={\bf Q}_c=(0.26,0) r.l.u.$. 
This spectrum (blue circles) is then fitted (blue line) summing the contribution 
of separate excitations: the orange line represents the elastic scattering of light 
(mostly due to surface effects), CDF (red line), phonon (mostly the bond-stretching phonon, green line), and its overtone (cyan line), while the broad continuum (magenta line) 
is usually attributed to paramagnons and particle-hole pair excitations. 
To fit the spectra, similarly to what has been reported 
in Ref. \onlinecite{arpaia-2023} with regard to the analysis of 
the RIXS data, 
all these excitations have been convoluted with a gaussian of width $30$ meV corresponding to the experimental resolution. Table \ref{tab:rixs-fit} reports the values of the fits for the
BSCCO, YBCO, and LSCO samples discussed in this section.

\begin{table}[]
    \centering
    \begin{tabular}{c|c|c|c|c|c}
      & & BSCCO  & YBCO & LSCO & \\
       \hline 
      & $W_{CDF} $  &3.9 &11.0 & 2.5 &\\
 CDF & $M [meV]$       &10.0 &8.0 &7.0 &\\
   & ${\bar \Omega} [meV]$ & 50 & 60 & 50 &\\ 
 & $\gamma$ & 1 & 1 & 1 &\\ 
   & & & & \\
       \hline 
       & $W_{PH}$  &5.3 &6.3 & 3.2 & \\
  PHONON & $\omega_{PH} [meV]$ &58.0 &45.0 &65.0 &\\
  & $\gamma_{PH} [meV]$ &9.0 &4.0 &9.0& \\  
& & & &   &\\
       \hline 
       & $W_{PH,OT} $  & 0.8 & 0.0 & 0.5& \\
 OVERTONE & $\omega_{PH,OT} [meV]$ & 140& - & 150& \\
 & $\gamma_{PH,OT} [meV]$ &18 & - & 18& \\  
 & & & & &\\
        \hline 
 & $W_{PM}$  &21.0 &20.0 &22.0 &\\
PARAMAGNON & $\omega_{PM} [meV]$ &90.0 &40.0 &40.0& \\
 & $\gamma_{PM} [meV]$ &110.0 &70.0 &70.0& \\
& $\Omega_{c} [meV]$  & 40 & 40 & 40 &\\
& $\Gamma_{c} [meV]$  & 0.1 & 0.1 & 0.1 &\\
 & & & & &\\
       \hline 
$\tilde{\lambda}_\omega$ & [meV]  &2.25   & 1.0 &3.0& \\
$\tilde{\lambda}_T$ & [$10^{-4}\Omega cm$]  &0.172   & 0.085 & 0.120 &
    \end{tabular}
    \caption{Parameters for fitting the individual contributions of the  RIXS spectra of BSCCO, YBCO, and LSCO
    as described in the text.}
    \label{tab:rixs-fit}
\end{table}

\begin{figure}[thb]
    \centering
    \includegraphics[width=1.0\linewidth]{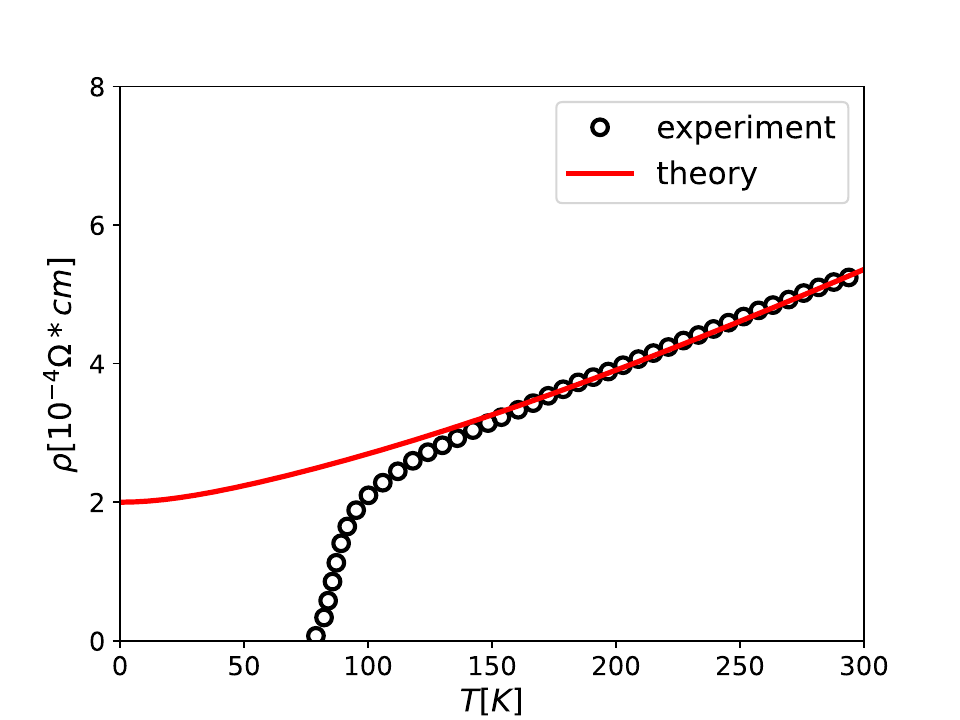}
    \includegraphics[width=1.0\linewidth]{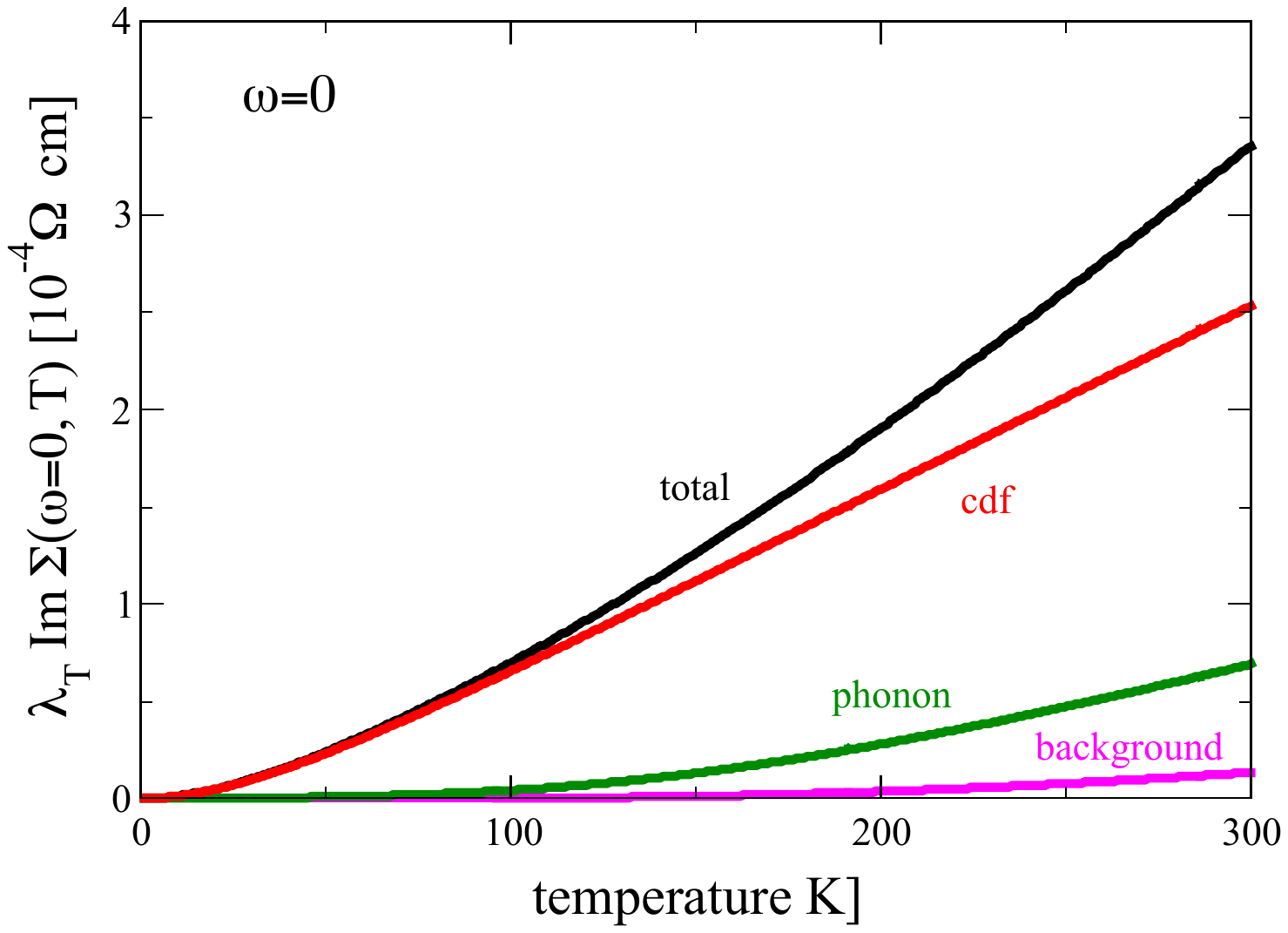}
    \caption{Top panel: DC resistivity data (circles) for an optimally doped Bi2212 sample ($T_c=87 K$) from Ref. \onlinecite{marconi-2012}.
    The fit from Eq. (\ref{eq:dc}) is shown by the red line with
    $\tilde{\lambda}_T=0.172\cdot 10^{-4}\Omega cm$ and $\rho_0=2.0\cdot 10^{-4}\Omega cm$. 
    Lower panel: Individual contributions from
    CDF (red), phonons (green) and background (magenta) to the total inelastic part $\lambda_T \Sigma''(\omega=0,T)$ (black).} 
    \label{bscco-rho}
\end{figure}

According to the previously discussed strategy, we now take all the
excitations (with their relative intensity) and insert them
into Eq. (\ref{sigma2}) to calculate
the imaginary part of the quasiparticle self-energy $\Sigma''(\omega,T)$.
We first fit the DC conductivity for an optimally doped Bi2212 sample \cite{marconi-2012} with $T_c=87 K$, close to the 
doping of the sample for which RIXS data have been analyzed
in Fig. \ref{fig-rixs} (a).  
From Eq. (\ref{eq:dc}) we find that $\tilde{\lambda}_T=0.172\cdot 10^{-4}\Omega cm$ and $\rho_0=2.0\cdot 10^{-4}\Omega cm$ 
yields an excellent fit, cf. Fig. \ref{bscco-rho}, to the data 
for $T > T_c$.

The lower panel in Fig. \ref{bscco-rho} reports the individual contributions to the temperature dependent resistivity. In general,
a mode with characteristic frequency $\Omega$ imprints a linear
temperature dependence onto the resistivity for $k_BT>\Omega$, whereas
FL behavior $\rho\sim T^2$ is recovered for $k_BT<\Omega$. For the
present case CDF's provide a linear temperature dependence for
$k_BT>\omega_{CDF}\sim M/\gamma
\sim 10$ $ meV \sim 100$ K, whereas the phonon 
contributes a Fermi liquid dependence to $\Sigma''(\omega=0,T)$
up to $\omega_{ph}=58 meV \sim 670 K$ ($1600 K$ for the overtone).
The temperature dependence from the background is also Fermi liquid like, however, it only contributes a minor fraction to the 
self-energy at $\omega=0$. Altogether, the temperature dependent
resistivity of optimally doped BSCCO is dominated by the linear-in-T
contribution from CDF's with a small FL liquid correction arising
from (bond-stretching) phonons.

We proceed by calculating for various $T$ the frequency dependent quasiparticle self-energy and we use it to fit the optical spectral scattering rate of a slightly overdoped sample
(doping $p=0.198$, $T_c=80 K$) from Ref. \onlinecite{park-2025}.

\begin{figure}[thb]
    \centering
    \includegraphics[width=1.0\linewidth]{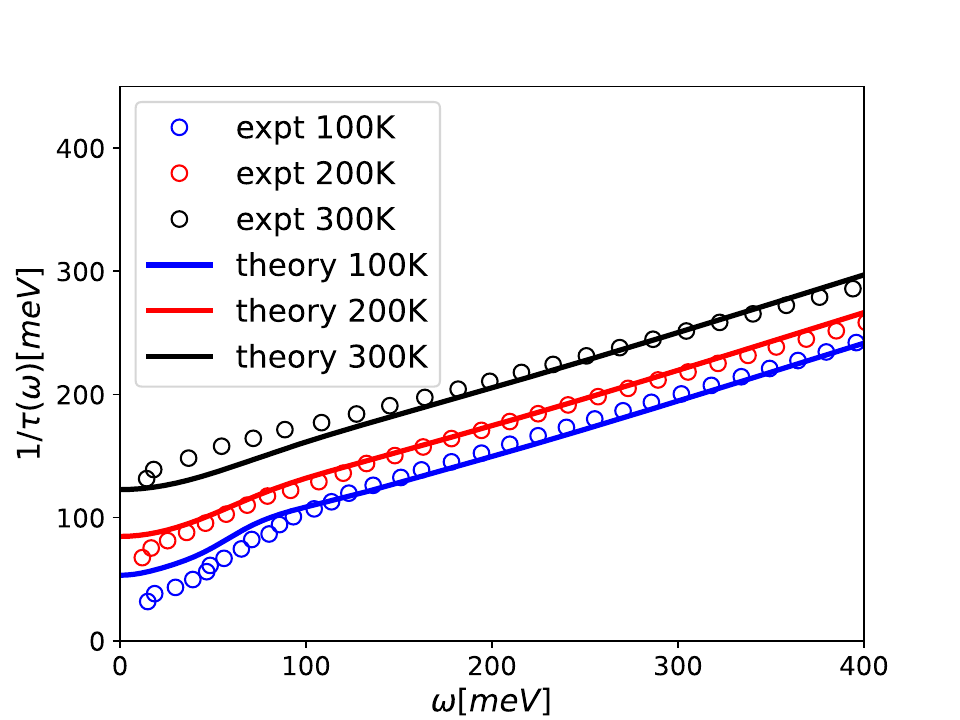}
    \includegraphics[width=1.0\linewidth]{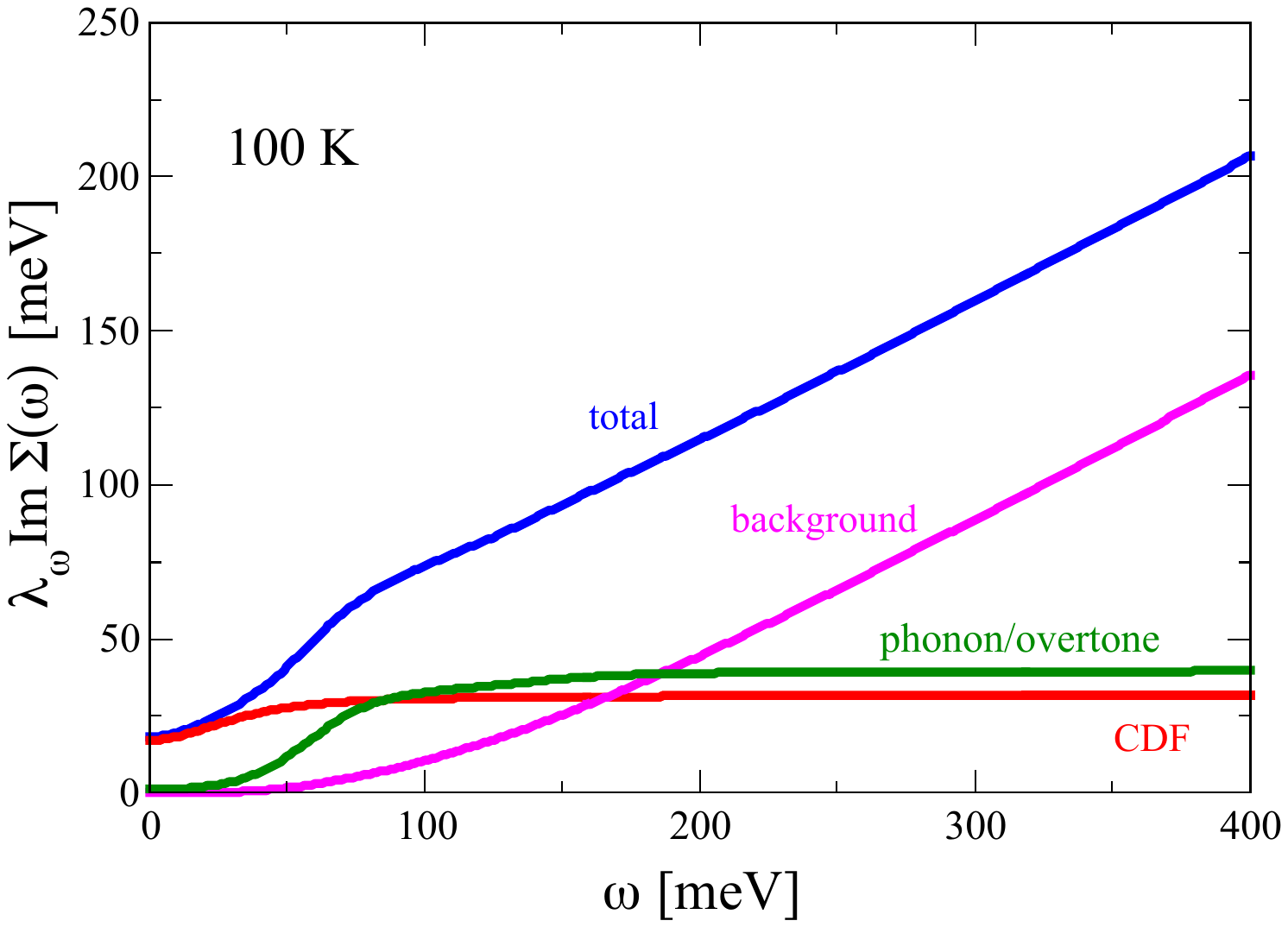}
    \caption{Top panel: Optical scattering rate for an optimally
    doped BSCCO sample ($T_c=82$ K) from Ref. \onlinecite{park-2025} at $100 K$ (black circles), $200 K$ (red circles), and $300 K$ (blue circles) compared to the theoretical fits (lines).
    Parameters: $\lambda_\omega=2.25 $ meV, $1/\tau_0=35 $ meV.
    Lower panel: Individual contributions to the $100 K$ fit of
    $\tilde{\lambda}_\omega \Sigma''(\omega,T)$(blue) from
    CDF (red), phonons (green), and background (magenta).}
    \label{bscco-optics}
\end{figure}
The result is reported in Fig.\ref{bscco-optics}a
and we find that the fits not only reproduce the linearity in the frequency dependence
of $\tau^{-1}(\omega)$, but also partially capture its low-frequency suppression. This is highly
non trivial since the data are taken at temperatures well above $T_c$ (particularly for those at $200$ K),
so that this suppression can hardly be related to the gap or to superconducting fluctuations. The reason for this suppression becomes apparent when we again split the total scattering rate into the individual
contributions from CDF, phonons, and background, cf. Fig. \ref{bscco-optics}b. First, we see that the linear-in-$\omega$
dependence is completely provided by the flat part of the background, similar to a MFL in the limit $\omega > k_BT$.
Excitations which are centered around some energy $\Omega$
provide a constant contribution to $1/\tau(\omega)$ for
$\omega>\Omega$ whereas they cannot scatter at lower energies.
In this regard the major contribution to the suppression of the
scattering rate at lower energies is due to the bond-stretching phonon at nearly $70 $ meV and
a smaller contribution is also coming from the CDF's.
A conserving calculation of the optical conductivity in the presence of CDW
was performed in Ref. \onlinecite{enss-2007}. Recently the same procedure was carried out with CDF in 
Ref. \onlinecite{spera-2026}.
This more complete treatment does not change our conclusions.
\begin{figure}
    \centering
    \includegraphics[width=1.0\linewidth]{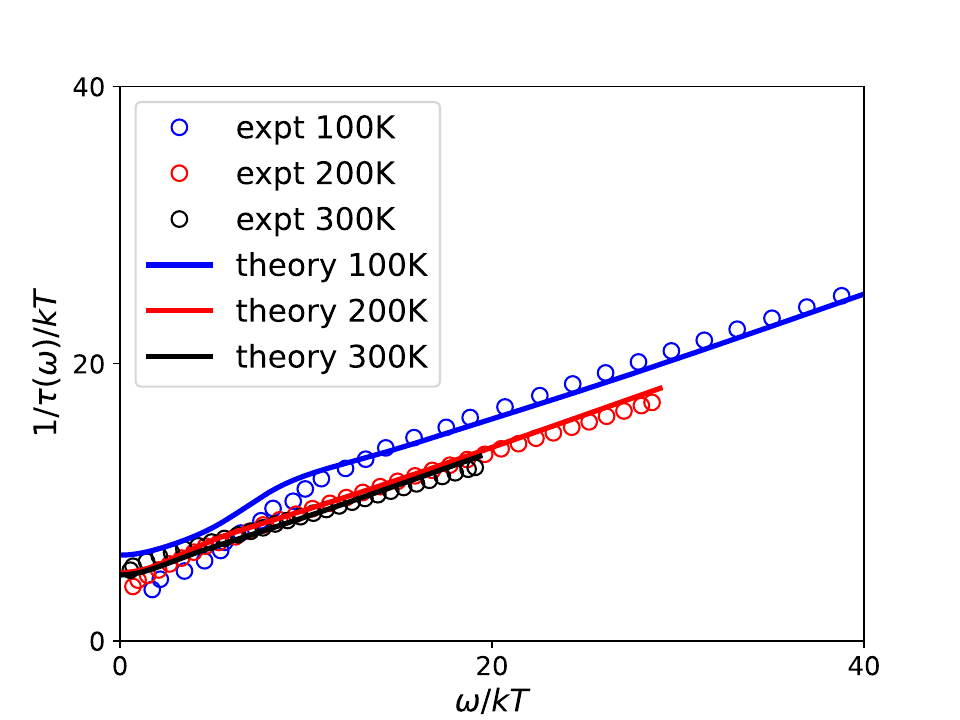}
    \caption{$\omega/T$ scaling of the optical scattering rate for an optimally
    doped BSCCO sample ($T_c=82$ K) from Ref. \onlinecite{park-2025} at $100 K$ (black circles), $200 K$ (red circles), and $300 K$ (blue circles) compared to the theoretical fits (lines).
    Parameters: $\tilde{\lambda}_\omega=2.25 $ meV, $1/\tau_0=35 $ meV.}
    \label{bscco-scaling}
\end{figure}
Finally, Fig. \ref{bscco-scaling} reports the scaling properties of the fitted optical scattering rates as compared to the
experimental data from Ref. \onlinecite{park-2025}. It is rather non-trivial that our fitted scattering rates show a very good scaling 
and agreement with the experimental data. It is important to stress again that, within our
approach, the linear-in-$T$ dependence is provided by the CDF's whereas the linear-in-$\omega$ dependence is coming from the background, their relative intensities being completely determined by the RIXS data. As has been shown in Ref. \onlinecite{SFLoptics-2024}
both scattering mechanisms can combine to an (approximate) scaling
form, whereas phonons obviously spoil the scaling properties
in the relevant frequency range.
Interestingly, however, the features due to phonons seem to be also present in real spectra
and improve the agreement with the data.
The issue of scaling is discussed in detail in Sec. \ref{sec:2d}.
%\textcolor{red}{ ELIMINATE THIS PART? I WOULD AGREE WITH CARLO...
%Approximating the background by a constant intensity, starting at frequency $\Omega_b$ the self-energy (scaled with $\beta=1/k_BT$)
%can be obtained $\beta\Sigma_b(\omega,k_BT) \sim \beta\omega-2ln(\beta\Omega_b)$ in the relevant frequency and temperature range $\beta\Omega_b <1$ and
%$\beta\omega>1$. Thus scaling is violated due to the low frequency
%cutoff $\Omega_b$ which leads to an increase of $\beta\Sigma_b(\omega,k_BT)$  with increasing temperature.
%On the other hand, approximating the CDF spectral function by
%a constant ranging from $\Omega_< < \omega < \Omega_>$ 
%and taking again the limit $\beta\Omega_< <1$, $\beta\Omega_> >1$, $\beta\omega>1$, one obtains $\beta\Sigma_{CDF}(\omega,k_BT) \sim
%\beta\Omega_> -\ln(1-e^{-\beta(\omega-\Omega_>)})$. Thus, in this case one finds a decrease of 
%$\beta\Sigma_{CDF}(\omega,k_BT)$ with increasing temperature which can (at least partially) compensate
%the opposite effect from the background.
%Therefore, in our scenario the critical doping for scaling
%is non-universal in the sense that it requires an optimal
%combination of background and CDF together with a minimum
%contribution of phonons.}

%=====================================
%=====================================

\subsubsection{YBCO and YCBCO}
\begin{figure}
    \centering
    \includegraphics[width=1.0\linewidth]{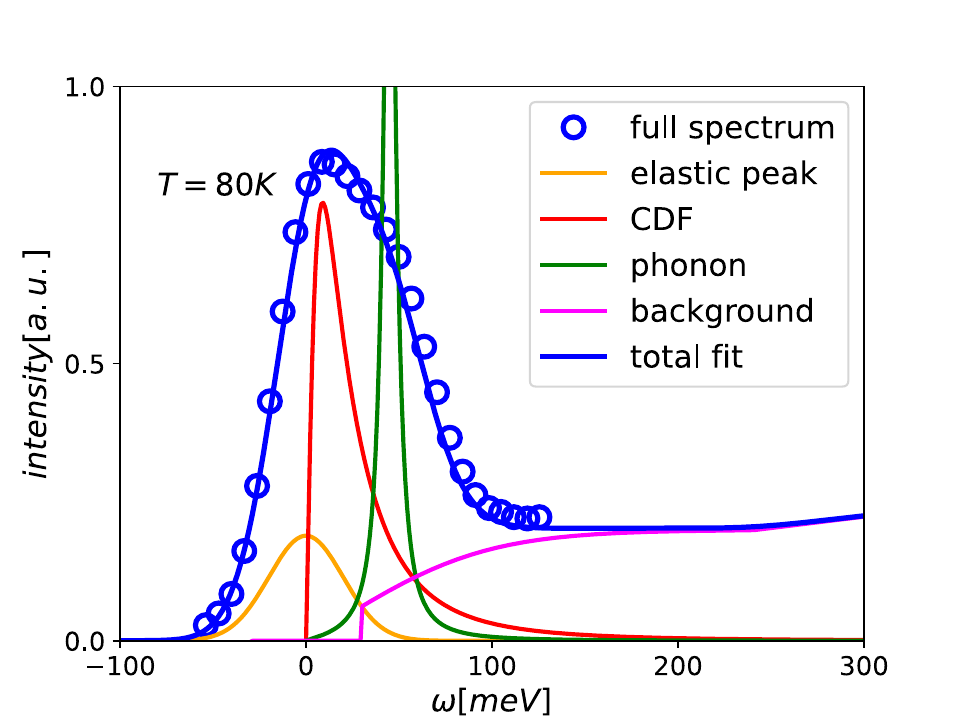}
    \caption{RIXS spectrum at ${\bf q}\approx {\bf q}_{CDF}=(0.31,0) [r.l.u.]$
    for a YBCO sample with doping $p=0.185$ and $T_c=85$ K from Ref. \onlinecite{arpaia-2023} (blue circles). Also shown is the decomposition into elastic (orange), CDF (red), phonon (green),  and background (magenta) contributions. }
    \label{ybco-rixs}
\end{figure}

\begin{figure}
    \centering
    \includegraphics[width=1.0\linewidth]{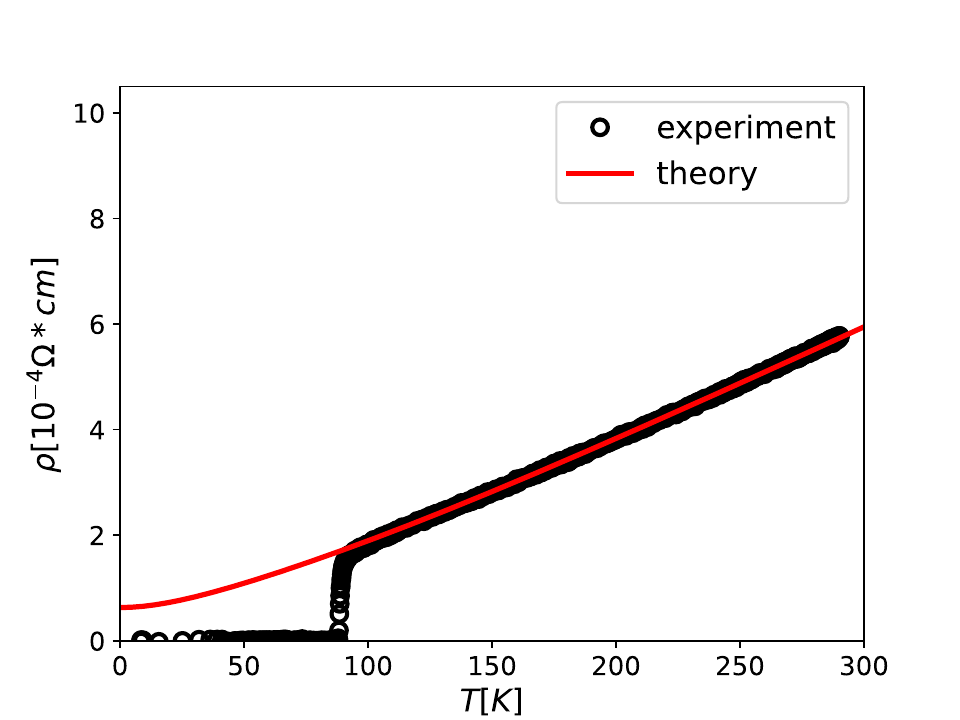}
    \caption{DC resistivity data (circles) for a YBCO sample ($T_c=85 K$) from Ref. \onlinecite{arpaia-2019}.
    The fit from Eq. (\ref{eq:dc}) is shown by the red line with
    $\tilde{\lambda}_T=0.085\cdot 10^{-4}\Omega cm$ and $\rho_0=0.68\cdot 10^{-4}\Omega cm$.}       
    \label{ybco-rho}
\end{figure}

The same procedure used above for BSCCO, is now adopted for YBCO.
Also in this case the RIXS spectrum of the excitations is 
taken from experiments [see Supplementary Fig. 9(b) of Ref. \onlinecite{arpaia-2023}] for a slightly overdoped sample with p=0.185 doping ($T_c=85$ K). This spectrum is reported in Fig.\ref{ybco-rixs} (a).
%We also considered a simpler RIXS spectrum
%where the buckling phonon (orange line in Fig. \ref{ybco-rixs} (a)) is neglected
%and only the much stronger bond-stretching phonon (red line!!!!!!!!!!!) is kept.
In this case resistivity data are available for the same sample \cite{arpaia-2019} which are shown in Fig. \ref{ybco-rho} together with the fit obtained from Eq.(\ref{sigma2}). Excellent agreement is obtained for the whole measured temperature range above $T_c$.

\begin{figure}
    \centering
    \includegraphics[width=1.0\linewidth]{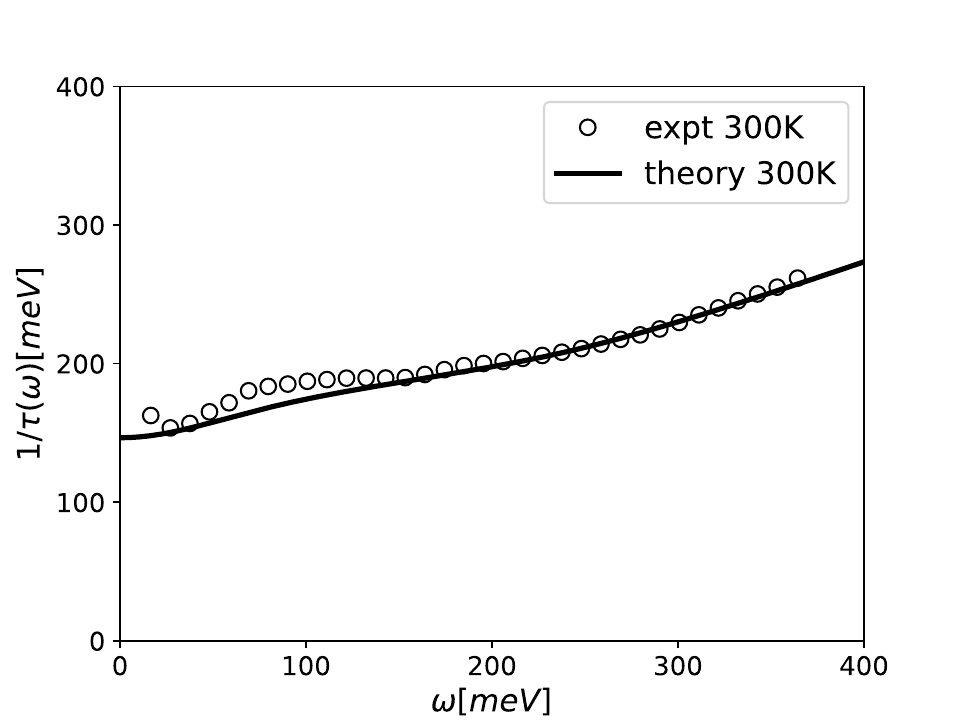}
    \caption{Optical scattering rate for a YCBCO sample (T$_c = 75$ K) from Ref. \onlinecite{uykur-2011} at 300K (black circles), compared
to the theoretical fit (black line). Parameters: $\tilde{\lambda}_\omega=1.0 $ meV, $1/\tau_0=10 $ meV.}
    \label{ybco-optics}
\end{figure}

Finally we used the RIXS spectra to fit the optical scattering rate of Ref. \onlinecite{uykur-2011} of a Y$_{1-x}$Ca$_x$Ba$_2$Cu$_3$O$_{7-\delta}$  
(YCBCO) sample where a substitutional amount $x=0.11$ of Ca was introduced to obtain an overdoped YBCO sample with $p=0.2$ and $T_c=75$ K. Despite the 
differences in doping and structure between the YBCO sample of RIXS and $\rho(T)$ 
and YCBCO used for optical data, at 300 K the fitting is quite satisfactory (black line
in Fig. \ref{ybco-optics}. 
We point out that in order to account for the change of slope in $1/\tau(\omega)$ at high energy an additional contribution
$C(\omega)(\omega-240 meV)/(250 meV*2J_{PM})$ has to be added to the
paramagnetic background, which is beyond the RIXS high-resolution data
reported in Fig. \ref{ybco-rixs}.
Moreover, it seems that there is also some additional weight at
$\sim 100 meV$, which becomes even more pronounced at lower temperature.
We comment on this issue in appendix \ref{ap:ybco}, however, we are not aware
of other optical scattering rate data for overdoped Y(C)BCO so that this feature obviously calls for further investigation.

%the feature at $\sim 100 meV$, which becomes even more pronounced at %lower temperature, is not captured by our fitting
%and could in principle be accounted for by adding a 'unknown' excitation
%to the RIXS spectrum. Since we are not aware
%of other optical scattering rate data for overdoped Y(C)BCO this feature %obviously calls for further investigation.

\subsubsection{LSCO}
Here we analyze optical and resistivity data for LSCO from Ref. \onlinecite{michon-2023} at doping $p=0.24$. For this very overdoped system we are not aware of high-resolution
RIXS spectra for LSCO, so that we proceed differently: We assume a RIXS spectrum using the BSCCO
data as a "starting point" model and adjust the intensities and the energies of the peaks to fit 
the optical scattering rates. Finally, we check the consistency by fitting the temperature dependence of the resistivity. 
\begin{figure}
    \centering
    \includegraphics[width=1.0\linewidth]{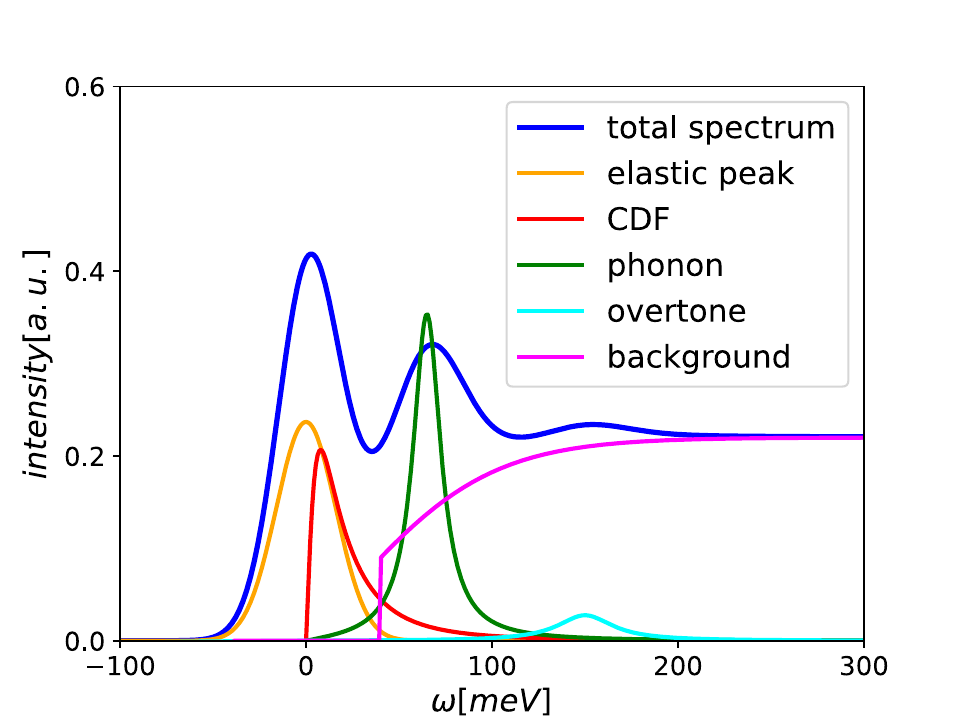}
    \caption{Hypothetical RIXS spectrum for slightly overdoped LSCO obtained by modifying the
    experimental data shown in Fig. \ref{fig-rixs}(a) for Bi2212. Orange, red, green, magenta, and  cyan lines refer to the decomposition into the contributions from elastic peak, CDF, phonon, background and overtone.}
    \label{lsco-rixs}
\end{figure}
Fig. \ref{lsco-rixs} reports the final appearance of the putative RIXS spectrum, where
we can again see the presence of an elastic peak (orange line), a red line marking the CDF 
peak with a characteristic energy $M/\gamma\approx 10$ meV, a bond-stretching
phonon at $\omega_{ph} \approx 65$ meV (green line), and a magenta line to describe 
a flattish continuum starting around $40$ meV representing the high-energy paramagnons and p-h pairs.

\begin{figure}
    \centering
    \includegraphics[width=1.0\linewidth]{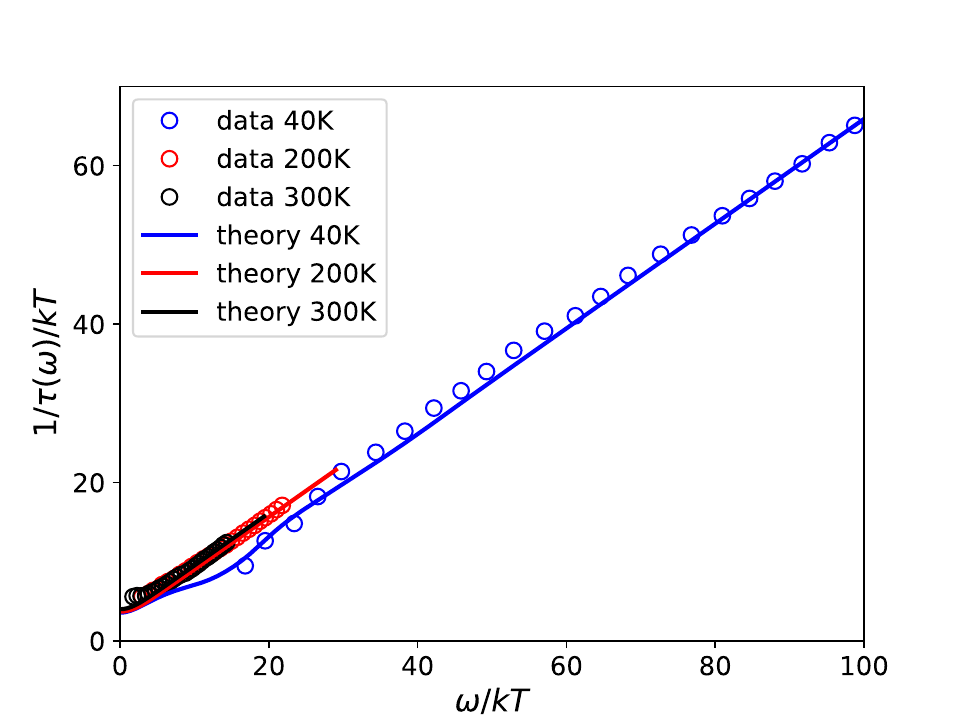}
    \caption{ %Optical scattering rate for a LSCO sample ($p=0.24$, $T_c=19 K$) 
    %from Ref. \cite{michon-2023}. Parameters: $\lambda_\omega=3.0 $ meV, $1/\tau_0=0 $ meV.}
    Scaled optical scattering rate for a LSCO sample with $p=0.24$ and $T_c=19 K$. The data are
    from Ref. \onlinecite{michon-2023}. Parameters: $\tilde{\lambda}_\omega=3.0 $ meV, $1/\tau_0=6$ meV.}
    \label{lsco-optics}
\end{figure}

\begin{figure}
    \centering
    \includegraphics[width=1.0\linewidth]{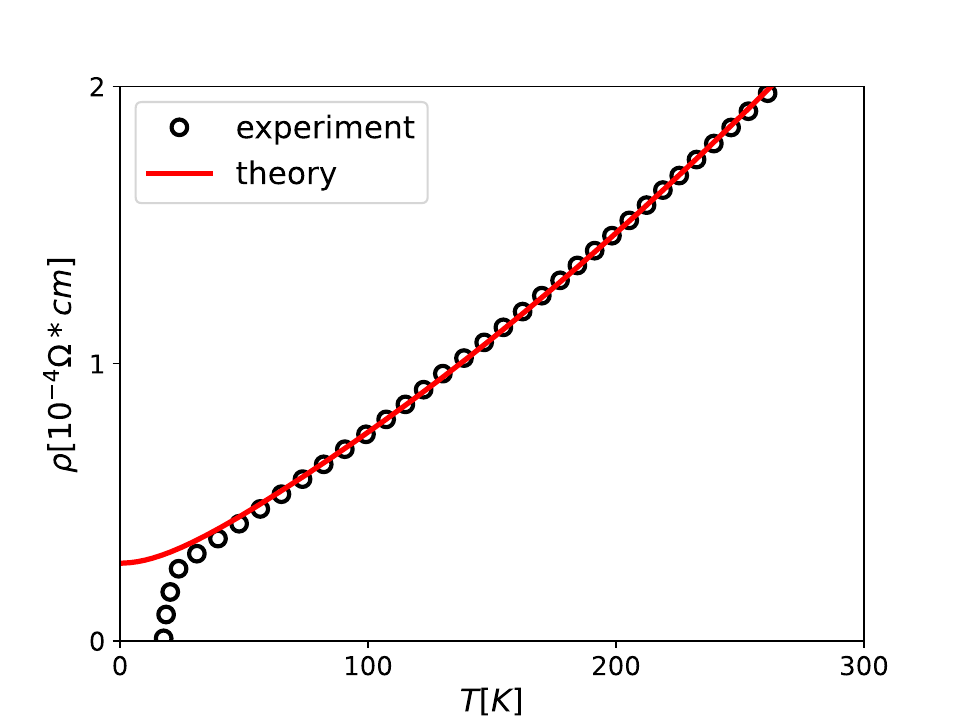}
    \caption{DC resistivity data (circles) for a LSCO sample ($p=0.24$, $T_c=19 K$) from Ref. \onlinecite{michon-2023}.
    The fit from Eq. (\ref{eq:dc}) is shown by the red line with
    $\tilde{\lambda}_T=0.120\cdot 10^{-4}\Omega cm$ and $\rho_0=0.28\cdot 10^{-4}\Omega cm$.}
    \label{lsco-rho}
\end{figure}

Fig. \ref{lsco-optics} demonstrates that computing the scaled optical scattering rate from 
this "hypothetical" RIXS spectrum using Eq. (\ref{eq:ac}) gives an
excellent account of the experimental data from Ref. \onlinecite{michon-2023}. 
It is worth remembering that we fitted all the data at different temperatures with just two parameters:
the overall dimensionless coupling $\lambda_\omega$ between all the bosons 
(the relative intensity is obviously encoded in the shape of the
RIXS spectrum) and the fermions and the
elastic scattering rate $1/\tau_0$. Of course the good fitting at the various different temperatures
also reproduces well the scaling of the scattering rate reported in Fig. \ref{lsco-optics}.
Thus we notice that the simple perturbative use in Eq. \ref{sigma2} of a standard RIXS spectrum
with CDF, phonons, and paramagnon quite similar to the one explicitly observed in other cuprates, 
fits well the optical data and their scaling feature
and also the resistivity data extracted from Fig. 2(a) of
Ref. \onlinecite{michon-2023}, see Fig. \ref{lsco-rho}.

\subsection{Raman spectroscopy}
Similar to the optical conductivity, Raman scattering in the normal state of high-T$_c$ cuprates 
allows for the investigation of the frequency dependence of the quasiparticle relaxation time. 
However, upon selecting incident and scattered light polarizations parallel ($B_{1g}$) and 
at $45^\circ$ ($B_{2g}$) to the Cu-O bond direction, one can additionally probe 
$\tau(\omega)$ in different regions of momentum space. 
This has been used in Ref. \onlinecite{caprara11} in order to distinguish the contributions 
from spin-density wave (SDW) and CDW scattering across the phase diagram of LSCO superconductors. 
Due to vertex cancellations it was shown that spin collective modes dominate the Raman 
response in the B$_{1g}$ channel whereas collective CDW modes determine the 
structure for $B_{2g}$ symmetry. Moreover, by substracting the constant background the 
collective modes become  
visible in the Raman spectra as low energy peaks which increase in intensity 
and soften with decreasing temperature. In this way it is
possible \cite{suppa05} to extract the critical behavior of the
collective modes in the respective channels.

Here we are concerned with Raman spectra for the overdoped samples 
for which the RIXS spectra were discussed in Sec. I.B and II.A
and which at low energy are dominated by CDF's rather than CDW's.
Together with the constant background arising from 
particle-hole paramagnon
excitations we will address the following question: Are the normal state Raman spectra 
in the different symmetry channels compatible with CDF excitations and a background which are
almost constant in momentum space? Since the Raman form factors for the different 
symmetries probe different regions in the Brillouin zone, a 
consistent description of the corresponding spectra is
significantly more constrained than the fitting of optical data only.

Here we evaluate the Raman response function in analogy to the
approach of Allen for the optical conductivity \cite{allen}, but taking into account the Raman vertex $g_{k,s}$, see appendix \ref{ap:raman}. For the considered symmetries we have
$\gamma_{k,B_{1g}}=(\cos(k_x)-\cos(k_y))/2$ and $\gamma_{k,B_{2g}}=\sin(k_x) \sin(k_y)$ arising from symmetry projections for the D$_{4h}$ point group
relevant for the cuprates \cite{dev07}. Since the matrix elements for Raman scattering 
in a correlated system involve a plethora of transitions between many body states\cite{dev07}
it is difficult to predict the relative intensities between the different symmetries. Instead we estimate
these contributions scaling the calculated Raman intensities in $B_{2g}$ symmetry by 
an additional factor of $\alpha$. From the Raman vertex $g_{k,s}$ one can define an energy dependent 
vertex function $N_s(\omega)$, see Eq. (\ref{eq:vertex}),
which depends on the underlying band structure. For the present calculations we choose a 
single-band tight-binding model with nearest- ($t$) and next nearest ($t'$) neighbor hoppings. 
We fix $t=300 $ meV and allow $t'/t$ to vary between the different
materials.

\begin{figure}
    \centering
    \includegraphics[width=1.0\linewidth]{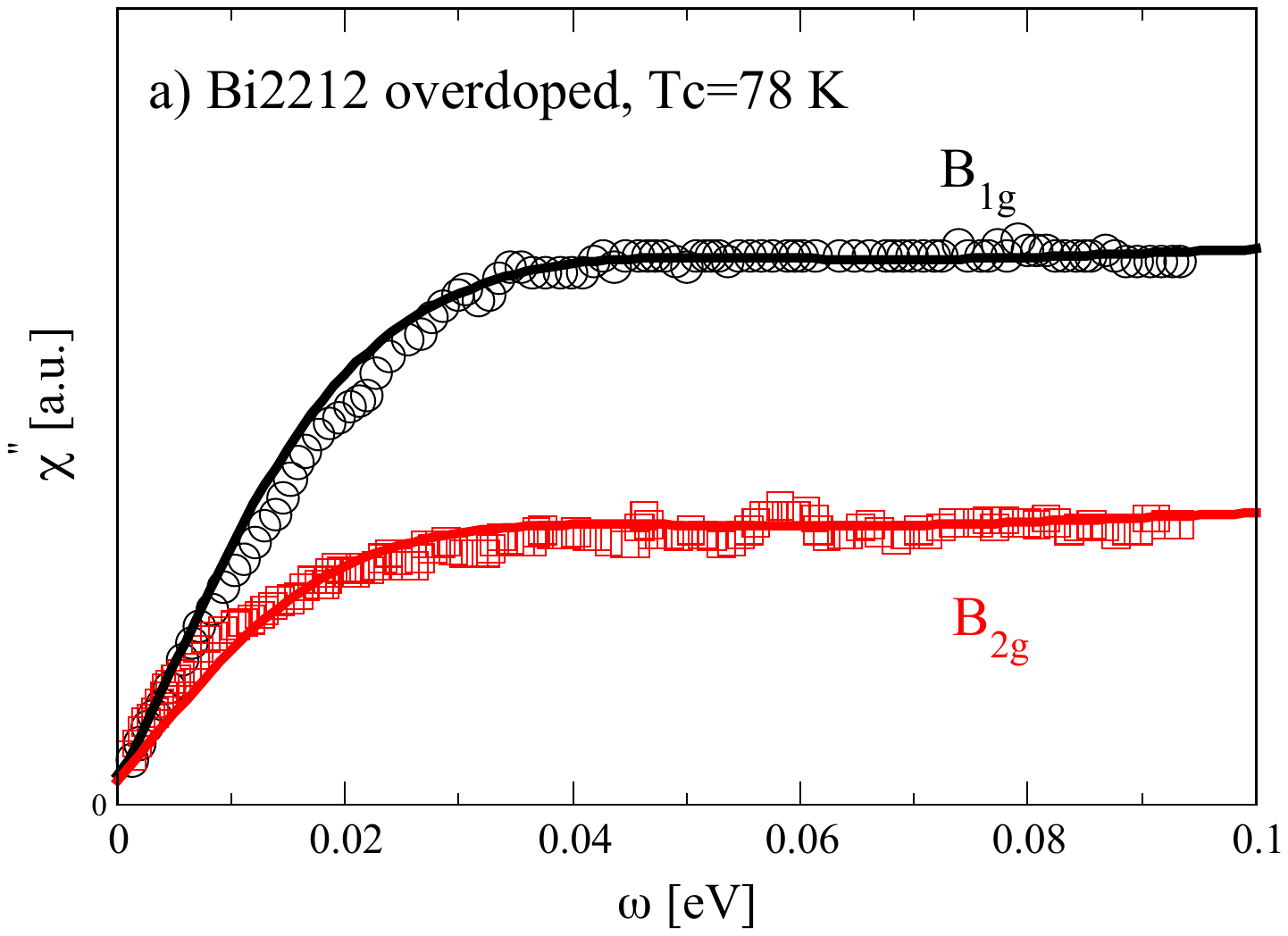}
    \includegraphics[width=1.0\linewidth]{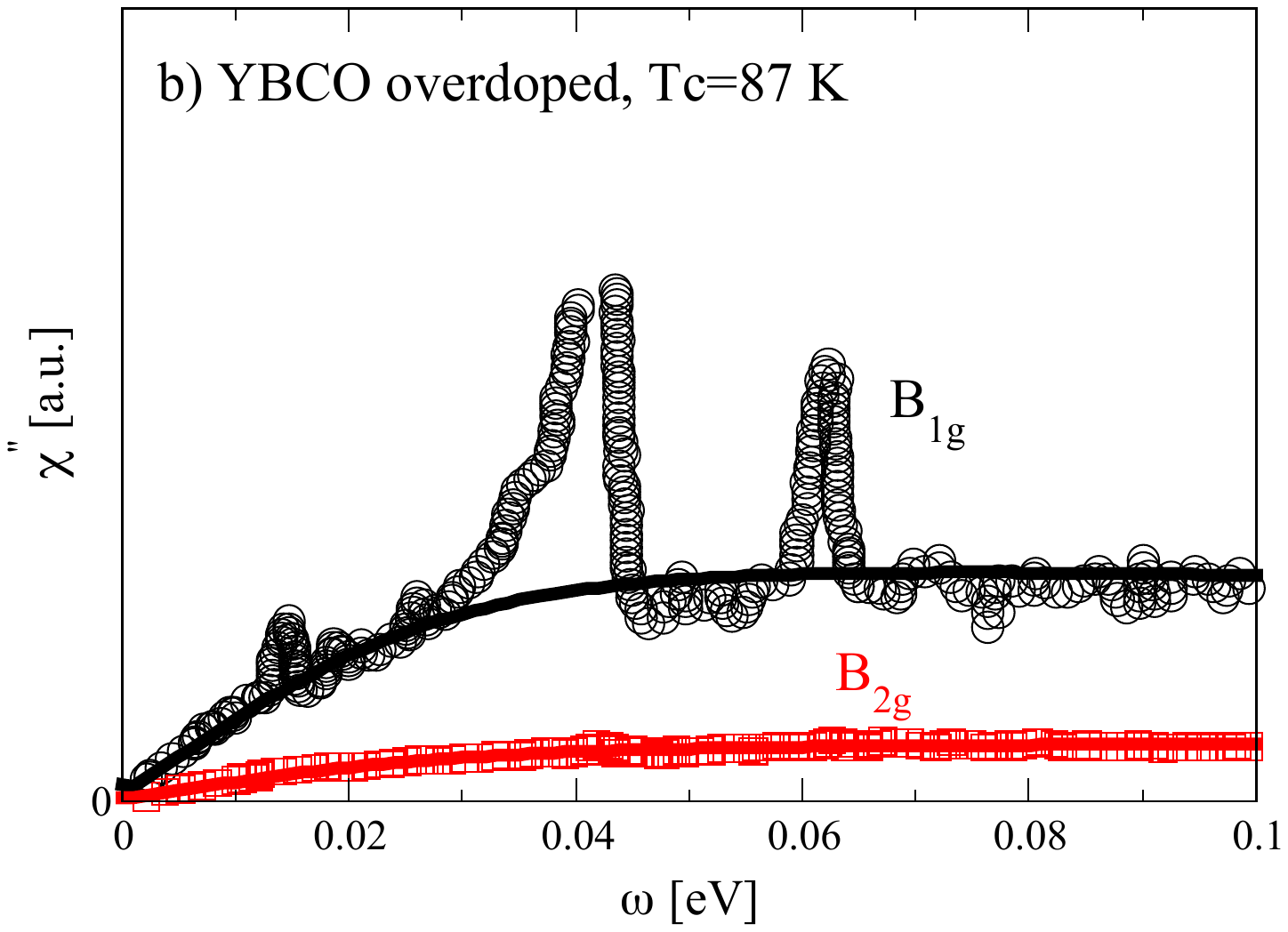}
    \includegraphics[width=1.0\linewidth]{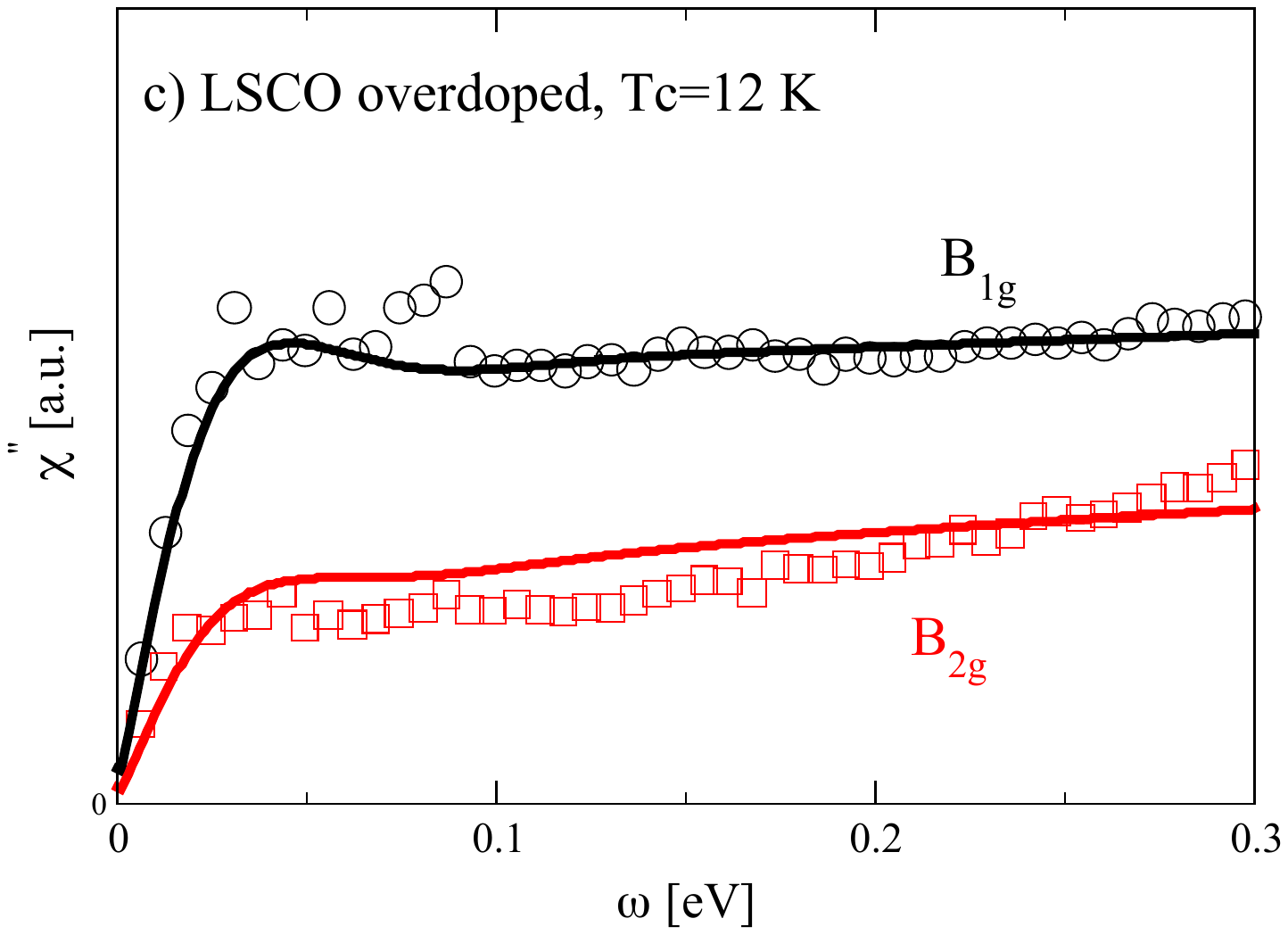}    
    \caption{Fit (full lines) and experimental data (circles) of $B_{1g}$ (black) and $B_{2g}$ (red) Raman spectra of Bi2212 \cite{venturini02} (a), YBCO \cite{opel00} (b) and LSCO \cite{caprara11} (c). In all spectra the B2g intensity is scaled 
    by an additional factor $0.8$. Parameters in (a): $t=300 $ meV, $t'/t=-0.4$, $1/\tau_0=10 $ meV, $\tilde{\lambda}_\omega=3.5 $ meV, $T=90 K$, $\alpha=0.8$, doping $p=0.2$. Parameters in (b): $t=300 $ meV, $t'/t=-0.3$, $1/\tau_0=18 $ meV, $\tilde{\lambda}_\omega=2 $ meV, $T=248 K$, $\alpha=0.45$, doping $p=0.185$. Parameters in (c): $t=300 $ meV, $t'/t=-0.2$, $1/\tau_0=30 $ meV, $\tilde{\lambda}_\omega=5 $ meV, $T=100 K$, $\alpha=0.8$, doping $p=0.25$.}
    \label{fig:raman}
\end{figure}

Fig. \ref{fig:raman} shows the Raman spectra for $B_{1g}$ and
$B_{2g}$ symmetry in Bi2212 \cite{venturini02}, YBCO \cite{opel00}, 
and LSCO \cite{caprara11} which are close in $T_c$ to the samples for which the RIXS spectra have been discussed in Sec. \ref{sec:sfl}. For all cases we obtain a very good fitting of the data, in particular for the case of Bi2212 (panel a).
For YBCO the $B_{1g}$ spectrum is dominated by two pronounced phonon modes,
one at $40 $ meV attributed to antisymmetric bending mode of in-plane oxygens along the c-axis \cite{feile89}, the other at $62 $ meV due to
c-axis vibration of the apical oxygen \cite{thomsen88,feile89}. While these are not captured by our underlying bosonic (RIXS) spectrum we can see that again the
electronic contribution can be well fitted in both symmetry channels.

\subsection{Some general features of the scattering rate: The issue of scaling}\label{sec:2d}
In section I.B.3 we already emphasized that the temperature and frequency 
dependencies of the imaginary part of the electron self-energy can be roughly described
by a simple analytic expression, Eq. (\ref{SFLapp}), which is valid for small 
$M/\gamma$ and closely resembles the MFL form of Eq.(\ref{MFLapp}).
%It was also anticipated that Eq. (\ref{SFLapp}) stems from two separates sources of scattering:
%the low energy CDF having $M/\gamma<T$ and the high energy paramagnon 
%for $\omega>(M/\gamma,T)$.

\begin{figure}
    \centering
    \includegraphics[width=1.\linewidth]{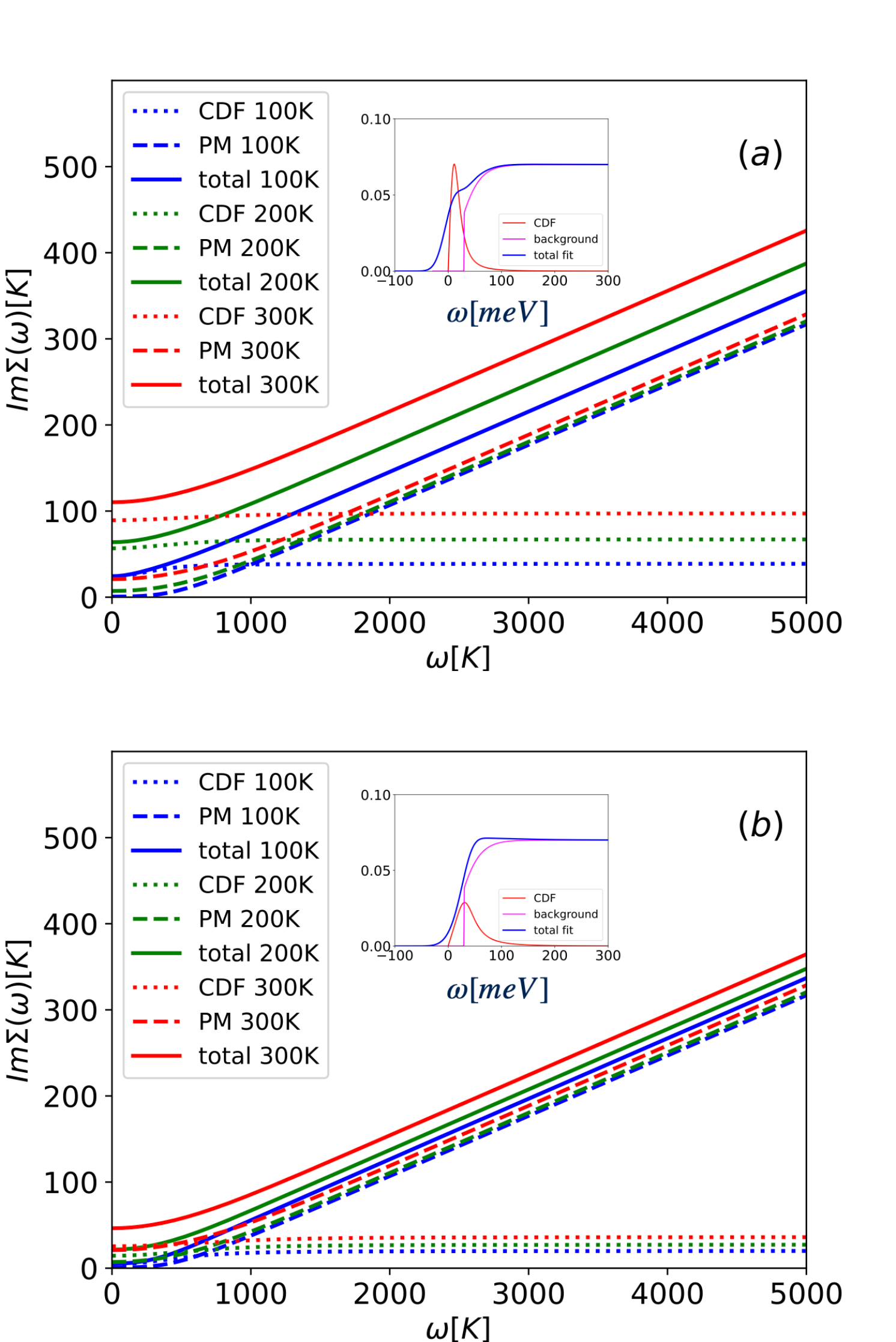}
    \caption{Quasiparticle scattering rate (equivalent to optical for momentum independent
    scattering)  due to CDF (dotted lines) and particle-hole (paramagnon) continuum.
    Solid lines report the total scattering rate.
    Panel (a) is for CDF of rather low characteristic energy $M/\gamma=10$ meV typical of 
    slightly overdoped samples. Panel (b) is for CDF of rather high characteristic energy $M/\gamma=30$ meV 
    typical of underdoped samples. }
    \label{opt-scatt-rate}
\end{figure}

In this subsection we aim at investigating the separate roles and effects
of the two sources of scattering leading to Eq. (\ref{SFLapp}): the low-energy CDF 
with $M/\gamma<T$ and the high energy paramagnon 
for $\omega>(M/\gamma,T)$
%In this subsection we aim at investigating the separate roles and effects
%of these two sources of scattering. 
In particular, we will show the relevance of
$M/\gamma$ in determining the presence or the (near) absence of 
temperature dependence in the linear high-energy part of the optical 
scattering rates. Fig. \ref{opt-scatt-rate} (a) 
shows $Im \Sigma(\omega)$ for a rather low value
characteristic CDF energy $M/\gamma =10$ meV. For this value, temperatures ranging from 100 K (blue lines)
to 300 K (red lines) are large enough to thermally excite the CDF. 
In this regime where $M/\gamma\lesssim T$ their statistics is roughly classical and the Bose function 
can be approximated as $b(\omega)\approx T/\omega$. Therefore, for frequencies 
above the characteristic CDF energy $M/\gamma$, the scattering rate due CDF (dotted lines)
saturates to constant in frequency values that have a nearly T-linear dependence $1/\tau\vert_{CDW}\sim T$.
On the contrary, the temperature dependence of the scattering rate due to paramagnons (typically starting at
higher energy above 30 meV) is instead much weaker due to the exponential 
suppression of the Bose and Fermi distributions in Eq. (\ref{sigma2}) (dashed lines).
Therefore the scattering rate due to paramagnons is nearly temperature independent and 
linear in $\omega$ because the paramagnon spectral weight has the form of a broad continuum:
$1/\tau\vert_{PM}  \sim \omega$.
Therefore, when CDF and paramagnon self-energies are added, the total
scattering rate turns out to be temperature dependent and
split into separate $\omega$-linear curves linearly separated in T (solid lines in Fig. \ref{opt-scatt-rate}).

Quite remarkably these simple combined occurrences, namely $\omega,T>M/\gamma$ of order 10 meV or less, and a
broad electronic continuum starting at $30$ meV up to $\sim 0.3$ eV, produce a total scattering rate of the form
\begin{eqnarray}
&&\left[\frac{1}{\tau(\omega,T)}\right]_{CDF}+\left[\frac{1}{\tau(\omega,T)}\right]_{PM}\sim \Lambda_T T+
\Lambda_\omega \omega \nonumber \\
&&=\Lambda_T T \left[1+\frac{\Lambda_\omega}{\Lambda_T}\left(\frac{\omega}{T}\right)\right].    
\end{eqnarray}
and provide a natural and simple explanation
for the seeming scaling properties of the optical scattering rate 
\begin{equation}
\frac{1}{T\tau(\omega/T)}\propto 1+a\omega/T
\label{lin-scal}
\end{equation}
with $a\equiv {\Lambda_\omega}/{\Lambda_T}$.

Some remarks are now in order on this relevant $\omega/T$ scaling issue.
First of all, the scaling
property of the optical scattering rate is not generic in cuprates,
as, e.g., the data of Ref. \onlinecite{uykur-2011} reported in Fig. \ref{bscco-scaling}  
display a rather poor data collapse. In fact, it requires $\alpha={\cal O}(1)$ whereas obviously no scaling is observed when $\Lambda_T\gg \Lambda_\omega$ or vice versa. 

Secondly, early experiments \cite{marel03} provided evidence for a power-law behavior 
$\sigma(\omega)\sim (1/\omega)^\alpha$, with a material and 
doping dependent exponent $\alpha<1$. Of course the simple combination 
in Eq. (\ref{lin-scal}) would be completely unable to account for 
scaling indices $\alpha$ different from one. As a matter of fact, however, the 
recent reanalysis of optics in Ref. \onlinecite{michon-2023} precisely shows that
 in La$_{2-p}$Sr$_p$CuO$_4$ (LSCO) at doping $p=0.24$, 
$1/(\tau T)$ is a linear function of $(\omega/T)$. This opens the way to our
simple, almost mundane, explanation of $(\omega/T)$-linear scaling in cuprates
in terms of the combined scattering due to low-energy (CDF) excitations and a high-energy
(paramagnon) continuum.  Of course, to challenge this seemingly MFL behavior and scaling without
any critical underlying mechanism, extended optical experiments
and (re)analysis of optical data in cuprates
other than LSCO are highly desirable.

%scaling behavior \cite{marel03,marel06,zaanen22,michon-2023}. Early 
%azrak94,marel03.timusk07}  More recently, it has been shown that such 
%an apparent power-law behavior is in fact 
%compatible with a $\omega$-linear scaling Ansatz for the inelastic scattering rate 
%leading to a unified theoretical description of the experimental data. This is 
%also compatible with an underlying (local) 
%self-energy which obeys $\omega/T$ scaling \cite{michon-2023}. In particular, it 
%has been claimed that a MFL structure \cite{varma89} 
%$\mathrm{Im}\,\Sigma(\omega)\sim (\omega/T) \coth(\omega/T)$, 
%can account for the optical conductivity data

Finally we emphasize 
the relevance of the low-energy $M/\gamma$ scale. This relevance
can be appreciated by noticing
that this scale was experimentally found to depend on doping
(see Figs. 4 (c) and (d) in Ref.  \onlinecite{arpaia-2023}).
In particular $M/\gamma$ substantially increases up to $M/\gamma\sim 25-30$ meV
in strongly underdoped cuprates. By taking such higher values, it is natural to
expect that in the same 100-300 K temperature range,
thermally exciting CDF becomes more difficult and the temperature
dependence of the scattering rates is heavily suppressed. Indeed, this is what one can
observe in the calculations of Fig. \ref{opt-scatt-rate} (b), where
the temperature dependence of the CDF scattering rate (dotted lines) is much weaker. As a consequence
also the total optical scattering rates have a much weaker temperature
dependence and all the $\omega$-linear parts of the curves are much closer. 
Interestingly, this agrees well with what is observed in optical spectral
rates, as reported in Figs. 7, 9, and 11 of Ref. \onlinecite{puchkov-1999}
and Fig. 1 of Ref. \onlinecite{hwang-2004} [see also
Ref. \onlinecite{timusk-2005} (Fig. 20) and references therein]. 

Throughout this section for the sake of simplicity, we kept the high temperature
value of $\gamma=\gamma_\infty$ equal to one for all dopings and materials. Of course for each material,
a different value of $\gamma_\infty$ could be reabsorbed by a rescaling of $M$. 
The temperature dependence of
$\gamma(T)$ that will be relevant below $T_c$ and is extensively discussed in the next sections
is actually expected to start quite smoothly above $T_c$. This leads to a mild smooth 
increase of $\gamma$, which would further improve the fittings of the resistivity above $T_c$.

To summarize, the fact the $M/\gamma$ is typically smaller than $T$ in optical experiments 
on slightly
overdoped samples opens the way to the linear scaling property of $1/(T\tau)\sim (\omega/T)$.
Moreover, the possibility of interpreting 
the weaker or stronger temperature dependence of $1/\tau(\omega,T)$ in terms of this single
low-energy parameter $M/\gamma$ 
stresses the fact that its finiteness is a relevant physical feature of these systems.

\section{The SFL scenario below $T_c$: Thermodynamics and transport}
Below T$_c$ strange metal properties are hidden within the SC phase.
The suppression of the latter requires strong magnetic fields that cannot be reached in RIXS facilities so that 
in this regime we are not able to deduce the
parameters entering the CDF propagator Eq. (\ref{eq:cdf}) with the
fitting procedure described in Sec. \ref{sec:2}.
Thus, we can only indirectly infer the specific features of these excitations (like, e.g., the CDF characteristic energy $M/\gamma(T)$) 
from transport and thermodynamic observables. 
In particular, we will show that the construction of
a coherent consistent picture requires a Landau damping parameter $\gamma(T)$ that grows logarithmically upon lowering the temperature.
This is the core of the SFL scenario, as it implies a reduction in the crossover scale $M/\gamma(T)$ toward the Fermi liquid.

%\textcolor{red}{The direct strategy of the previous Section, where we inserted 
%the measured features of the scattering excitations
%in the calculation of the spectral and transport properties, is not viable below $T_c$. 
%This is because, in order
%to reveal the strange metal properties in this regime, superconductivity has to be
%suppressed by strong magnetic fields, that cannot be reached in RIXS facilities.
%Therefore we can only indirectly infer the specific features of these excitations
%(like, e.g., the CDF characteristic energy $M/\gamma(T)$) and try to construct 
%a coherent consistent picture comparing the non trivial transport and thermodynamic quantities.}

%\textcolor{red}{PLEASE COMMENT ON THIS COMMENT....
%A clarification is now in order about our choice of $\gamma$. In the previous sections
%we were only relying on RIXS experiments, where the CDF energy scale %$M/\gamma$ is
%directly accessible in high-resolution experiments. This does not allow to 
%separately access the value of $\gamma$, which was kept constant and arbitrarily set to one. In the following
%we will try to extract from various experiments  a generic expression of $\gamma(T)$,
%which will turn out to be of order one, with $\gamma(T\sim 100K) \sim 1.5-5.0$. This means
%that the values of $M$ have to be adjusted accordingly to keep %$M/\gamma(T\sim100\,K)$ 
%at the same value extracted by fitting RIXS spectra in the $T>T_c$ section.}

\subsection{Specific heat}
\subsubsection{The fermionic contribution and the quasiparticle mass enhancement}
\begin{figure}
    \centering
    \includegraphics[width=1.0\linewidth]{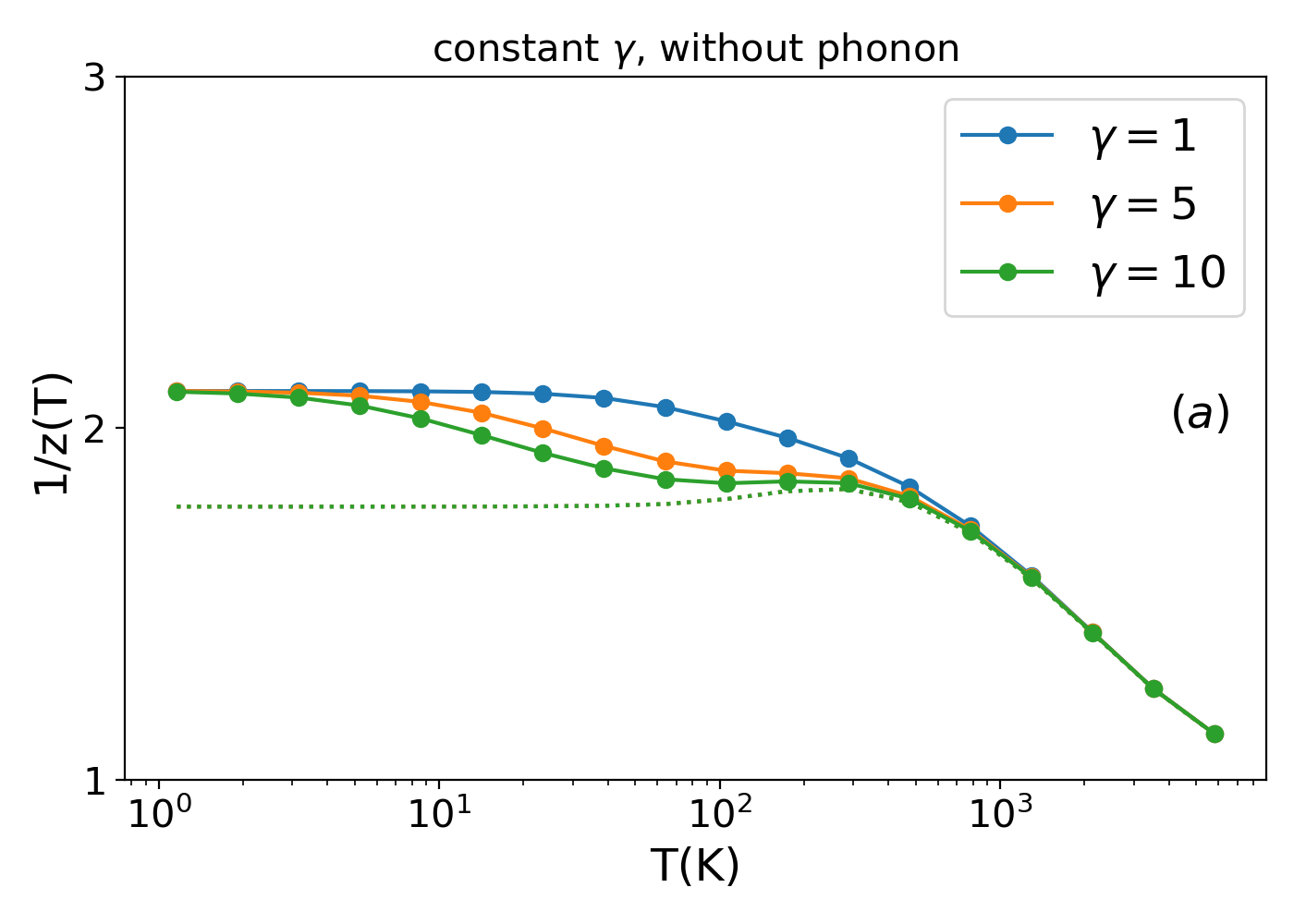}
    \includegraphics[width=1.0\linewidth]{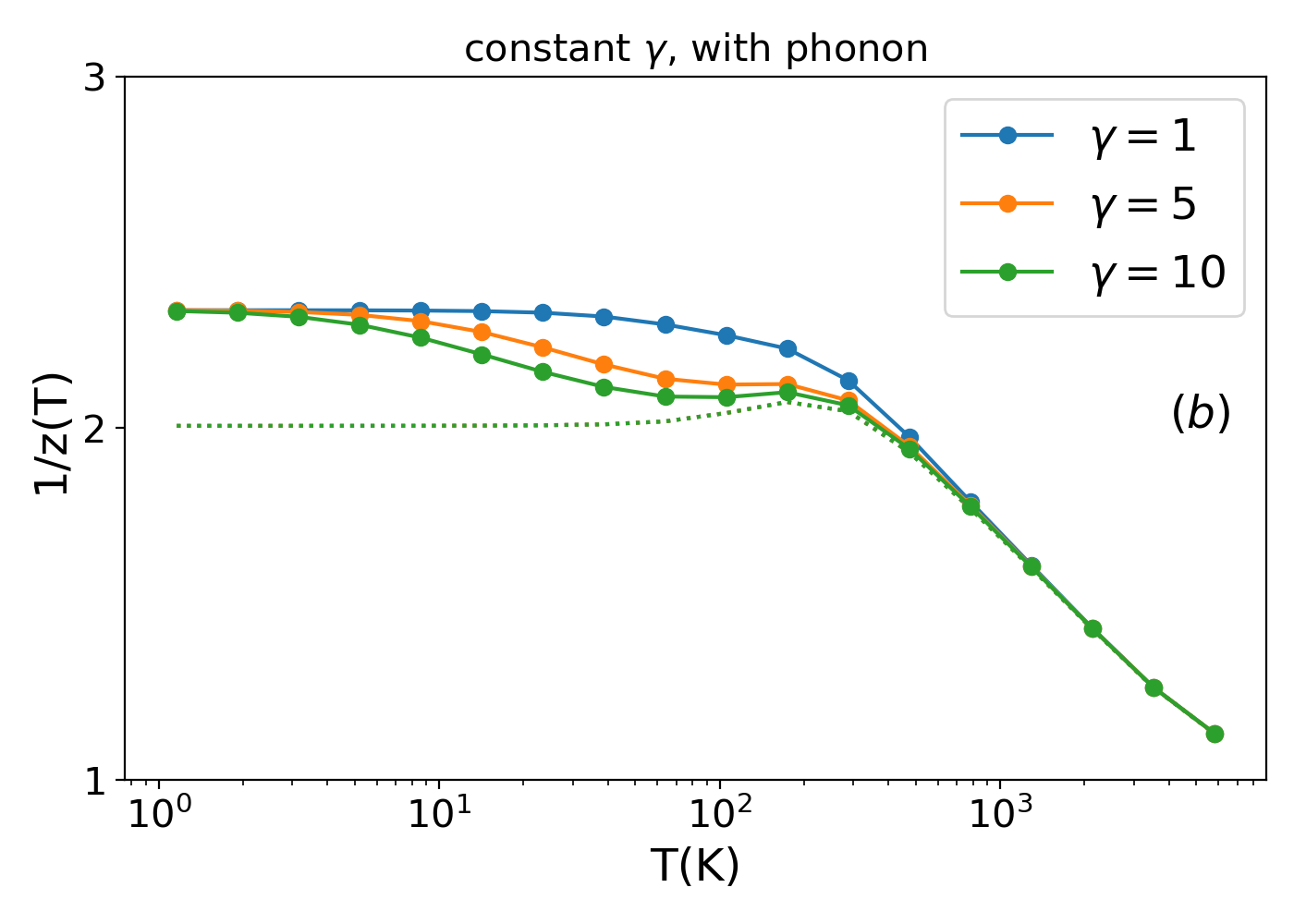}
    \includegraphics[width=1.0\linewidth]{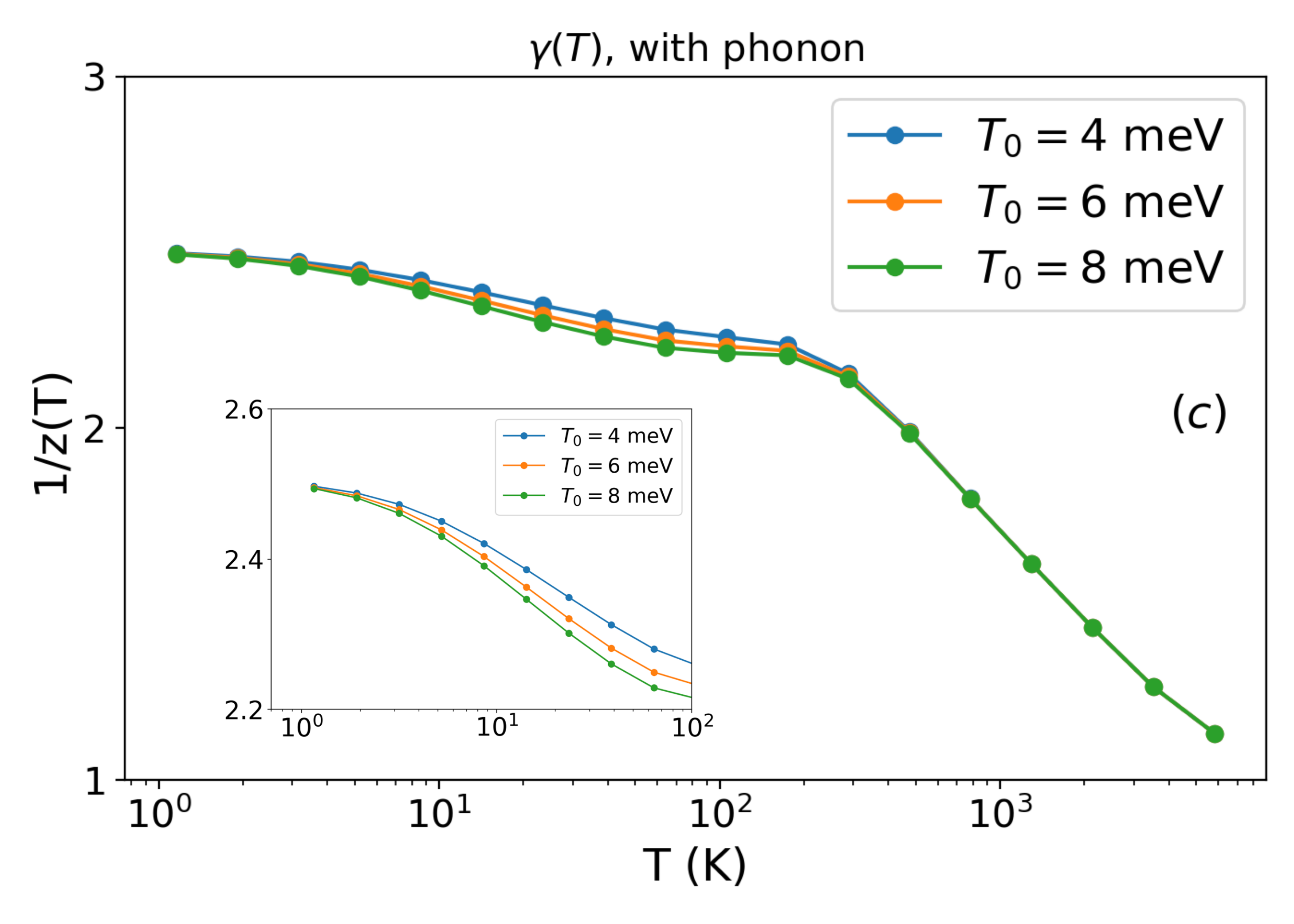}
    \caption{Temperature dependence of the mass enhancement factor $1/z(T)$. In (a) only the 
    CDF and the paramagnon continuum are considered. 
    The $\gamma$ damping
    factor is kept constant at three fixed values $\gamma=1$ (blue line), $\gamma=5$ (orange line), 
    and $\gamma=10$ (green line). (b) The phonon peak at $65$ meV is included. In panels (a) and (b), the contribution due to paramagnons, and the net contribution of phonon and paramagnons, are shown in dotted green lines respectively.
    (c) CDF, paramagnon continuum, and $65$ meV phonon are all considered, with the
    damping parameter $\gamma=1+3.5\log(1+\frac{T_{0}}{T})$ for three different values of $T_0$. The inset shows the magnified region, where $1/z(T)$ shows a logarithmic rise.
    The parameters correspond to those for LSCO from Table~\ref{tab:rixs-fit}.
    }
    \label{fig-z-T}
\end{figure}

The explanation of the T-linear specific heat of metals $C_V=\alpha T$, 
in agreement with the Sommerfeld model of a
gas of free fermions, was one of the major successes of the Landau Fermi liquid theory,
where $\alpha\propto N^*\propto m^*$, with $N^*$ and $m^*$ being the  density of states and the
mass of the quasiparticles respectively. 
This makes even more remarkable the observation of a logarithmically divergent $C_V/T$ coefficient
$C_V/T\propto \log (T_0/T)$ in Ref. \onlinecite{michon-2019}. A similar behavior is also observed 
in some heavy fermion systems like, e.g., Ce$_2$Cu$_{6-x}$(Au,Si)$_x$\cite{vonloehneysen-1994,vonloehneysen-2007}.
The divergence of the specific heat coefficient finds a natural interpretation within the MFL 
theory, where the quasiparticle mass $m^*=m/z$ ($m$ is the electron mass possibly 
dressed by strong electron-electron correlation effects and
the quasiparticle residuum  on the Fermi surface vanishes logarithmically $z=1/\log(T_0/T)$).
On the other hand, within the SFL, the quasiparticle mass also follows a logarithmic increase
mimicking the MFL behavior (see inset in Fig. \ref{fig-z-T}(c) below), but eventually saturates below a non universal temperature
scale $T_{sat}\sim T_{FL}$. Thus, at $T=0$ a {\it finite}
renormalization $m^*/m=1+ \lambda$ is obtained  with $\lambda$ being the 
dimensionless coupling between the CDF and the fermionic quasiparticle
$\lambda=g^2N^*/M$ and similar for the coupling to the other excitations.
Fig. \ref{fig-z-T}(a) reports $1/z(T)$
due to CDF and paramagnons, for parameters $M=18$ meV, $\bar{\Omega}=50$ meV, but varying the damping parameter $\gamma$. We also choose the dimensional parameter (mentioned in Sec.~\ref{sec:2}) $\tilde{\lambda}=6$ meV.
%This taken to be 0.5 and it corresponds to the
%$W_{CDF}$ coupling reported in Tab. \ref{tab:rixs-fit}. The couplings with
%the other excitations are adjusted according to the corresponding ratio %in the RIXS intensities,
%also reported as $W_\alpha$ in Tab. \ref{tab:rixs-fit}.
Therefore, were this mass renormalization the only ingredient which
determines the temperature dependence of the total specific heat
in the low-temperature regime, the SFL would predict a behavior of the form
\begin{equation}
    C_V^{F}/T\propto \log\left[\frac{AT_{sat}}{T_{sat}+T}\right]
     \label{cv-fer}
\end{equation} 
with $A=\exp(m^*/m)$. However, since this eventually saturates to the 
constant value $\sim (1+\lambda+(W_{PH}/W_{CDF})\lambda_{PH}+(W_{PM}/W_{CDF})\lambda_{PM}$, it
cannot account for the strong variation of $C_V/T$ experimentally observed even for $T<T_{sat}$.

Of course,  $\frac{1}{z(T)}=1-\frac{\partial \Sigma'(\omega,T)}{\partial \omega}\vert|_{\omega=0}$.
Now, since by the Kramers-Kronig relations $\Sigma'$ corresponds to the integral of $\Sigma''$ over all
frequencies, the value of $1/z(T)$ depends on all excitations at all energy scales, although 
it changes its behavior at different temperatures depending on the specific excitations at energies $\omega\sim T$. Then, owing to the composite set of excitations present in cuprates
at different energy scales, we now analyze separately their effect in the various temperature ranges.
We start by considering the paramagnon continuum. Owing to its (nearly constant) extension
up to high energies, these excitations determine the behavior of $1/z(T)$ at large temperatures,
as it can be seen in Fig. \ref{fig-z-T} for $T\gtrsim 300$ K.
By lowering the temperature the effect of CDF becomes more important and starts at 
higher temperature for smaller values of $\gamma$ owing to the fact that the CDF spectral weight 
is peaked at the characteristic value $M/\gamma$. For the smallest values of $M/\gamma$ it is
clearly visible that the mass enhancement due to the paramagnon continuum tends to level forming a plateau
to which the low frequency CDF add their enhancement at lower temperatures. The effect of
phonons at $\omega\sim 65$ meV as observed in RIXS spectra is considered in (b). Clearly a
mass enhancement due to phonons takes place at intermediate temperatures between those governed
by paramagnon and CDF scattering.  Finally we consider the possibility that the damping parameter
logarithmically increases lowering T leading to the `shrinking' of the CDF energy $M/\gamma(T)$.
This naturally leads to a CDF-mediated intermediate curve interpolating the mass enhancement 
described by the curves in (a). The resulting curves are in (c) for different values
of the temperature $T_0$ below which $\gamma(T)$ starts growing according to the form
$\gamma(T)= \gamma_\infty +\gamma_0\log(1+T_0/T)$, with $\gamma_0=3.5$ and $\gamma_\infty=1$. 
The reason why this phenomenological form is taken will be clear after reading the next subsection.

Notice that over a rather extended temperature range from 2-3 K up to 40-50 K the
mass enhancement factor grows logarithmically and in this range, therefore $1/z$ bears a similar 
behavior to the MFL case. However, this increase is small and cannot account for the much
larger experimentally observed increase of the specific heat coefficient.
 
\subsubsection{The bosonic contribution and the CDF softening}
It is natural (actually almost mandatory) 
to assume that the experimentally observed abundant collective CDF
excitations can give a Bosonic (B) contribution
to the specific heat coefficient $C_V^B/T$. 
In particular, it was found\cite{caprara-2022} that CDF
can contribute to the specific heat with a term proportional to the damping coefficient
\begin{eqnarray}
 C_V^B/T=\frac{1}{T}\frac{\partial}{\partial T} \int_0^\infty d\omega b(\omega) \rho_B(\omega)\omega \nonumber \\
 \approx
 k_B^2\frac{\gamma}{3\nu}\log(1+\pi\nu/M)\equiv {\cal{C}}\gamma(T,p)
 \label{cv-bos}
\end{eqnarray}
where $ \rho_B(\omega)$ is an effective bosonic spectral density (see Ref. \onlinecite{caprara-2022} for details)
 and $\nu$ is the electronic scale related to the stiffness of the CDF fluctuations
 around ${\bf Q}_c$ [cf. Eq. (\ref{eq:cdf})].
As discussed in Sect. I.B we stress that in the SFL theory the logarithmic 
divergence of $C^B_V/T$ does not stem from a vanishing of the mass $M$ in the logarithmic
term of Eq. (\ref{cv-bos}) as it
customarily occurs at the QCP in the Hertz-Millis theories, but it is rather {\it assumed}
as the intrinsic behavior of the damping term $\gamma=\gamma(T,p)$ appearing as a factor in Eq. (\ref{cv-bos}).

In Ref. \onlinecite{caprara-2022} we already fitted the low-energy logarithmically increasing specific heat
coefficient of Nd-LSCO and Eu-LSCO \cite{michon-2019}. From the fitting we found an
increase of $\gamma(T)\sim \ln(T_0/T)$ by lowering $T$. It was also found that 
this increase was large enough to lower the Fermi liquid scale $\omega_{FL}=M/\gamma(T)$
thereby extending T-linear resistivity down to $T\sim 1$ K. 

Here we apply the same procedure by using a different (but equivalent)
 more physically transparent 
 parametrization of the temperature and doping dependence of $\gamma(T,p)$
 \begin{equation}
 \gamma(T,p)=\left[\left[\gamma_\infty+\gamma_{0}\log(1+T_{0}/T\right]^{-1}+v\vert p-p^*\vert \right]^{-1},
 \label{gammaT}
 \end{equation} 
 which is valid below some saturation temperature of order of $T_0$, which cannot be determined by this low temperature
 fitting procedure.
\begin{figure}
    \centering
    \includegraphics[width=1.\linewidth]{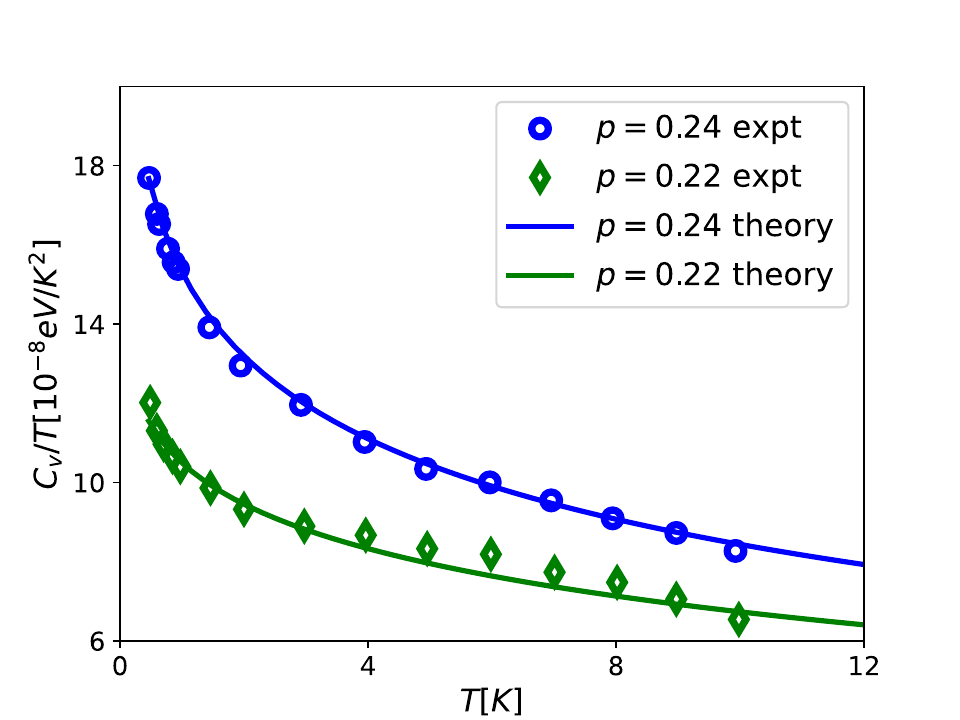}
    \caption{The fitting (blue and green solid lines) of the low-temperature specific heat coefficient $C_{v}/T$ of Nd-LSCO for two doping levels: $p=0.22$ 
    (green open diamonds) and $p=0.24$ ((blue open circles). The fitting curves are the bosonic (CDF) contribution, with a temperature dependent 
    $\gamma(T)$, according to Eq.~\ref{gammaT}.}
    \label{fig-cv-fit}
\end{figure}
Figure \ref{fig-cv-fit} reports the resulting fit obtained by considering only the bosonic 
contribution of overdamped CDF. The other excitations (phonons and paramagnons)
have much too high energies to contribute to the specific heat at low temperatures
and, moreover, phonons are generically subtracted from the experimental data reported
in Ref. \onlinecite{michon-2019}. 
$\gamma_{\infty}$, $\gamma_0$, $T_{0}$, $p^{*}$, and $v$
are determined by the fit as $\gamma_\infty=1.$, $\gamma_{0}=3.1$, $T_{0}=100$ K, $p^{*}=0.24$, and $v=1.5$.
 This expression leads to the fitting of Fig. \ref{fig-cv-fit}
 with no other dimensional factors [i.e. ${\cal{C}}=10^{-8}$ eV/K$^2$ in Eq. (\ref{cv-bos}).
%It is clear from the fit that a consistent scenario seems possible where the logarithmic increase of the
%damping slows down the bosonic collective fluctuations thereby explaining the increase of the 
%low-temperature specific heat. 

Of course also fermions contribute to the total
specific heat. In particular, the contribution of 
both fermions and bosons near a QCP in a metal was extensively investigated with similar findings
in Ref. \onlinecite{chubukov-2023} (see, e.g., the close similarity between 
Eq. (\ref{cv-bos}) here and Eq. (47) in Ref. \onlinecite{chubukov-2023}).
However, from the previous subsection it is clear that this latter contribution,
although logarithmic in an intermediate temperature range, eventually saturates at low 
temperature and can only contribute a minor increase. Since this can easily be
reabsorbed by adjusting the parameters of the bosonic contribution, in order to minimize the
fitting parameters, we choose to disregard the fermionic contribution.

The main point to assess now is whether or not the abundant CDF degrees of freedom 
revealed in the specific heat also contribute to the transport quantities like heat 
conductivity, the Seebeck coefficients, and, of course, the T-linear resistivity
when superconductivity is suppressed. This is the issue addressed in the next
subsections. In order to do this we adopt from the above
specific heat fitting a generic and simple form for $\gamma(T,p)$,
which is reasonably valid at different dopings $p\sim p^*$ and for different materials.
To this purpose we take a generic form
$\gamma(T)= \gamma_\infty +\gamma_0\log(1+T_0/T)$ with $\gamma_\infty$ and $\gamma_0$
of order one and $T_0\sim 100-200$ K.

%%%%%%%%%%%%%%%%%%%%%%%%%%%%%%%%%%%%%%%%%%%%%%%%%%%%%%%%%%%%%%%%%%%%%
%%%%%%%%%%%%%%%%%%%%%%%%%%%%%%%%%%%%%%%%%%%%%%%%%%%%%%%%%%%%%%%%%%%%%

\subsection{Heat conductivity}
\begin{figure}
    \centering
    \includegraphics[width=1.\linewidth]{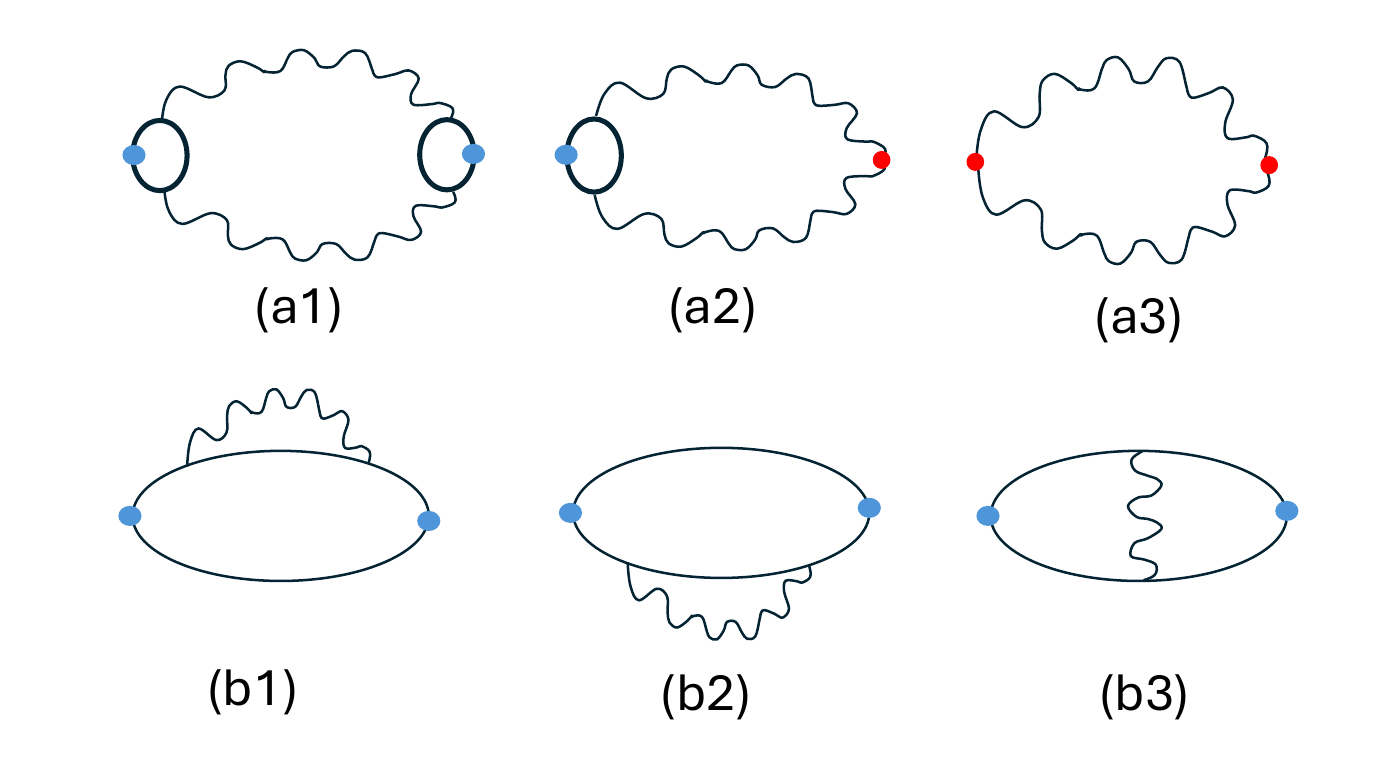}
    \caption{Diagrams for the heat transport: (a) diagrams involve the CDF-drag mechanism, while
    (b) diagrams are the usual diagrams based on quasiparticle heat transport. Wavy lines represent
    CDF propagators and solid lines represent fermionic quasiparticles. Blue dots are the quasiparticle
    heat current vertices and red dots are the corresponding vertices for the CDF.}
    \label{fig-heat-transport}
\end{figure}

According to the SFL scenario, the increase in low-temperature specific heat in the strange metal region  marks the
presence of a large amount of low-energy bosonic degrees of freedom which might also contribute to the
heat conductivity $\kappa$. 
%Differently from the MFL theory, where these degrees of freedom have a 
%fermionic character, our scenario is based on the presence of CDF, i.e. collective bosonic degrees of freedom.
This is why, besides the usual mechanism based on dressed electronic degrees of freedom $\kappa_e$, here we
also consider heat transport due to CDF $\kappa_{CDF}$. 
Fig. \ref{fig-heat-transport} reports the diagrams involved in the heat current-heat current response
relevant when a temperature gradient is applied to the system to measure the heat conductivity.
The electronic part is given by the diagrams in Fig.\ref{fig-heat-transport} (b1-b3),
where the blue dots represent the heat current carried by fermionic quasiparticles whose propagators
are the black solid lines. Fig. \ref{fig-heat-transport}, also reports the 
diagrams with bosons contributing to heat transport and the red dots represent the heat current carried by 
CDF whose propagators are the black wavy lines. 
Formally, this latter mechanism is similar to the one based on
phonons carrying heat in solids\cite{ziman}: the phonon heat transport (diagram (a3) in Fig. \ref{fig-heat-transport})
and the so-called phonon drag mechanism (diagrams (a1,a2) in Fig. \ref{fig-heat-transport}).
One should remember that CDF are rather local degrees of freedom
which involve low energy excitations of both small and large momenta. This is why they do not suffer the
typical kinematic restriction of low-energy acoustic phonons being limited to low momenta when
involved in low-temperature transport phenomena.

The heat conductivity $\kappa$ is a tensor, but in $C_4$ symmetric system 
is a scalar $\kappa=\kappa_{xx}=\kappa_{yy}$  given by 
\begin{equation}
    \kappa=\kappa_F+\kappa_B=\lim_{\Omega\to 0} \frac{1}{\Omega}\left[ Im \chi^{hh}_F(\Omega)+Im \chi^{hh}_B(\Omega) \right]
\end{equation}
Since now on, we will discard the indexes of the space directions.
The diagrams for the CDF-drag contribution 
are given by expressions like
\begin{eqnarray}
    &&\chi^{hh}_{B\alpha,\beta}(\Omega_l)= \\
    && T\sum_{\omega_n,{\bf q}} \Lambda_x^\alpha({\bf q},\omega_n,\Omega_l)\Lambda_x^\beta({\bf q},\omega_n,\Omega_l)
    D(\omega_n+\Omega_l)D(\omega_n) \nonumber
\end{eqnarray}
where $\omega_n$ and $\Omega_l$ are bosonic Matsubara frequencies and 
$\Lambda^\alpha$ being either a fermionic loop with a heat current vertex 
along the $x$ direction (see Fig.\ref{fig-heat-transport} (a1,a2)) or 
a bosonic heat current vertex (see Fig.\ref{fig-heat-transport} (a3)). The fermionic loop vertex is given by
\begin{eqnarray}
     &&\Lambda_x^F({\bf q},\omega_n,\Omega_l)= \\
     &&=T\sum_{n,{\bf k}} \frac{k_x}{m}\xi_{\bf k}G({\bf k},\epsilon_n+\Omega_l)G({\bf k}+{\bf q},\epsilon_n+\omega_n)
     G({\bf k},\epsilon_n) \nonumber
\end{eqnarray}
This vertex enters the response function at finite external frequency and is therefore regular
for any $q_x$. The complete calculation is reported in a separate paper fully devoted to the Seebeck effect
in cuprates \cite{mirarchi-2026}. 
Since the relevant frequencies for the bosons 
are much smaller than the fermionic ones, this has been evaluated at $\omega_n\sim 0$.
Here we simply report the final result,
\begin{equation}
     \Lambda_x^F({\bf q},\Omega_l)\approx-\frac{C(T)}{q_x}\left(1-\frac{|\Omega_l|}{\sqrt{k_F^2q_x^2+\Omega_l^2}}    \right)
\end{equation}

The $C(T)$ coefficient vanishes identically when the fermionic spectrum is particle-hole symmetric. The 
calculation of the loop in the presence of a linear dependence of the fermionic density of states
involves the quantity $\tilde{N}(\xi)\equiv(1/2) \sum_{\bf k}(k_x^2+k_y^2)\delta(\xi-\xi_{\bf k})$.
Expanding around the chemical potential ($\xi=0$), 
$\tilde{N}(\xi)=\tilde{N}(0)+\tilde{N}'(0)\xi$ the fermionic loop calculation 
involves integrals of the form 
\begin{eqnarray}
  C(T)&\propto & \int d\xi \partial_\xi f(\xi)\xi [\tilde{N}(0)+\tilde{N}'(0)\xi] \nonumber \\
&=& N'(0)\int d\xi \xi^2 \partial_\xi f(\xi)\propto T^2.
\label{ferm-vert-T}
\end{eqnarray}
giving a $T^2$ dependence to the heat-current fermionic vertex.

The bosonic heat current vertex also is reported in Ref.\onlinecite{mirarchi-2026}
and is given by
\begin{equation}
     \Lambda_x^B({\bf q},\Omega_l)=\omega_{\bf q} 
     \frac{\partial}{\partial q_x} \omega_{\bf q}\approx \overline{v}^2 q_x
\end{equation}
Here we assume that the bosonic heat current is due to the energy of the mode 
$\omega_{\bf q}\sim M+\nu|\qvec-{\bf Q}_c|^2$ times its
group velocity $v_{\bf q}=\nabla_{\bf q} \omega_{\bf q}$ subsequently
replaced by its average over the weak mode dispersion
$\langle \omega_{\bf q}v_{\bf q}\rangle \sim \overline{v}\omega_{\bf \overline q}$. 
As it usually occurs for phonons in metals, one expects the group velocity 
of CDF to be rather small 
in comparison to the Fermi velocity of the quasiparticles. 
From the fact that CDF involve a tight mixing of optical phonons and electronic excitations
\cite{becca-1996,fedele-2026}, 
a typical sound velocity is a rough upper estimate of the group velocity. 
On the other hand
CDF at low temperature become long lived with a lifetime $\tau_{CDF}\sim \gamma(T)$ tending to grow at
low temperature and partially compensate the small velocity. In general, however, 
it is rather difficult to provide a quantitative estimate of the relative importance 
of the various contributions.  

Once this bosonic heat current is introduced
in the expression of the vertices, the response is evaluated by neglecting the weak
dispersion of the CDF in the denominator of their propagator with Matsubara frequency
\begin{equation}
    D_{CDF}({\bf q},\omega_n)\approx \frac{1}{M+\gamma|\omega_n|}.
\end{equation}
This is the low frequency limit  version of Eq. (\ref{eq:cdf}).

The imaginary part of the response functions (a1)-(a3) in Fig. \ref{fig-heat-transport} are
evaluated by passing to the spectral representation of the bosonic lines and analytically continuing the
external Matsubara frequency $(i\Omega_l \to \Omega+i0^+)$ obtaining
\begin{eqnarray}
    &&Im\chi^{hh}_{B\alpha,\beta}(\Omega)= \\
    && \int\frac{dy}{\pi} \frac{\gamma y}{M^2+(\gamma y)^2}
\int\frac{dz}{\pi} \frac{\gamma z}{M^2+(\gamma z)^2}\times\nonumber \\
    && 
\left[b(y)-b(z)\right]\delta(y-z-\gamma \Omega)\sum_{\bf q} 
\Lambda_x^\alpha({\bf q},\Omega)\Lambda_x^\beta({\bf q},\Omega) \nonumber 
\end{eqnarray}
Where $\alpha$ and $\beta$ can be either $B$ or $F$.
Owing to the regular behavior of the vertices, the interesting dependence on $\gamma$
arises from the integration of the two Bose propagators.
The dummy variables inside the spectral integrals can suitably be rescaled by $\gamma$,
while the fermionic and bosonic vertices do not scale with $\gamma$.
Taking the small $\Omega$ limit introduces a derivative of the Bose function
\begin{eqnarray}
    &&\frac{Im\chi^{hh}_{B\alpha,\beta}(\Omega)}{\Omega}\propto  \label{gamma12}\\
    && -\tilde C_{\alpha\beta}(T)\int\frac{dz}{\pi} \left(\frac{\gamma z}{M^2+(\gamma z)^2)}\right)^2
\left[\frac{\partial b(z)}{\partial z}\right]\nonumber 
\end{eqnarray}
Disregarding the additional temperature dependence coming from the fermionic heat-current vertices
encoded in the $\tilde C(T)$ coefficient, this expression has two limiting 
behaviors: it is proportional to $\gamma T$ when $M<\gamma T$ , while it behaves as $(\gamma T)^2$
in the opposite limit $M>\gamma T$. For simplicity we can schematize the full behavior 
of the above integral with an interpolating function
\begin{equation}
 F(M,\gamma T)\equiv \frac{\gamma^2T^2}{M^3\sqrt{M^2+\eta \gamma^2T^2}}   
 \label{f-function}
\end{equation}
$\eta$ is of order one and it is chosen to match closely the exact expression of the  integral in Eq.(\ref{gamma12}).
As far as the whole temperature dependence is concerned, 
fermionic loops introduce a $T^2$ dependence [see Eq. (\ref{ferm-vert-T})], while no
additional $T$ factors appear when a boson heat-current vertex is present.
In summary, calculating the diagrams of Fig. \ref{fig-heat-transport} (a),
one finds that the first  two have a subleading temperature dependence
depending on the $M/\gamma T$ ratio in the function $F(M,\gamma T) $.
Specifically (a1) goes like $T^4F(M,\gamma T)$ and 
(a2) goes like $T^2F(M,\gamma T)$, while the last (a3) term
goes like $F(M,\gamma T)$. Therefore, assuming $\gamma(T)\sim \log(1/T)$, the 
heat conductivity due to CDF-drag processes vanishes at low temperature at least
like $T\log(T_0/T)$ or $T^2\log^2(T_0/T)$. 

The usual fermionic contribution should also be considered, $\kappa_F=K_F(T)T$
with $K_F(T)\sim m^* \sim m/z(T)$. $\kappa_F/T$ will therefore mirror the 
behavior of the fermionic effective mass due to the quasiparticle residue $z(T)$.
In particular there will be a low-temperature intermediate logarithmic behavior
eventually saturating to a constant value at the lowest temperatures
according to Eq. (\ref{cv-fer}) and Fig. \ref{fig-z-T}:
The fermionic contribution to the thermal conductivity coefficient $\kappa_F/T=K_F$ is
finite at the lowest temperatures. 
 On the other hand the CDF contribution to thermal conductivity coefficient 
 vanishes. 
Therefore, remarkably, the vanishing of $\kappa_{B}$ at low temperature and the ultimately
Fermi liquid character of the fermionic quasiparticles at $T=0$ warrant that the Wiedemann-Franz law
is satisfied, in agreement with experiments in Nd-LSCO \cite{michon-2018}.

In summary the total heat conductivity does not display any singular behavior,
but the long lifetime effect of the CDF could be revealed by studying
\begin{equation}
    \frac{1}{T} \left(\frac{\kappa}{T} -K_F(T\sim 0) \right)=K_{B}\gamma^2\propto \log^2\left(\frac{1}{T}\right)
\end{equation}
In this case one could hope to highlight 
a low temperature range where a $\log^2(1/T)$ is hidden in heat transport. 

%%%%%%%%%%%%%%%%%%%%%%%%%%%%%%%%%%%%%%%%%%%%%%%%%%%%%%%%%%%%%%%%%%%%%
%%%%%%%%%%%%%%%%%%%%%%%%%%%%%%%%%%%%%%%%%%%%%%%%%%%%%%%%%%%%%%%%%%%%%

\subsection{Seebeck coefficient}
\begin{figure}
    \centering
    \includegraphics[width=1.0\linewidth]{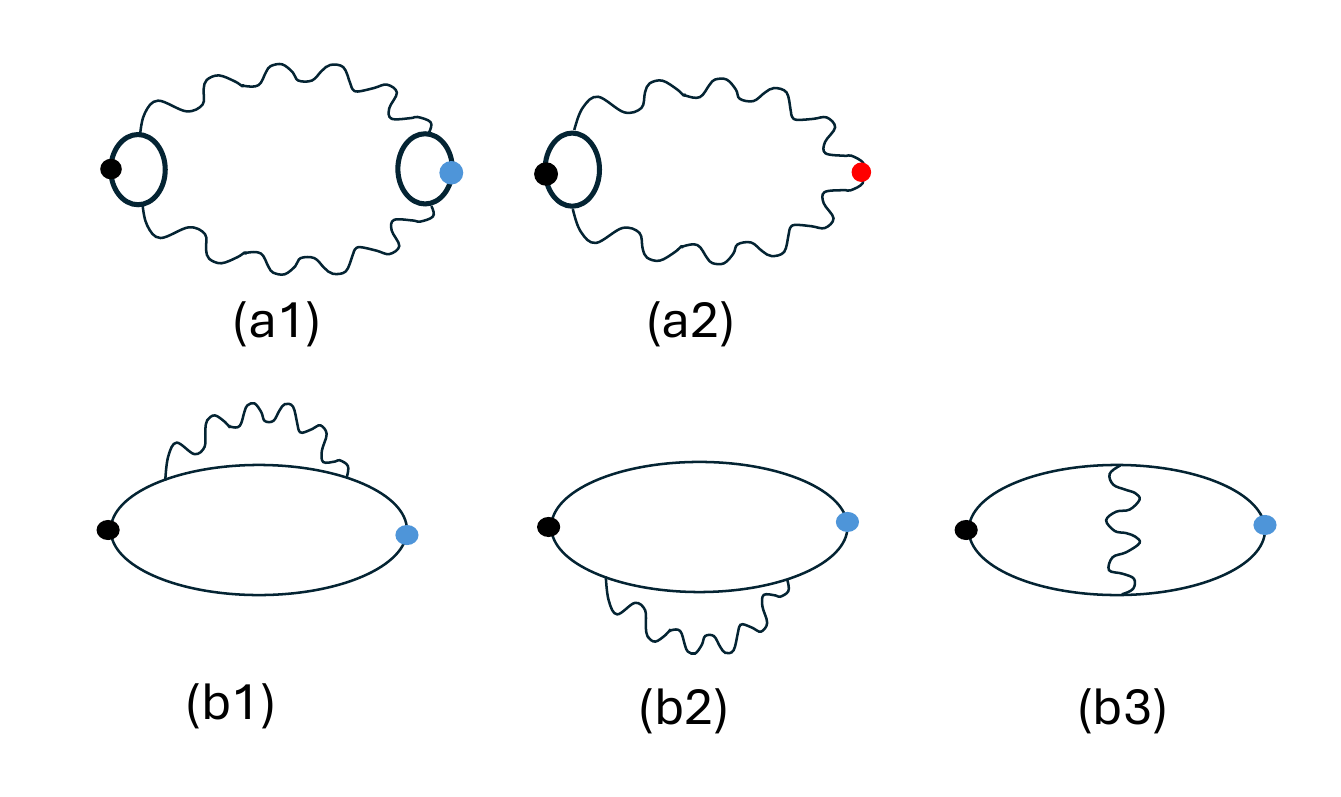}
   \caption{(a) Diagrams for the charge-current--heat current response function: 
   (a) diagrams involve the CDF-drag mechanism, while
    (b) diagrams are the usual diagrams based on quasiparticle heat transport. Wavy lines represent
    CDF propagators and solid lines represent fermionic quasiparticles. Black dots are the quasiparticle
    charge current vertices, blue dots are the 
    quasiparticle heat current vertices, and red dots are the corresponding vertices for the CDF.
}
    \label{fig-seebeck-diag}
\end{figure}

Thermoelectric power and the related Seebeck coefficient $S$ are also important physical quantities providing
information both on the amount of degrees of freedom that can be 
excited and on their mobility to transport heat and charge.
The first  point obviously raises the issue of the possible logarithmic increase of $S/T$ 
as a consequence of the
increase of the specific heat previously discussed. However, differently from the electronic 
contribution to the Seebeck coefficient, it is crucial to address the issue of how the neutral bosonic CDF 
can couple to the electric field involved in the mixed heat current-charge current 
response determining $S$. This very same
issue is also present in standard solid state theory whenever a contribution of phonons is present in the
thermoelectric phenomena. Again, like for heat conductivity, this occurs via a
phonon-drag mechanism\cite{ziman}. Specifically,
while the phonons can respond to a temperature gradient, they are also coupled to electrons, which in turn
respond to an electric field. 
Obviously the same processes can be at work if phonons are replaced by CDF
and it is worth remembering that CDF are rather local degrees of freedom.
Therefore do not suffer the
typical kinematic restriction of low-energy acoustic phonons in low-temperature transport phenomena.
The Feynman diagrams of the response function for these processes are 
reported in Fig. \ref{fig-seebeck-diag} (a) and have a similar structure as the first two
reported in Fig.  \ref{fig-heat-transport}(a). Of course, in the present case
one of the blue dots representing the heat current in the fermionic loops
is replaced by a black dot representing a charge current vertex coupled to an external electrostatic
potential gradient.
A similar treatment as in the case of the heat conductivity can be adopted
for the calculation of the temperature and $\gamma$ dependencies arising from the
fermion loops and boson propagators. Noticeably, now the current vertex does not lead to the vanishing of the
fermion loop in the particle-hole symmetric case and it gives a finite contribution even at $T=0$.
Therefore, diagram (a1) will behave as $T^2F(M,\gamma T)\sim \gamma T^3$ or $\gamma^2T^4$
depending on the different behavior of $F$ when $M/\gamma$ is larger or smaller than $T$. 
Diagram (a2) will behave as 
$F(M,\gamma T)\sim \gamma T$ or $\gamma^2T^2$. They obviously also differ by specific
factors: the second diagram contains a red heat current dot containing the CDF group velocity,
which in the first diagram should be replaced by the Fermi velocity times the square of the CDF-fermion couplings.

As long as the more standard purely electronic contributions represented by the diagrams of Fig. 
\ref{fig-heat-transport} (b1)-(b3) are concerned, various remarks are in order. First of all, the vertex correction 
diagram (b3) is negligible for a weakly momentum dependent CDF mode. Second, if the system is particle-hole symmetric,
also the diagrams with self-energy corrections vanish. In real systems, a non constant electronic density of states
introduces a violation of the particle-hole symmetry, making the corresponding contribution finite.
The detailed analysis was carried out in Ref. \onlinecite{mirarchi-2026}, where
also the role of the CDF momentum was considered. The main result is that all these
fermionic contributions are sub-leading with respect to those involving the phonon-drag mechanism 
[(a1)-(a2) in Fig. \ref{fig-heat-transport}]. Of course other models involving a skewed self-energy
could be considered, where the fermionic contribution becomes no longer sub-leading\cite{georges-2021}.

The Seebeck coefficient is then given by
\begin{equation}
    \frac{S}{T}=-\frac{1}{e\sigma T^2}\lim_{\Omega\to 0} \frac{Im\chi^{hc}(\Omega)}{\Omega}
\end{equation}
$\chi^{hc}$ being the mixed heat current-charge current response obtained from
the dominant (a2) diagram of Fig.  \ref{fig-heat-transport} and $\sigma$ being the d.c. conductivity,
which in the presence of disorder has a finite value at $T=0$.

\begin{figure}
    \centering
   \includegraphics[width=1.0\linewidth]{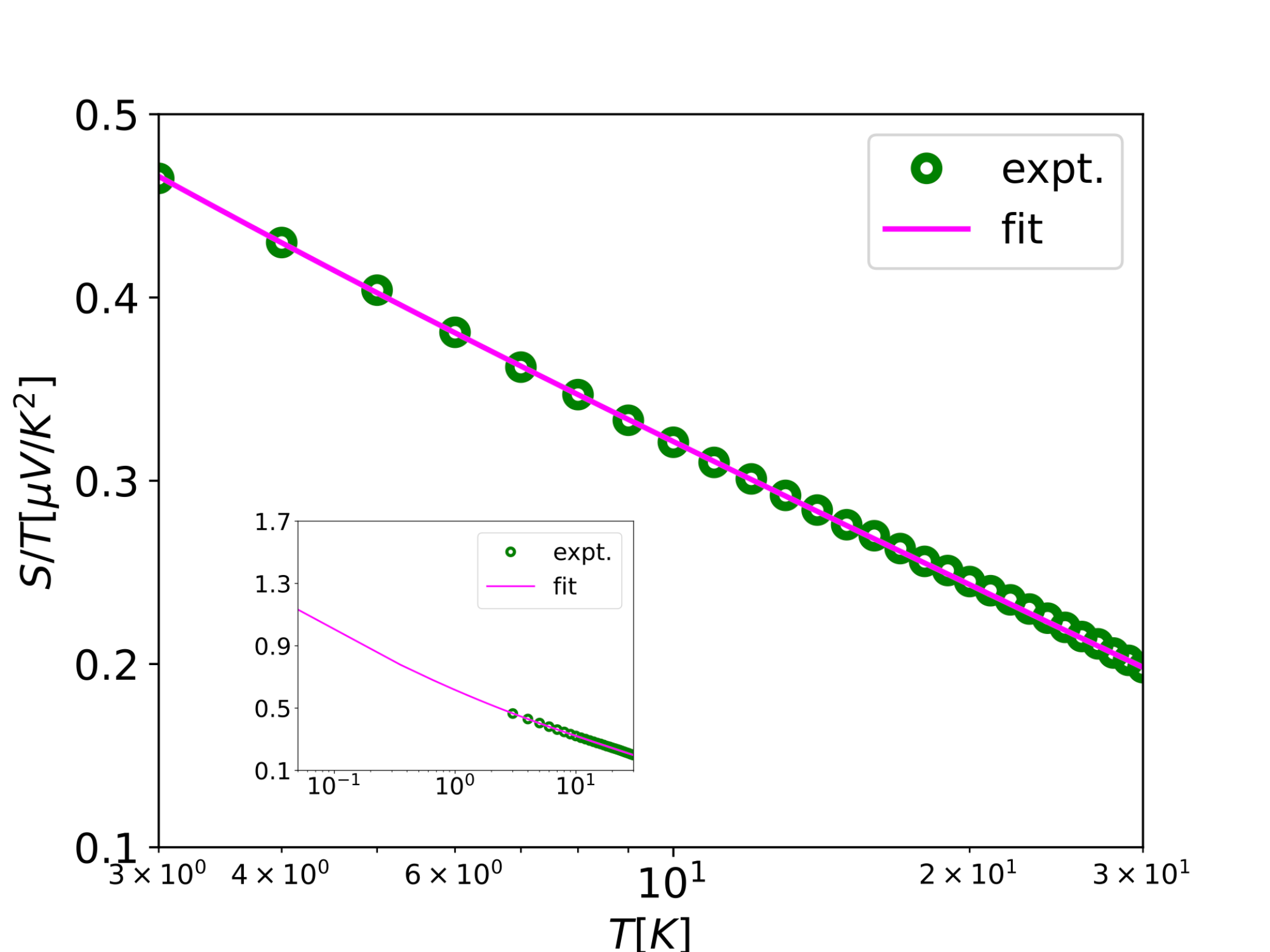}
   \caption{Comparison between experimental data taken from Ref.\,\onlinecite{gourgout-2022} 
   (circles) and our theoretical calculation 
(solid magenta line), with $S/T={\cal A}\gamma^2/{M}\sqrt{M^2+\eta \gamma^2T^2}$ and parameters  $\eta=0.37$, $M=200$ K, 
   ${\cal A}=212$ $\gamma_\infty=4$, $\gamma_0=1.28$, $T_0=200$ K (cf. Eq. (\ref{seebeck-fit}).
The inset displays the same data and fit, but on a broader temperature range to
emphasize the curvature of the theoretical curve, which at low T depends quadratically on $\gamma(T)$.
}
    \label{fig-seebeck-fit}
\end{figure}

Obviously the fitting of $S/T$ not only involves the use of $\gamma(T)$ as determined from
the specific heat of the CDF, but also the CDF capability of carrying energy 
and distribute it to traveling charged electronic quasiparticles. This naturally implies that additional fitting parameters 
should be introduced encoding the specificity of the CDF-quasiparticle coupling, the
Fermi velocities from the fermionic current vertices, the CDF group velocity
from the heat bosonic vertex, and so on. All these different quantities, however, can be clumped 
into one single prefactor for the whole CDF contribution to $S/T$.
Specifically, by using the interpolating expression of 
$F(M,\gamma T$ [Cf. Eq. (\ref{f-function})] we find
\begin{equation}
    \frac{S}{T}={\cal{A}}\frac{\gamma^2}{M}\frac{1}{\sqrt{M^2+\eta \gamma^2T^2}}.
    \label{seebeck-fit}
\end{equation}
By taking the same temperature dependence of $\gamma$ as determined by 
the specific heat fitting in Eq. (\ref{gammaT}), i.e. a form like
$\gamma(T)= \gamma_\infty +\gamma_0\log(1+T_0/T)$
choosing $M=200$ K, and adjusting the various parameters as ${\cal{A}}=212$,
 $\eta=0.37$, $\gamma_\infty=4.$, $\gamma_0=1.28$, $T_0=200$ K,
we obtain the fit in Fig. \ref{fig-seebeck-fit} showing a rather accurate
description of the data\cite{note-fit}.

Remarkably the seeming logarithmic dependence of $S/T$ is well captured, but extending to lower 
temperatures, where no data are available, our expression (see inset), displays a clear upward
curvature. This is related to the $\gamma^2$ asymptotic behavior. Obviously experiments
at lower temperatures would be highly desirable to test this prediction.

%%%%%%%%%%%%%%%%%%%%%%%%%%%%%%%%%%%%%%%%%%%%%%%%%%%%%%%%%%%%%%%%%%%%%
%%%%%%%%%%%%%%%%%%%%%%%%%%%%%%%%%%%%%%%%%%%%%%%%%%%%%%%%%%%%%%%%%%%%%

\subsection{Resistivity}
The analysis of transport above $T_c$ carried out in Sect. II, clearly showed that 
CDF have low enough energy to account for T-linear resistivity down to $T_c$.
On the other hand there is clear evidence that T-linear resistivity persists
below $T_c$ when superconductivity is suppressed by a strong magnetic field\cite{boebinger-1996,legros-2019}.
Apparently linearity persists down to temperatures much lower than $T_c$  and 
$\omega_{FL}=\omega_{CDF}=M/\gamma$ if $\gamma$ stays constant. However, 
from the analysis of the specific heat and of the Seebeck effect we phenomenologically obtained a
temperature dependence of the damping coefficient $\gamma(T)$. From this we can infer the
temperature dependence of the FL scale $\omega_{CDF}(T)=M/\gamma(T)$. In turn this
energy scale rules the temperature below which the CDF  no  longer have a classical character. 
Since, however, by lowering the temperature $\gamma$ keeps growing, $\omega_{FL}$ 
correspondingly decreases opening the possibility for the  resistivity to keep its T-linear
dependence down to substantially low temperatures.  It is indeed quite remarkable that 
the slope in temperature of the scattering rate depends on the dimensionless coupling
$\lambda_T$ and {\it it does not depend on} $\gamma$. 
$\gamma(T)$ only enters the temperature scale $T_{FL}\equiv\hbar\omega_{FL}/k_B$.
Then the T-linearity of $\rho(T)$ depends on specific details related to 
how fast or slowly $\gamma(T)$ grows when $T$ is lowered. 

\begin{figure}
    \centering
    \includegraphics[width=1.\linewidth]{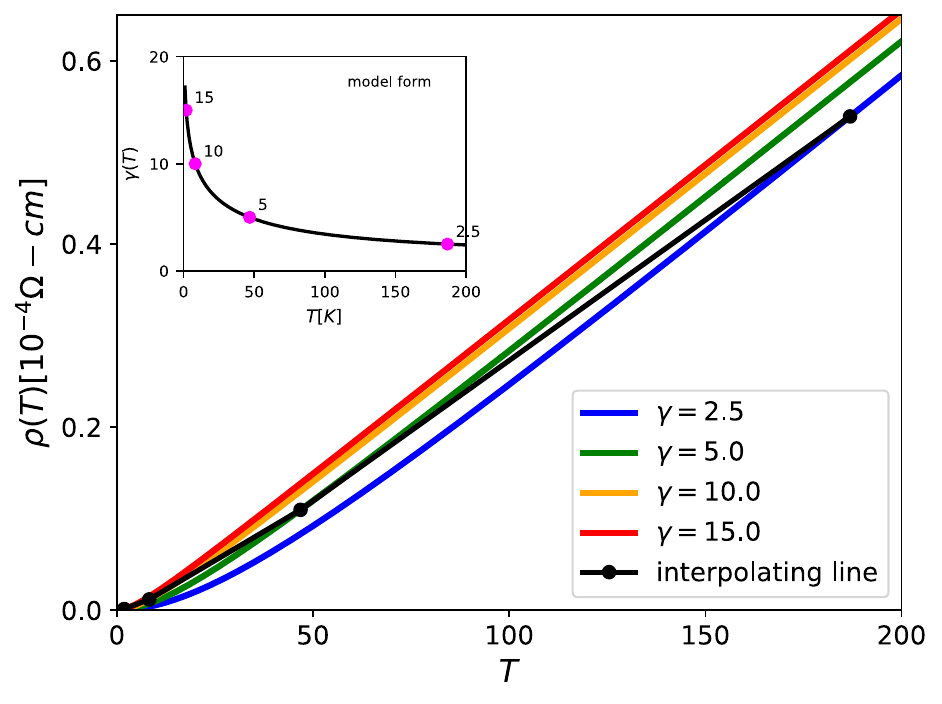}    
    \caption{
    The calculated resistivity within the SFL scenario for constant 
    $\gamma=\{2.5,5.0,10.0,15.0\}$ and also a  temperature-dependent $\gamma(T)=1.0+3.5\log(1+(100.0/T))$.
    The inset displays the continuous curve, chosen as a model for the latter. We notice that while the constant 
    $\gamma$ curves display a Fermi-liquid like curvature below $T \sim M/\gamma$ (here $M=20$ meV), 
    the black line, representing the varying $\gamma$ case, interpolates among them. The line is 
    constructed with points of the same color chosen from the $\gamma(T)$ curve. To 
    display the data in physical units, we have chosen $\lambda_{T}=0.5\cdot10^{-4} \Omega cm$.}
    \label{rho_T-gamma_T-1}
\end{figure}

We first consider the case in which $\gamma(T)$ stays nearly constant down to a few tens of Kelvin and
then grows rather abruptly (although logarithmically according to the previous subsections).
Fig. \ref{rho_T-gamma_T-1} schematically reports the $\rho(T)$ curves if $\gamma$ is assumed to be
temperature independent. Clearly, when $\gamma=2.5$ (blue line) the FL crossover occurs
at $T_{FL}\sim 40$ K, when $\gamma=5.0$ (green line) the FL crossover occurs
at $T_{FL}\sim 20$ K, when $\gamma=10.$ (orange line) the FL crossover occurs
at $T_{FL}\sim 10$ K,  and finally, when $\gamma=15.$ (red line) the FL crossover occurs
at $T_{FL}\sim 5-7$ K. Notice also that the high temperature slope does not depend on $\gamma$.
Now, take $\gamma(T)= \gamma_\infty +\gamma_0\log(1+T_0/T)$
being the continuous curve in the inset (with $\gamma_\infty=1.0$, $\gamma_0=3.5$,
and $T_0=100$ K). Of course the resulting
resistivity will smoothly interpolate between the previous curves. This is piecewise 
discretized by the black line for illustrative purpose. Quite noticeably we find that
the resistivity has both a linear and quadratic behavior 
\begin{equation}
    \rho(T)=\rho_0+AT+BT^2.
\end{equation}
This is precisely the behavior found in the extended overdoped regime considered in Ref. \onlinecite{ayres-2021}.
Thus the misterious strange metal phase above $p^*$ with $T-T^2$ mixed behavior finds a simple interpretation
in terms of a CDF-based scattering mechanism with an increasing $\gamma(T)$.

\begin{figure}
    \centering
    \includegraphics[width=1.\linewidth]{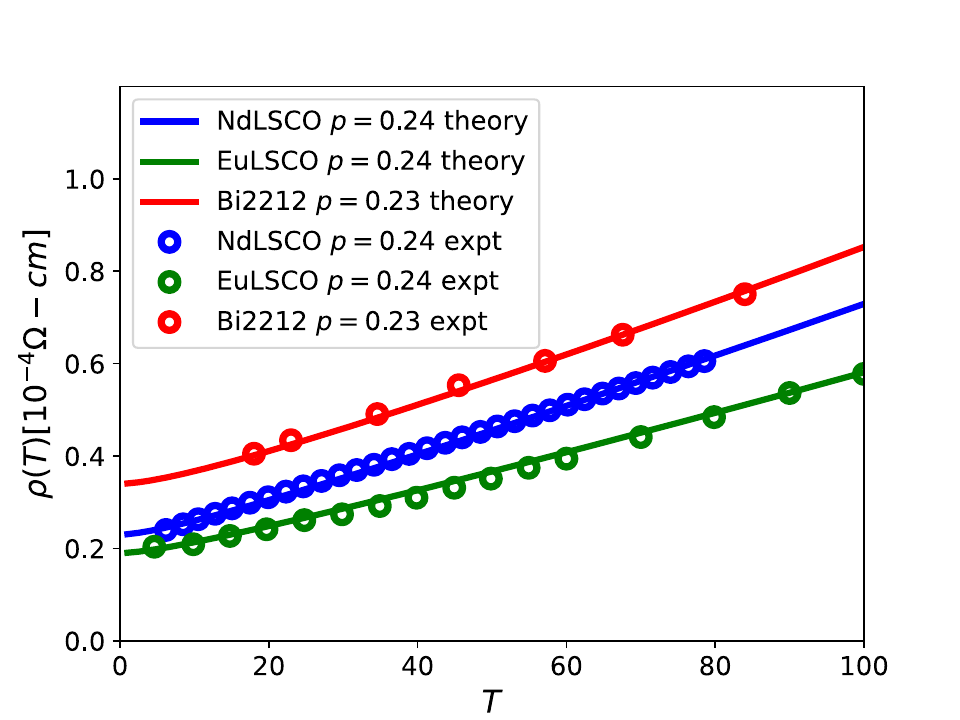}    
    \caption{The fitting of the low-temperature resistivity for Nd-LSCO ($p=0.24$), Eu-LSCO ($p=0.24$) and Bi-2212 ($p=0.23$), under the application of strong magnetic fields, with theoretical calculations, assuming a temperature-dependent $\gamma(T)=1+3.5\log(1+(100.0/T))$, as shown in Fig.~\ref{rho_T-gamma_T-1}. We take $M=18,18,25 meV$ for the three materials respectively. The fit parameters are $\lambda_T=\{0.83, 0.65, 1.28\}\cdot 10^{-4}\Omega cm$ and $\rho_0=\{0.23,0.19,0.34\}\cdot 10^{-4}\Omega cm$ for the three cases respectively.}
    \label{rho_T_H}
\end{figure}

On the other hand, when $\gamma(T)$ starts growing at somewhat higher temperatures
of order 100 K or so, the smooth crossover makes the $T^2$ contribution nearly negligible and
the linear resistivity keeps essentially the same slope and purely linear character from the
highest down to the lowest temperatures. This is what is visible in Fig. \ref{rho_T_H},
which reports the resistivity fitting for the same 
Nd-LSCO (blue circles and line) and Eu-LSCO 
(green circles and line) samples investigated in Sec. III.A.2 for the specific heat
(cf. Fig. \ref{fig-cv-fit}). Since for these samples we lack a direct RIXS measurement 
of the spectral density of the scattering mediators, we keep using the same 
$D''(\omega)$ reported in Fig. \ref{lsco-rixs} for LSCO, with the only change
that now $\gamma$ is no longer set to one,
but is instead a  function $\gamma(T)$ similar to the one used to fit the
specific heat. This therefore establishes an internal 
consistency between the form of $\gamma(T)$ 
as extracted from specific heat and Seebeck coefficients and the linear behavior of the
resistivity down to the lowest temperatures as directly measured under strong magnetic fields.
Notice also that the value of $M=18$ meV is chosen here in order to have $M/\gamma\sim 6$ meV
at high temperatures of 100-200 K, when $\gamma\sim 2.5-3.5$.

A similar linear resistivity fit is obtained for a Bi2212 sample reported in
Ref. \onlinecite{legros-2019} (red data and line): the scattering mediator spectral
density is the same as the one used in Sect. II.A, that we keep as generically 
representative of the scattering excitations for all Bi2212 slightly overdoped samples. 
On the other hand, since we lack low temperature specific heat data for BSCCO suggesting a 
temperature dependence of $\gamma(T)$, we use 
Eq. (\ref{gammaT}) with $p=p^*$ as a prototypical form valid for this case too.

Some final remarks are now in order regarding the low temperature regime
below the lowest experimental temperatures in Fig. \ref{rho_T_H}

First of all, it is obvious that if $\gamma$ increases only logarithmically 
there will surely be a temperature below which the condition
 for the classical behavior of CDF ($T>M/\gamma(T)$) is violated.
Below this temperature the T-linearity of $\rho$ will stop. 
This will occur at a temperature
$T_{FL}=M/\gamma(T_{FL})$. This last condition, however, 
is non universal insofar it depends on
$M$ and on the specific temperature and doping
dependence of $\gamma(T,p)$, (see the parameters $T_0$, $A,\, B,\,C$  in Eq. (\ref{gammaT})).
Another possibility could be that at low temperatures the logarithmic 
behavior $\gamma(T)\sim \log(T_0/T)$ crosses over to a power-law growth 
$\gamma(T)\sim (T_0/T)^\delta$ with $\delta\geq 1$.
Since this would also lead to a violation of the third principle of thermodynamics
with a finite or diverging value of the specific heat at zero temperature
$C_V=\gamma(T)T$, this means that a non-equilibrium state takes place.
In this case the classical character of the CDF
could persist down to $T=0$ with the formation of a glassy state at 
$T\gtrsim  0$\cite{ketmaier-2026}.

%%%%%%%%%%%%%%%%%%%%%%%%%%%%%%%%%%%%%%%%%%%%%%%%%%%%%%%%%%%%%%%%%%%%%
%%%%%%%%%%%%%%%%%%%%%%%%%%%%%%%%%%%%%%%%%%%%%%%%%%%%%%%%%%%%%%%%%%%%%
%%%%%%%%%%%%%%%%%%%%%%%%%%%%%%%%%%%%%%%%%%%%%%%%%%%%%%%%%%%%%%%%%%%%%
%%%%%%%%%%%%%%%%%%%%%%%%%%%%%%%%%%%%%%%%%%%%%%%%%%%%%%%%%%%%%%%%%%%%%

\subsection{Magnetoresistance}
Magnetoresistance is another important transport quantity that provides valuable information
on the nature of the metallic state. In particular it is well known that standard metals 
obeying the Fermi liquid paradigm display the so-called Kohler's rule \cite{ashcroft,luo02}:  
magnetoresistance is quadratic
in the magnetic field and has a simple scaling property 
\begin{equation}\label{kohler}
     \frac{\rho(H,T)-\rho(0,T)}{\rho(0,T)} \equiv 
   \frac{\Delta \rho(H,T)}{\rho(0,T)}
   =F(H \tau(T))
 %     \nonumber \\
 %  =F\left(  \frac{H}{\rho(0,T)}  \right)
 %  \propto \left( \frac{H}{\rho(0,T)} \right)^2
\end{equation}
with $F$ being a function only depending on the geometrical 
configuration and on the material. Eq. (\ref{kohler})
implies that the material is characterized by a momentum independent scattering rate $\tau$ so that for a temperature independent charge density Boltzmann theory allows to recast the r.h.s of Eq. (\ref{kohler}) as $F\left(  \frac{H}{\rho(0,T)}  \right) \propto \left( \frac{H}{\rho(0,T)} \right)^2$
up to lowest order in the field ('Kohler's plot').

In cuprates it became apparent early on \cite{harris}
that Kohler's plot is strongly temperature-dependent,
and therefore two different scattering rates or anisotropic scattering needs to be introduced \cite{hussey03,hussey08}. More recently, it was also
demonstrated that Kohler's rule is violated with regard to the field dependence: particularly at low temperature, in the region where
strange metal behavior is observed, the magnetoresistance displays a linear dependence 
on $H$ together with the transport scattering time having a peculiar scaling
T and H dependence\cite{ayres-2021,giraldo-gallo-2018}
$\tau^{-1}\propto \sqrt{(ak_BT)^2+( b\mu_BH)^2}$ 
($k_B$ is the Boltzmann constant and $\mu_B$ is the Bohr magneton) with 
$a$ and $b$ dimensionless constants of the same order. 

This scaling form of the magnetoresistance would put in jeopardy the
validity of the whole SFL scenario since in this latter the T-linear resistivity
simply arises from the nearly classical character of the CDF, in turn due
to the possibility of expanding the Bose function, which is a function
of $\omega/T$ only. Therefore, within this scheme, there is no reason 
why the T-linear resistivity should be linked to a H-linear magnetoresistance
as instead implied by the above $T-H$ scaling form. 
However, the very existence of such a scaling function   
in magnetoresistance was questioned by experiments in LSCO samples\cite{ataei-2022},
where a different scenario was proposed.
The crucial point made in Ref. \onlinecite{ataei-2022} is that the
scattering rate can be decomposed into an isotropic
inelastic scattering (responsible for the T-linear resistivity)
and a strongly anisotropic elastic scattering
(mostly effective at low $T$ where the inelastic scattering becomes subdominant),
\begin{eqnarray}
    \frac{1}{\tau(\phi,T)}=c\left[ \frac{1}{\tau_0}+
    \frac{1}{\tau_{aniso}}\vert \cos(2\phi)\vert^\nu\right]+\alpha k_BT/\hbar
    \label{ataei}
\end{eqnarray}
where $\nu$ is a phenomenological index ruling the degree of elastic scattering anisotropy.
The Chambers’ solution to the Boltzmann
equation \cite{chambers} with this composite scattering time
is valid to all orders in $H$ and it is found that in the limit of large fields the magnetoresistance scales
as $\lim_{H\to\infty}\rho(H,T) \sim H$. ~\cite{hussey22}. Therefore
this composite scattering leads to a crossover from quadratic to linear-in-H
behavior of the magnetoresistance with increasing $H$ where the
crossover field scales with the isotropic component $c/\tau_0+\alpha k_B T/\hbar$.~\cite{caprara24}.

\begin{figure}
    \centering
    \includegraphics[width=1.\linewidth]{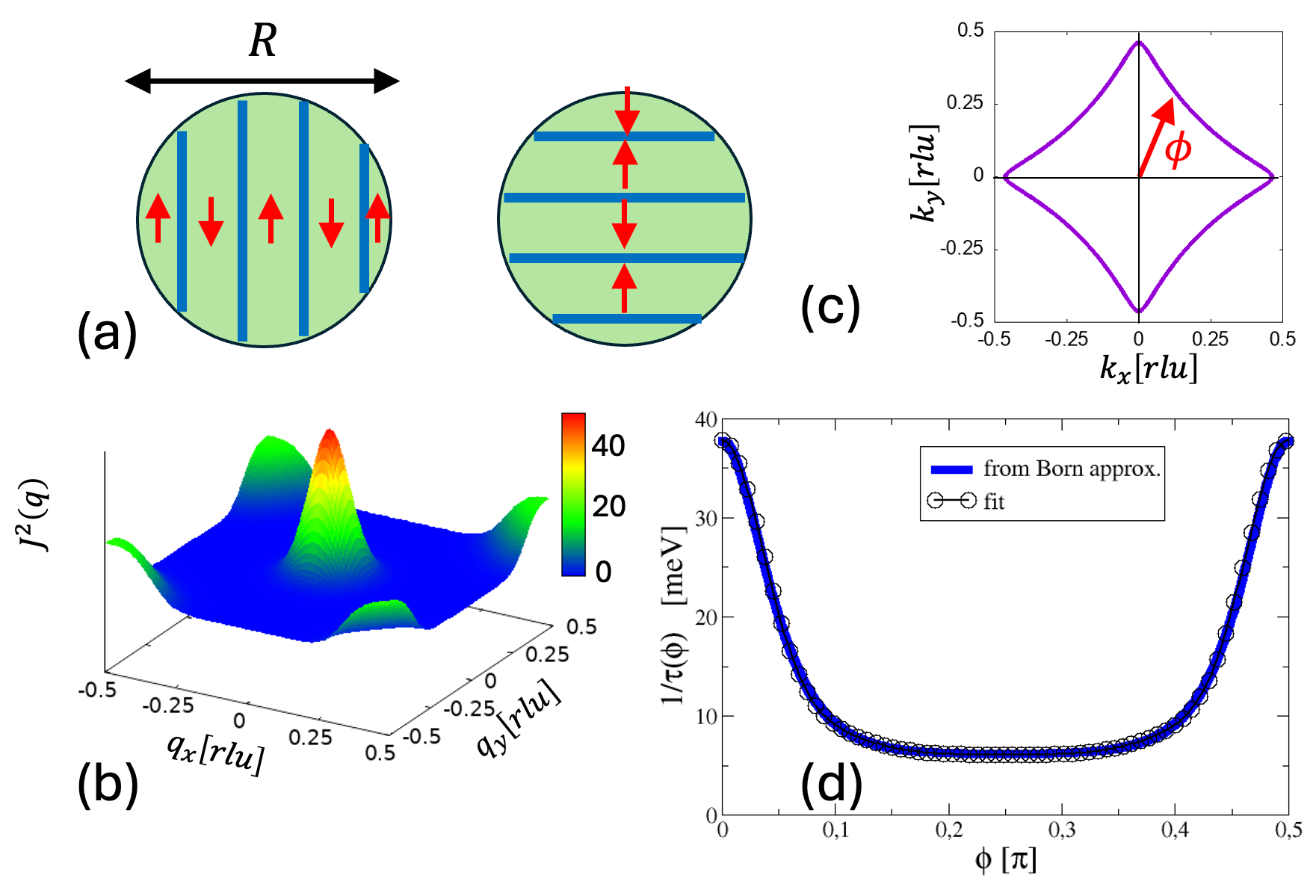}
    \caption{(a) structure of the static stripe puddles with diameter $R$. Arrows indicate the sign of the AF order parameter. (b) Fourier transformed potential $|V^c(q)|^2+|V^s(q)|^2$, Eqs. (\ref{eq:vcfour},\ref{eq:vsfour}). (c) Definition of the angle $\phi$ entering the definition of the scattering rate $1/\tau(\phi)$ along the Fermi surface. (d) Scattering rate obtained from 
    Eq. (\ref{eq:selfpot}) $1/\tau=\text{Im} \Sigma(E_F)$ (blue) compared with the parametrization (circles)
    $1/\tau [meV]=6.1+25.62\cos^{20}(2\phi)+6.1\cos^4(2\phi)$.
    Parameters: $t'/t=-0.1364$, $t''/t=0.0682$, doping $p=0.25$, $L=Ls=2.5$, $g_s=g_c\equiv g$ with
    $c_p g^2=12.2 (meV)^2$. The puddle diameter $R$ can be estimated from the  full width at tenth of maximum of Eq. (\ref{eq:pud}) as
    $R\approx 2L\sqrt{ln(10)}\approx 8$ in units of the lattice spacing.}
    \label{aniso-scatt}
\end{figure}

This scenario is fully compatible with the 
SFL theory, which attributes the isotropic scattering to the dynamical
character of the CDF. The strongly anisotropic elastic scattering can instead 
be naturally attributed to an inhomogeneous distribution of CDW static puddles.
While deep in the overdoped region such puddles could hardly be present,
the strange metal phase where H-linear magnetoresistance is observed, is
only weakly overdoped and not too far from the CDW-QCP. Therefore, even mild static disorder
might be effective in pinning and making static a non negligible amount of such extended
scattering centers. In the following we
investigate this scenario specifically for LSCO
and Nd-LSCO, investigated in Ref. \onlinecite{ataei-2022},
by introducing nanoscale puddles
having an internal stripe structure with antiphase domain walls, cf. Fig. \ref{aniso-scatt}. 
Charge carriers can be scattered both from the puddles itself but also from the internal 
stripe modulation, where for the latter the spin contribution by far exceeds the 
charge component which amplitude is rather small.~\cite{sei02,abba05}  
Recent inelastic neutron scattering experiments\cite{hayden-2026} have indeed revealed
the presence of quasistatic disordered spin stripes, which in LSCO and Nd-LSCO
are well known to be intertwined with charge ordering or charge fluctuations.
On a lattice, a single stripe puddle, centered around the origin, can be modeled by the following hamiltonian
\begin{equation}
\hat{H}_{pud}=\sum_n \left\lbrace V^{c}(R_n) \hat{n}(R_n)
+V^{s}(R_n) \hat{S}^z(R_n)\right\rbrace
\end{equation}
with
\begin{eqnarray}\label{eq:pud}
V^c(R_n)&=&\frac{g_c}{\pi L^2} e^{-\frac{x_n^2+y_n^2}{ L^2}} \\
V^s(R_n)&=&\frac{g_s}{\pi L_s^2} e^{-\frac{x_n^2+y_n^2}{ L_s^2}} (-1)^{x_n+y_n}\cos(Q_s x_n) \nonumber \,.
\end{eqnarray}
Here $g_c$ and $g_s$ characterize the strength of the
respective potentials and $L_s$ is the stripe correlation length which for simplicity we have taken equal along 
and perpendicular to the stripe. The wave vector of the antiferromagnetic (AF) order modulation 
is $Q_s=2\pi/8$ which in Eq. (\ref{eq:pud}) is defined along the $x$-direction, 
however, below we will average over both $x$ and $y$ orientations.
In the limit of weak scattering, the Born approximation for the self-energy 
associated with a random distribution of puddles (concentration $c_p$)
reads
\begin{equation}
\Sigma(\vec{k},\omega)=c_p \sum_q \left\lbrace |V^c(\vec{q})|^2+|V^s(\vec{q})|^2\right\rbrace G^0(\vec{k}-\vec{q},\omega)\label{eq:selfpot}
\end{equation}
where the Fourier-transformed potential
\begin{eqnarray}
V^c(q)&=& g_c e^{-\frac{1}{4}q^2 L^2} \label{eq:vcfour}\\
V^s(q)&=& \frac{1}{2} g_s
\left\lbrace e^{-\frac{1}{4}(q_y-\pi)^2 L_s^2}
e^{-\frac{1}{4}(q_x-\pi\pm Q_s)^2 L_s^2} \right. \nonumber \\
&+&\left. e^{-\frac{1}{4}(q_x-\pi)^2 L_s^2}
e^{-\frac{1}{4}(q_y-\pi\pm Q_s)^2 L_s^2}
\right\rbrace  \label{eq:vsfour}
\end{eqnarray}
takes into account puddles with stripe orientations along both $x-$ and $y-$ direction, see Fig. \ref{aniso-scatt}. The non-interacting
Green's function $G^0=1/(\omega-\mu-\varepsilon_k)$
depends on the electronic dispersion $\varepsilon_k$ for which we include nearest ($t$), next-nearest ($t'$), and next-next nearest ($t''$) neighbor hopping, respectively.

 The potential Eqs. (\ref{eq:vcfour},\ref{eq:vsfour}) induce a highly anisotropic static scattering rate
$1/\tau_k=Im \Sigma(\vec{k},\omega=0)$
due to the local density of states $N(k)$ which is small (large) in the nodal (antinodal) regions. Whereas $V^c(q)$ is pronounced around $q=0$ and therefore picks up a (convoluted) contribution of $N(k)$ along
the Fermi surface, $V^s(q)$ only contributes around the antinodal regions where the associated $Q_{AF}\pm Q_s$ scattering is close to a nesting vector.
Therefore, contrary to the usual forward scattering induced
by extended circular and internally homogeneous puddles (i.e. the $V^c(q)$ contribution), the present internally structured puddles are much more effective in generating a strongly anisotropic scattering rate along the Fermi surface. 

\begin{figure}
    \centering
    \includegraphics[width=1.0\linewidth]{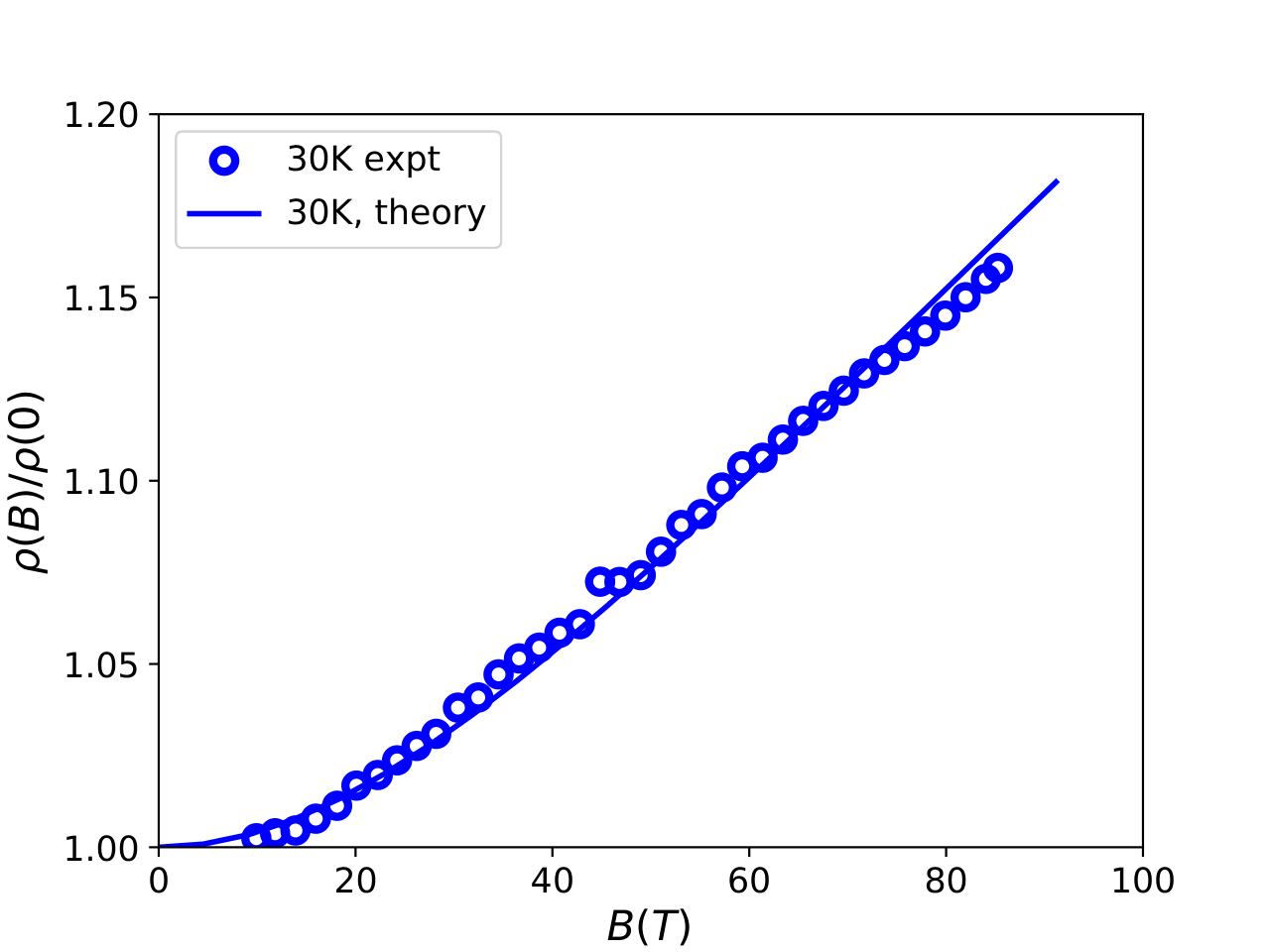}
    \caption{Magnetoresistance of Nd-LSCO at doping $p=0.24$ (circles) measured at temperature $T=30 K$ (from Ref. \onlinecite{ataei-2022}
    as a function of the magnetic field (in Tesla).
    The solid line is the result from Boltzmann theory using the scattering rate shown in Fig. \ref{aniso-scatt}d and band structure parameters reported in the caption to Fig. \ref{aniso-scatt}.}
    \label{MR}
\end{figure}

In order to fix the parameters entering Eqs. (\ref{eq:vcfour},\ref{eq:vsfour}) we compute the
magnetoresistance within a Boltzmann equation approach
and compare with experimental data from Ref. \onlinecite{ataei-2022}, see Fig. \ref{MR}.
It turns out that in order to fit the data we have to
invoke
puddles with diameter $\sim 8$ unit cells which cover $\sim 2$ charge stripes and one period of AF spin order modulation.
This will also limit the stripe correlation length to a 
similar scale which can be expected with regard to the large doping
close to the termination of the spin stripe phase.~\cite{gaulin21}
The parametrization of the scattering rate $1/\tau$ resulting from the fit, which is shown 
in Fig. \ref{aniso-scatt}(d), bears some similarity with the phenomenological form 
Eq. (\ref{ataei}) proposed in Ref. \onlinecite{ataei-2022} with the difference that 
for the present model a sum of two $\cos^\nu(2 \phi)$
functions with different $\nu$ is needed to account for the anisotropy.

We have thus shown that in order to account for the 
linear-in T and linear-in H behavior of the transport scattering time at large fields 
in the strange metal phase one does not necessarily have to rely on scaling arguments. 
Rather, the corresponding experiments can also be explained on the basis of a scenario 
where the linear-in T behavior due the inelastic scattering from CDF's is 
supplemented by a strongly anisotropic elastic scattering from stripe puddles which, 
when properly treated in a Boltzmann approach for strong fields, leads to the observed linear-in H 
behavior of the magnetoresistance.

One should point out that the argument above relies on the experimental data from Ref. \onlinecite{ataei-2022}
which for LSCO and Nd-LSCO demonstrate a field dependence of the magnetoresistance only for the magnetic field 
applied perpendicular to the CuO$_2$ planes. Similar findings have also been reported for the three 
layer compound Tl-2223.~\cite{shi25} This supports the point of view that it is 
predominantly the magnetic moment from the electron's motion in the plane which couples to the field.
On the other hand, earlier experiments on overdoped Tl2201 and Bi2201 \cite{ayres-2021} 
suggest that the magnetoresistance is comparable for all field
orientations being more in line with a Zeeman coupling of the electron spin to the 
magnetic field \cite{chung24}. Therefore, it is possible that both orbital and Zeeman 
coupling contribute to magnetoresistance, with their respective strengths depending on material and doping.

Another point to note is that our derivation of the scattering rate relies on the plausible assumption
of the existence of small stripe puddles which can
produce a strongly anisotropic elastic scattering rate.
In fact, there is no one-to-one correspondence between the momentum dependence of the scattering potential $V(q)$ 
and the anisotropy of $1/\tau(\phi)$ so that we cannot exclude that other shapes of $V(q)$ 
lead to similar scattering rates. 
However, we also considered a similar type of domains with an elliptic elongated shape and internal charge modulation
(not spin) such as those that have been 
observed in underdoped YBCO\cite{comin-2015}. Assuming that a small amount of
such inhomogeneities might survive at low temperature for $p\sim p^*$, we checked whether the elongation might 
induce a stronger anisotropy. Although this helps, we have checked that a $V^c(q)$ only approach 
(i.e. finite size puddles from e.g. out-of plane impurities) is not sufficient to generate the 
necessary anisotropy which instead can be achieved by including the internal (short-range) stripe structure.

\section{Discussion and conclusions}
\label{sec:discuss}
As mentioned in Sect. I.A, this work is the natural completion of the long story
constructed along the years around the relevance in cuprates of charge ordering tendency and CDW-QCP
both for pairing and for anomalous normal state properties. Besides the long-standing 
proposal of the possible relevance of dynamical CDW both for the d-wave 
particle-particle pairing\cite{perali-1996}, 
and as possible source of pseudogap,\cite{andergassen-2001,hashimoto-2010}
the present work focuses on the role of CDF, which are the only low energy excitations in the 
strange metal region. 
In the particle-particle framework
 CDF could as well play a role in pairing due to their dynamical retarded character. 
% This should be
%investigated within an Eliashberg scheme since it might occur that the instantaneous residual
%quasiparticle electronic repulsion selects the $d$-wave channel and the screening effects 
%weaken it in order to be overcome by the dynamical/retarded pairing mediated by the CDF.
%This issue has still to be investigated.

The main concern of the present work is on the strange metal properties 
and the crucial role played by CDF.
The main aim is to show that within the SFL scheme, using the RIXS spectra, 
we could account for all the experimental challenges posed by the strange metal 
in the V shaped region of the $T-p$ phase diagram around the doping $p^*$ (see Fig. \ref{phase-diag}). 
 The RIXS spectra of Figs. \ref{fig-rixs} and \ref{ybco-rixs}
(and the similar spectrum for LSCO `guessed' in Fig. \ref{lsco-rixs})
have been extensively used in this paper to illustrate the success of the new approach of the SFL 
in matching the experiments. These experiments range from resistivity, to optical conductivity, and Raman
in various families (BSCCO, YBCO, LSCO), above $T_c$ in section II.

The specific heat, heat conductivity, Seebeck coefficient, resistivity, and magnetoresistance below $T_c$ 
in the presence of strong magnetic field to suppress superconductivity are then discussed in section III.
We essentially find four main results supporting the SFL scenario: i) CDF are an experimentally well
established scattering mechanism at low energy, which can 
account for the strange metal properties; ii) the finite energy scale $\omega_{CDF}=M/\gamma(T)$
is the crucial handle tuning the crossover to the low temperature FL behavior, the strong or small
temperature dependence of the optical scattering rate, and its scaling violation; iii) when $M/\gamma$
is smaller than T, linear $\omega/T$ scaling is naturally accounted for
bearing a strong similarity with MFL theory; iv) the logarithmic increase in the
dissipation parameter $\gamma(T)$ accounts for the low-T transport and thermodynamic strange metal properties.

Regarding point i), above $T_c$ we made the paradigmatic change of referring to fluctuations with finite small correlation length, 
rather than nearly critical modes (CDW or AFM fluctuations). Our simple new paradigm is therefore based on the presence 
of almost local dynamical low energy noncritical CDF,  which are so broad (small correlation lenght $\xi$)
that, when used as mediators of scattering of Fermi quasiparticle, produce nearly uniform scattering on the 
entire Fermi surface. Moreover, their characteristic frequency $\omega_{CDF}=M/\gamma$ 
is of the order of $T_c$ or less, as experimentally measured by high resolution RIXS experiments. This leads to a scenario
as simple as a FL with Landau quasiparticles scattering with thermally excited broad-in-momentum
low-energy collective excitations.

%\textcolor{red}{LEAVE OR OUT? THE DISCUSSION IS ALREADY LONG. DO WE REALLY NEED THIS PARAGRAPH?}

%\textcolor{red}{
%Besides T-linear resistivity, another puzzling issue of transport in cuprates 
%is related to the temperature dependence of the
%Hall angle. Specifically it was found\cite{ong-1991} that a $T^2$ behavior of the
%Hall angle cotangent is hardly reconciled with a T-linear resistivity
%(see, e.g., Ref. \onlinecite{coleman-1996} and references therein). 
%Actually, an analysis of several experimental works seems to indicate
%that ${\rm cotan} (\theta_H) \sim T^2$ is not an universal behavior
%and it does not occur in the strange metal region, but is rather characteristic
%of weakly underdoped cuprates, which will be the focus of a subsequent %work.
%This is why we did not address this intriguing issue here.
%}

Regarding point ii),  first of all $M/\gamma$ is the scale below which FL properties should be restored.
However, the doping dependence of this quantity is also relevant to determine qualitative differences in the 
optical scattering rate of underdoped and overdoped cuprates. Specifically it determines the strong or weak
temperature dependence of the optical scattering rates in overdoped or underdoped samples respectively (see
Fig. \ref{opt-scatt-rate}). This may also
shed light on the old observation\cite{tallon-2001} that the pseudogap crossover temperature $T^*$ is related
to a temperature independent, but doping dependent energy scale: $M/\gamma$ might well be this scale for $T\gtrsim T_c$
and its analysis in this framework will be the topic of a forthcoming work.

As for $\omega/T$ scaling in the optical scattering rate [point iii)], within the SFL scenario, we also found 
that scaling does not occur because of the diverging
correlation length of some quantum critical low energy excitation. 
In Sec. II.C we showed that $\omega/T$ linear
scaling simply arises for $T\gtrsim M/\gamma$ as the combined effect of low energy CDF scattering
and a broad paramagnon continuum. We also notice that these ingredients to obtain $\omega/T$ linear 
scaling in dynamical properties are quite general as long as strong correlations induce a broad continuum
of particle-hole excitations and low energy excitations are abundant due to the proximity to any generic
QCP as in heavy fermions or in other strongly correlated systems.
Of course further analysis of optical data (as well as other dynamical spectroscopic quantities)
in cuprates and other systems are in order to test our proposal that this scaling may
simply arise from the cooperative effect of a low energy excitation and a broad continuum.

Finally, regarding point iv), we here extend and systematize with experimental comparisons the idea that
an increase of dissipation (i.e. of lifetime) for the CDF accounts for the anomalous 
logarithmic contributions to specific heat, Seebeck coefficient, and magnetoresistance in cuprates below $T_c$.
Specifically, around the doping  $p^*$,  an increase of 
the dissipation parameter $\gamma(T)$ induces a decrease of $\omega_{CDF}$ down to the lowest measured T 
without any divergence of the charge density correlation length. 
If $\gamma$  increases logarithmically,  the SFL scenario accounts both for
a purely linear $\rho\sim T$ down to low $T$, or for a mixed $\rho\sim AT+BT^2$ over extended doping regions.
%A logarithmic growth of $C_V/T$ and the Seebeck coefficient $S(T)/T$ are also interpreted within the SFL theory.
Moreover, this logarithmic form of $\gamma$ may be obtained by assuming that a two-dimensional CDF can decay 
in slow diffusive particle-hole pairs\cite{grilli-2023}. The formation of a CDF glassy phase is
also currently investigated\cite{ketmaier-2026}. In any case, whatever the dissipation mechanism is, if 
$\gamma$ increases by lowering T at doping $p^*$, all the fluctuating droplets of size $\xi$ 
tend to relax with longer and longer relaxation time $\tau_q$ $(1/\tau_q=[M+\nu(q-q_c )^2]/\gamma)$ 
giving rise to `anomalous'  slowing down and resulting in an almost persistent (possibly non equilibrium)  
state, correlated in time  and not in space.

In this framework we propose that the logarithmic increase of the specific heat coefficient has a bosonic
origin and that the fermionic contribution, although displaying a partial logarithmic growth,
saturates to a finite value and it is not quantitatively sufficient. We also propose that extending further
at lower temperature the Seebeck coefficient measure, a quadratic logarithmic contribution $\gamma(T)^2$ 
should become visible. We hope that this expectation will be experimentally tested. 

Regarding the $H/T$ scaling in magnetoresistance, we extend the analysis of Ref. \onlinecite{ataei-2022}
introducing a model of scattering both inelastic and isotropic due to CDF and 
elastic and anisotropic due to extended puddles
of more static intertwined CDW and spin, i.e. stripe, order. 
We find in Sec. III.E a good agreement between experiments and the expectations of our model.

A final remark on shot-noise experiments.
At low T the SFL scenario involves fluctuations, which become increasingly slower
the more $T$ is lowered. These fluctuations are therefore highly populated even at low temperature
and can mask the particle-like features of charge transport in shot-noise experiments: 
it is well known that phonon scattering washes out the shot-noise features in nano-bridge transport 
experiments when temperature is increased or long bridges are considered. In the present 
scenario, even at very low $T$ the quasiparticles move in a highly dissipative medium of
very low energy excitations that act like `phonons' in standard systems. This is why
measuring the current noise  at low T under strong magnetic fields to suppress superconductivity,
we should expect a strong suppression of the Fano factors similarly to what has recently been observed
in a heavy fermion YbRh$_2$Si$_2$ strange metal\cite{natelson-2023}. 

Our SFL scenario provides a simple alternative to the fancier approaches which 
couple {\it \`a la} Yukawa the fermion quasiparticles with low-energy nearly local dynamical boson degrees of freedom.
This class of models strongly rely on disorder to transform a nearly critical low-momentum scattering
(e.g. nematic) into momentum-dissipating nearly local excitations, which acquire the properties of a
SYK model (see, e.g., Refs. \,\onlinecite{sachdev23_1}-\onlinecite{schmalian-2024}, \onlinecite{valentinis-2026} 
and references therein). 
Similarly, our SFL scenario also relies on semiclassical nearly local fluctuations as the main source of
scattering. However, the SFL scenario does not need strong disorder because the CDF provide a large
momentum scattering mechanism, largely involving umklapp scattering processes 
and effectively dissipating momenta.

Generally speaking our approach follows the principle of Occam's razor: Instead of relating 
the strange metal properties of cuprates to some more or less exotic quantum criticality, 
we refer to experimentally observed excitations,
namely CDF, phonons, and paramagnons. We then feed them in a simple first order perturbative
calculation finding that, although non critical, CDF can be
responsible for the strange metal behavior, and that together with the broad paramagnon 
continuum  they account for the observed (approximate) scaling features. On top of this, we also
notice that experimentally observed kinks, scaling violations and other features of 
the spectra can be explained by experimentally determined scales like $M/\gamma$, 
coupling to phonons and so on.

%\textcolor{red}{Importantly, despite the proximity to the CDW-QCP providing abundant low energy CDF,
%the low energy scattering below $T_c$ 
%does {\it not} arise from a form of local criticality, but rather on
%mechanism(s) of increasing dissipation of the CDF.  MAYBE THIS IS TOO REPETITIVE: WE SAID THIS ALREADY MANY TIMES...}
%Still, our scenario was filtered along the years passing trough the quantum criticality paradigm. 
%In the search of a critical strong mediator of scattering, we suggested that near optimal doping, 
%in the absence of superconductivity, the strongly correlated FL would easily be unstable for CDW formation 
%at a hidden QCP near optimal doping (see Fig. \ref{phase-diag}) much before its discovery. Critical fluctuations,
%however, mediate non-FL behavior in very limited regions of the Fermi surface only (hot points), 
%thus producing Fermi liquid behavior altogether. This motivated the search of dynamical charge density 
%fluctuations sufficiently near $p_{CDW}$ but not too near $(p\geq p^*>p_{CDW)}$. In this way,  
%these fluctuations have  low characteristic energy  and a finite small correlation 
%length and broad momentum distribution to be mediators of isotropic scattering. Their characteristic 
%energy $M/\gamma$ allowed us to introduce the temperature $T_{FL}$  above which strange metal occurs 
%which experimentally turned out to be of the order of $T_c$ or less. 
%Then, to explain the strange metal properties also below $T_c$ we  extended this paradigmatic 
%change of CDF with finite small correlation length, rather 
%than critical modes.

At the end of day, we have obtained a complete scenario for the cuprates associated to their  
marked tendency to charge ordering
for doping close (but not too much) to the hidden CDW-QCP  and based on the CDF near $p^*$, the
end point of the pseudogap onset line $T^*(p)$.  The question naturally arises whether $p^*$ is the putative 
QCP with the vanishing of the $M/\gamma$ energy scale when $\gamma\to \infty$. In any case, at low temperature 
the SFL scenario, fully accounting for the present experimental situation, simply describes a system of FL 
quasiparticles strongly interacting with highly populated dynamical nearly local CDF, which become 
increasingly slower the more T is lowered. We hope that this proposal will stimulate further
experimental efforts to challenge and possibly improve this theoretical scheme.

\vspace{2 truecm}

\noindent {\bf Acknowledgments}

\noindent
We acknowledge interesting discussions with Riccardo Arpaia, Giacomo Ghiringhelli,
Lucio Braicovich, Sergio Ciuchi, and Davide Valentinis.

G.S.  acknowledges support from the DFG under SE806/20-1. 
M.G. and S.B. acknowledge financial support from the PNRR MUR Project No. PE0000023-NQSTI, and 
specifically the project “Topological Phases of Matter, Superconductivity, and Heterostructures” 
Partenariato Esteso 4-Spoke 5 (Grant No. PE4221852A63A88D).  
S.C. and M.G. acknowledge financial support from the Ateneo Research Projects:
“Models and theories from anomalous diffusion to strange-metal behavior”
(Grant No. RM12218162CF9D05),
“Non-conventional aspects for transport phenomena and non-equilibrium statistical mechanics” (Grant No. RM123188E830D258),
“Elementary excitations at the origin of glassy or hexatic behavior in low dimensional system, at and out of equilibrium” (Grant No. RM124190C54BE48D).
S.C. acknowledges financial support from the Ateneo Research Project:
“Interplay of phonons and charge collective excitations in cuprate high-$T_c$ superconductors”  (Grant No. RP125199B9FDBFE4).

%%%%%%%%%%%%%%%%%%%%%%%%%%%%%%%%%%%%%%%%%%%%%%%%%%%%%%%%%%%%%%%%%%%%%%%%
%\begin{figure*}
%\vspace{-1 truecm}
%\includegraphics[angle=0,scale=0.7]{prova-nico.eps}
%\includegraphics[angle=0,scale=1.]{./Caprara_fig1.eps}
%\vspace{0.5 truecm}
%\caption{(Color online) 
%}
%\label{fig1}
%       fig. 1:
%\end{figure*}
%%%%%%%%%%%%%%%%%%%%%%%%%%%%%%%%%%%%%%%%%%%%%%%%%%%%%%%%%%%%%%%%%%%%%%%%.

\appendix

\section{The contribution of 'off-shell' scattering to the conductivity}
\label{app:opt}
In Sec. \ref{sec:sfl} we have fitted resistivity within a Drude
type approach $\sigma(T) \sim \tau(T)$ and the optical relaxation
by identifying it with the (inverse) self-energy arising from the
coupling of quasiparticles to the bosonic spectrum as obtained from RIXS. Within the framework of Boltzmann theory such simplified approach is valid for (a) a momentum independent scattering and (b) scattering processes which are confined to $E_F$. Prerequisite (a) is justified from the broadness of the CDF in momentum space, cf. Sec. \ref{ssec:cdf}, whereas here we will investigate the validity of (b).

Since the self-energy is momentum independent one can, after
performing a Kramers-Kronig transformation for the real part
$\Sigma_1(\omega)$, evaluate the optical conductivity\cite{allen}
\begin{equation}\label{allen}
  \sigma(\omega+i\eta)=\frac{i\kappa}{\omega}
  \int d\omega' \frac{f(\omega')-f(\omega'+\omega)}{\omega-\Sigma(\omega'+\omega+i\eta)+\Sigma^*(\omega'+i\eta)}\,.
\end{equation}
For a layered superconductor $\kappa=e^2 t/(\hbar d)$ with
$t$ related to the kinetic energy and $d$ is the interlayer spacing.
If in Eq. (\ref{allen}) $\omega$ is measured in $eV$ then
$\kappa={\cal O}(1) \left\lbrack eV/(k\Omega cm)\right\rbrack$.
The DC conductivity is then obtained in the limit $\omega\to 0$
as
\begin{equation}\label{allenw0}
\sigma_1(0)=2\beta \kappa\int_0^\infty d\omega' \frac{\frac{\partial f(\omega')}{\partial\omega'}}{2\Sigma_2(\omega')}
\end{equation}
where subscripts $1,2$ denote real- and imaginary part, respectively.
For the fit we set $\Sigma_2(\omega)=\rho_0 + \lambda_T \Sigma^{RIXS}_2(\omega)$ where $\Sigma^{RIXS}_2(\omega)$ is obtained
from Eq. (\ref{sigma2}).
In the limit $T\to 0$ the derivative of the Fermi function only 
picks up the contribution at $\omega'=0$ and the resistivity
$\rho \sim 1/\Sigma^{RIXS}_2(0)$ reduces to the form used in Sec.
\ref{sec:sfl} to fit the experimental resistivity data.

\begin{figure}
    \centering
    \includegraphics[width=0.8\linewidth]{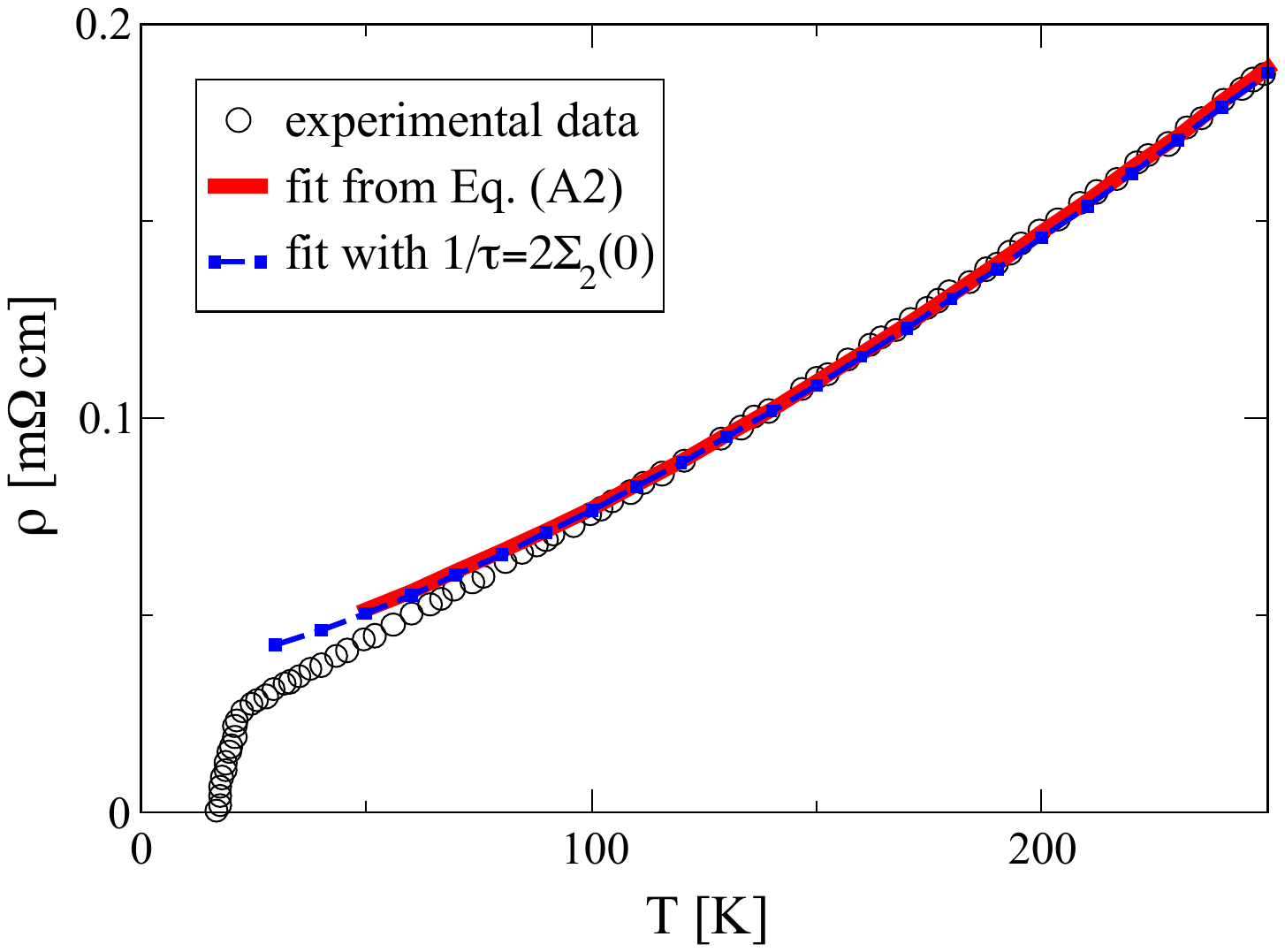}
    \caption{DC resistivity data (circles) for a LSCO sample
($p = 0.24, Tc = 19K$) from Ref. \onlinecite{michon-2023} 
The fit from Eq. (\ref{allenw0}) is shown by the red line with 
$\lambda_T = 0.01 \cdot 10^{-3}\Omega cm$ and
$\rho_0 = 0.037 \cdot 10^{-3}\Omega cm$. The blue dashed line corresponds to the fit $\rho \sim 1/\Sigma^{RIXS}_2(0)$ already shown in Fig. \ref{lsco-rho} with $\lambda_T = 0.0111 \cdot 10^{-3}\Omega cm$ and $\rho_0 = 0.037 \cdot 10^{-3}\Omega cm$.}
    \label{ybco-rixs-supp}
\end{figure}

As exemplified for the case of LSCO, Fig. \ref{ybco-rixs-supp} demonstrates that the fits with and without the
inclusion of off-shell contributions account equally well for the temperature dependent resistivity \cite{michon-2023} when one considers a difference
in the coupling of $\sim 10\%$.

\begin{figure}[thb]
\includegraphics[width=8cm,clip=true]{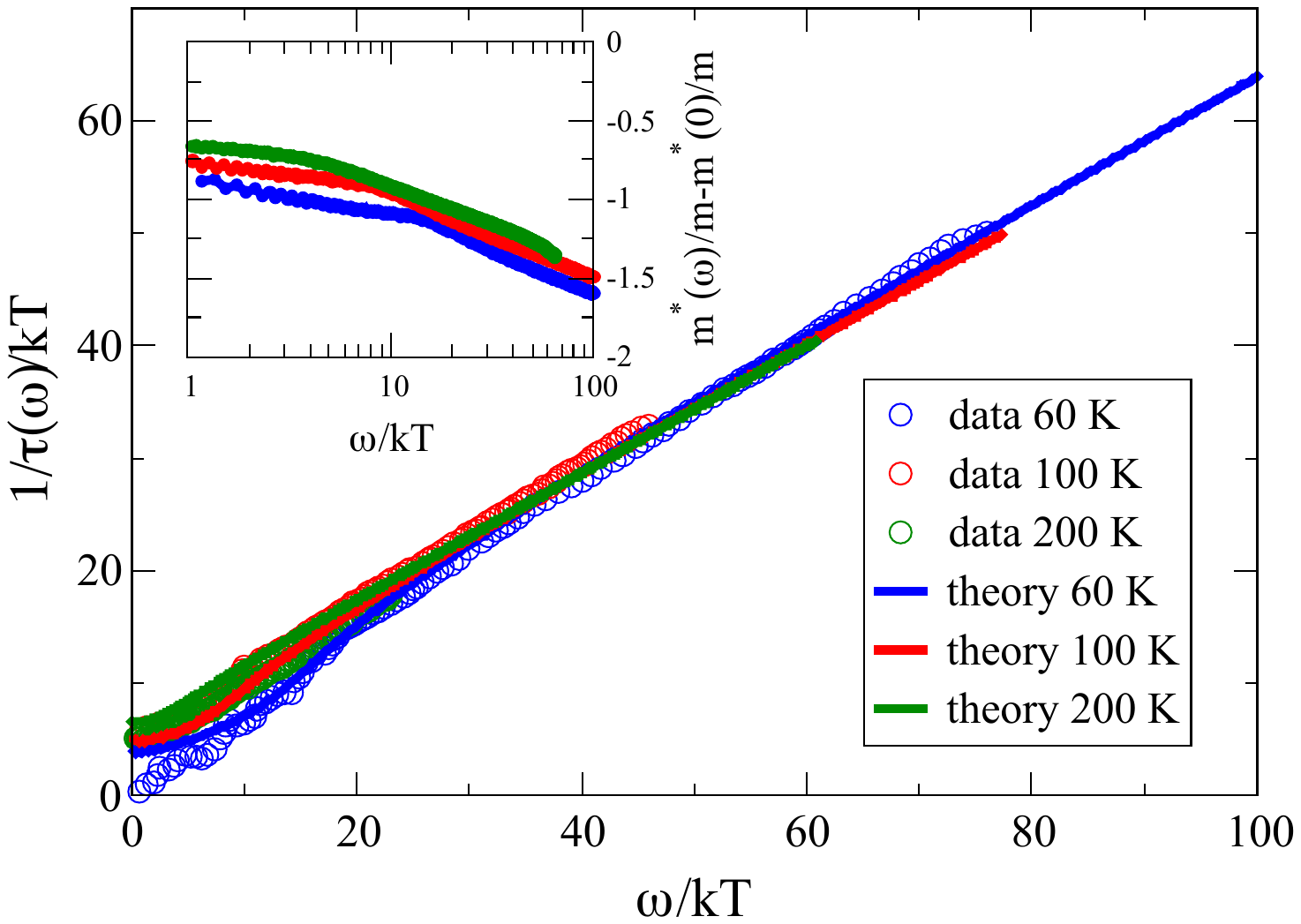}
\caption{Main panel: Scaling of the optical scattering rate
  obtained from Eq. (\ref{eq:tau}) and compared with experimental data for LSCO \cite{michon-2023}. Inset: optical mass enhancement calculated from Eq. (\ref{eq:mass}.)
  Parameters: $\lambda_{\omega}=2.5 $ meV, $1/\tau_0=0 $ meV, $\kappa= 1 eV/(k\Omega cm)$.}
\label{figscal1}                                                   
\end{figure}  

Similarly, also the on-shell fits of the frequency dependent data in Sec. \ref{sec:sfl}
can be substantiated via Eq. (\ref{allen}.
In particular, we calculate the optical relaxation time
$\tau(\omega)$ and optical mass enhancement according to
\begin{eqnarray}
  \frac{\hbar}{\tau(\omega)}&=& \kappa \mbox{Re}\left\lbrack\frac{1}{\sigma(\omega)}\right\rbrack \label{eq:tau}\\
  \frac{m^*(\omega)}{m}&=&-\kappa \mbox{Im}\left\lbrack \frac{1}{\hbar\omega\sigma(\omega)}\right\rbrack \,, \label{eq:mass}
\end{eqnarray}
which are shown in Fig. \ref{figscal1}
for the case of LSCO.
The optical relaxation (main panel) displays similar scaling behavior than the self-energy
in Fig. \ref{lsco-optics} and an equally good agreement wit experimental data \cite{michon-2023} when the coupling
is renormalized by $\sim 15\%$.
In fact, in the small frequency limit and for $k_BT\to 0$ $1/\tau$
is solely determined by $\Sigma''(\omega)$,
cf. Eq. (\ref{allenw0}).
Also the approximate collapse of the mass enhancement (see inset to Fig. \ref{figscal1}) is obtained from Eq. (\ref{eq:mass}) and
compares well with the corresponding experimental data (cf. Fig. 5d in Ref. \onlinecite{michon-2023}) 
although significant noise in the latter puts in jeopardy a direct comparison.

\section{Raman response from a local self-energy}\label{ap:raman}
We calculate the Raman response function 
\begin{equation}
r(i\omega_\mu)=-\frac{1}{\beta}\sum_{k,\nu} \gamma_{k,s}^2 G(k,i\omega_\nu
+i\omega_\mu)G(k,i\omega_\nu)
\end{equation}
where $\gamma_{k,s}$ defines the Raman vertex for symmetry $s=B_{1g},B_{2g}$ 
and the self-energy in the Green's function
\begin{equation}
G(k,i\omega_\mu)=\frac{1}{\omega_\mu-\varepsilon_k-\Sigma(\omega_\mu)}
\end{equation}
is obtained from Eq. (\ref{sigma2}) with the real part calculated via the Kramers-Kronig relation.
Upon introducing the spectral function $A_k(\omega)=-(1/\pi) Im G(k,i\omega_\mu \to \omega+i\delta)$ and performing the Matsubara sum the response function can be written as
\begin{eqnarray}
r(i\omega_\mu)&=&-\int d\epsilon N_s(\epsilon)\int d\omega_1 \int d\omega_2 A_k(\omega_1) A_k(\omega_2)\nonumber \\
&\times& \frac{f(\omega_2)-f(\omega_1)}{\omega_2-\omega_1+i\omega_\mu}\,.
\end{eqnarray}

\begin{figure}
    \centering
    \includegraphics[width=0.7\linewidth]{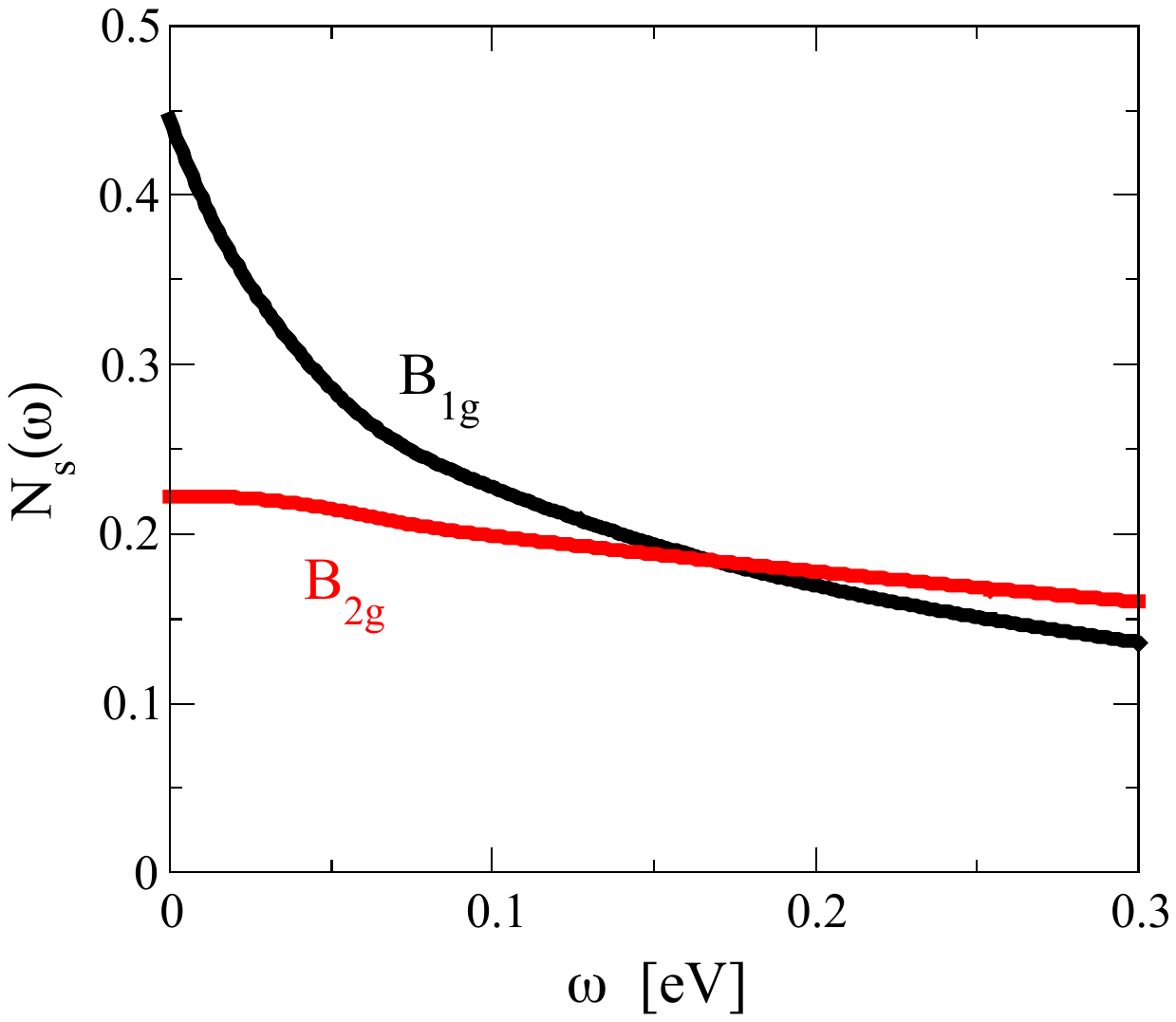}
   \caption{Vertex function $N_s(\omega)$ for $B_{1g}$ and B$_{2g}$ symmetry. 
   Parameters are those for YBCO used in Fig. \ref{fig:raman}b.}
    \label{fig-vertex}
\end{figure}
Here we have defined the spectral function projected on symmetry $s$
\begin{equation}\label{eq:vertex}
N_s(\epsilon)=-\frac{1}{\pi}\sum_k Im \frac{\gamma_{k,s}^2}{\omega-\varepsilon_k+i\delta}
\end{equation}
and $f(\omega)$ denotes the Fermi function.
Since the Raman form factor for B$_{1g}$ symmetry $\gamma_{k,B_{1g}}$ picks up
spectral weight from the antinodal regions, $N_{B_{1g}}$ is significantly
larger than $N_{B_{2g}}$ at low energy due to the contribution from the 
large DOS around the saddle points at $(\pm \pi,0)$ and $(0,\pi)$, respectively.
At higher energies, the Fermi surface is closed and moves away from the antinodes
so that $\gamma_{k,B_{1g}}$ becomes increasingly ineffective leading to a crossover of the Raman vertex functions.

The integral over $\epsilon$ involves four contributions
\begin{eqnarray}
&&\int d\epsilon N_s(\epsilon)\left\lbrack
\frac{1}{\epsilon\pm\Sigma'(\omega_1)-\omega_1\pm i\Sigma''(\omega_1)} \right. \nonumber \\
&\times& \left.
\frac{1}{\epsilon\pm\Sigma'(\omega_2)-\omega_2 \pm i\Sigma''(\omega_2)}\right\rbrack
\end{eqnarray}
from which only those contribute where the two poles are located in opposite complex half planes.
For the imaginary part of the Raman response function one
finally finds
\begin{eqnarray}
r''(\omega)&=& \int d\omega' \left\lbrack f(\omega')-f(\omega'+\omega)\right\rbrack \tilde{N}_s(\omega,\omega')\nonumber \\
&\times& Im \frac{1}{\omega-\Sigma(\omega+\omega')+\Sigma^*(\omega')}
\end{eqnarray}
with
\begin{eqnarray}
\tilde{N}_s(\omega,\omega')
&=&\frac{1}{2}\left\lbrack  N_s(\omega'-\Sigma^*(\omega')) \right. \nonumber \\
&+& \left. N_s(\omega+\omega'-\Sigma^*(\omega+\omega') \right\rbrack \, ,
\end{eqnarray}
and $\Sigma^*(\omega)$ denotes the complex conjugate
self-energy.
In the spirit of Allen's approximation \cite{allen} one can
also perform an expansion for small temperatures and frequencies which reads
\begin{equation}
r''(\omega)=Im \frac{N_s(\omega-\Sigma^*(\omega))}{\omega/Z-2i\Sigma''(\omega)}
\end{equation}
with the mass renormalization $1/Z=1-\partial\Sigma'(\omega)/\partial\omega|_{\omega=0}$.

\section{Optical scattering rate from prototypical schematic excitations}\label{ap:ybco}
 In this Appendix we qualitatively discuss how excitations having different 
spectral weight shapes give rise to different optical scattering rate.
 \begin{figure}
    \centering
    \includegraphics[width=0.9\linewidth]{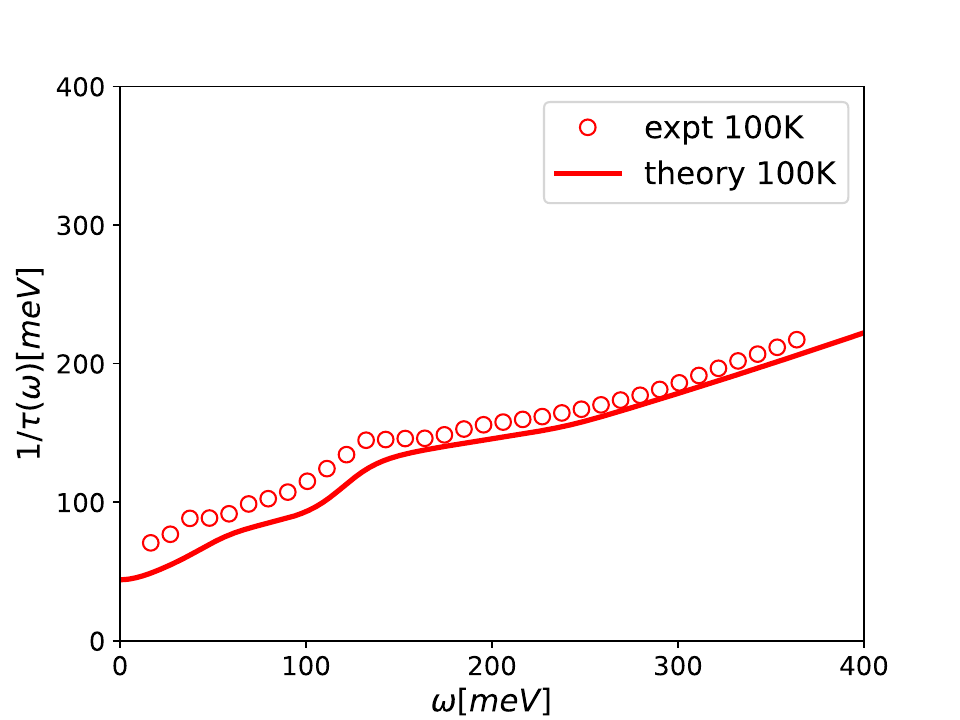}
    \caption{Optical scattering rate for a YCBCO sample (T$_c = 75$ K) from Ref. \onlinecite{uykur-2011} at 100K (red circles), compared to the theoretical fit (red line). The parameters $\lambda_\omega=1.0 $ meV, $1/\tau_0=10 $ meV
are the same as in Fig. \ref{ybco-optics}.}
    \label{ybco-optics-app}
\end{figure}
First of all we can notice that at frequencies much larger than the temperature
the expression for the imaginary part of the quasiparticle self-energy
$\Sigma''$ can be simplified as
\begin{equation}
\label{sigma2-simple}
  \Sigma''(\omega,T<<\omega)\propto \int_0^\omega d\omega' D''(\omega')
\end{equation}
Therefore, if $D''=W\delta(\omega-\omega_0)$ (like, e.g., an Holstein phonon)
then $\Sigma''\sim W\theta(\omega-\omega_0)$, with the step somewhat smoothened 
by T. 
On the other hand, if $D''=W\min(1,\omega/\omega_0)$, 
like, e.g., the continuum of excitations revealed at high frequencies in the RIXS 
spectra reported in this work, then 
 $\Sigma''\sim W \omega^2$ for $\omega<\omega_0$ and then smoothly crosses over a linear dependence
 $\Sigma''\sim W \omega$ for $\omega>\omega_0$. 

 This simple exercise provides a possible clue to interpret the optical scattering
 rate data on YCBCO at 100 K (see Fig. , a lower temperature than the data reported in Fig. \ref{ybco-optics}.
 With respect to the 300 K data one can see that a new sharp kink in $1/\tau(\omega)$ 
 is present at $\omega_0\approx 130$ meV. According to the above discussion, 
 this kink could be interpreted in terms of a  rather narrow excitation
 in $D''(\omega)$ at this energy.  Specifically we needed to add to the RIXS spectra of Fig. \ref{ybco-rixs}
 a Lorentzian peak with the following parameters: $W_X=12.0$, $\omega_X=120$ meV, $\gamma_X=3.0$.
If confirmed, since this energy is too high
 to be attributed to phonons, 
 this kink could suggest the presence of a magnetic excitation which arises
 or becomes narrower by lowering the temperature from 300 K down to 100 K.
 Further experiments would be desirable to investigate this occurrence.  

\bibliographystyle{apsrev}
\bibliography{references}

\begin{thebibliography}{119}
\expandafter\ifx\csname natexlab\endcsname\relax\def\natexlab#1{#1}\fi
\expandafter\ifx\csname bibnamefont\endcsname\relax
  \def\bibnamefont#1{#1}\fi
\expandafter\ifx\csname bibfnamefont\endcsname\relax
  \def\bibfnamefont#1{#1}\fi
\expandafter\ifx\csname citenamefont\endcsname\relax
  \def\citenamefont#1{#1}\fi
\expandafter\ifx\csname url\endcsname\relax
  \def\url#1{\texttt{#1}}\fi
\expandafter\ifx\csname urlprefix\endcsname\relax\def\urlprefix{URL }\fi
\providecommand{\bibinfo}[2]{#2}
\providecommand{\eprint}[2][]{\url{#2}}

\bibitem[{\citenamefont{Martin et~al.}(1990)\citenamefont{Martin, Fiory, Fleming, Schneemeyer, and Waszczak}}]{martin90}
\bibinfo{author}{\bibfnamefont{S.}~\bibnamefont{Martin}}, \bibinfo{author}{\bibfnamefont{A.~T.} \bibnamefont{Fiory}}, \bibinfo{author}{\bibfnamefont{R.~M.} \bibnamefont{Fleming}}, \bibinfo{author}{\bibfnamefont{L.~F.} \bibnamefont{Schneemeyer}}, \bibnamefont{and} \bibinfo{author}{\bibfnamefont{J.~V.} \bibnamefont{Waszczak}}, \bibinfo{journal}{Phys. Rev. B} \textbf{\bibinfo{volume}{41}}, \bibinfo{pages}{846(R)} (\bibinfo{year}{1990}), \urlprefix\url{https://link.aps.org/doi/10.1103/PhysRevB.41.846}.

\bibitem[{\citenamefont{Fournier et~al.}(1998)\citenamefont{Fournier, Mohanty, Maiser, Darzens, Venkatesan, Lobb, Czjzek, Webb, and Greene}}]{greene98}
\bibinfo{author}{\bibfnamefont{P.}~\bibnamefont{Fournier}}, \bibinfo{author}{\bibfnamefont{P.}~\bibnamefont{Mohanty}}, \bibinfo{author}{\bibfnamefont{E.}~\bibnamefont{Maiser}}, \bibinfo{author}{\bibfnamefont{S.}~\bibnamefont{Darzens}}, \bibinfo{author}{\bibfnamefont{T.}~\bibnamefont{Venkatesan}}, \bibinfo{author}{\bibfnamefont{C.~J.} \bibnamefont{Lobb}}, \bibinfo{author}{\bibfnamefont{G.}~\bibnamefont{Czjzek}}, \bibinfo{author}{\bibfnamefont{R.~A.} \bibnamefont{Webb}}, \bibnamefont{and} \bibinfo{author}{\bibfnamefont{R.~L.} \bibnamefont{Greene}}, \bibinfo{journal}{Phys. Rev. Lett.} \textbf{\bibinfo{volume}{81}}, \bibinfo{pages}{4720} (\bibinfo{year}{1998}), \urlprefix\url{https://link.aps.org/doi/10.1103/PhysRevLett.81.4720}.

\bibitem[{\citenamefont{Daou et~al.}(2009)\citenamefont{Daou, Doiron-Leyraud, LeBoeuf, Li, Lalibert{\'e}, Cyr-Choini{\`e}re, Jo, Balicas, Yan, Zhou et~al.}}]{taillefer09}
\bibinfo{author}{\bibfnamefont{R.}~\bibnamefont{Daou}}, \bibinfo{author}{\bibfnamefont{N.}~\bibnamefont{Doiron-Leyraud}}, \bibinfo{author}{\bibfnamefont{D.}~\bibnamefont{LeBoeuf}}, \bibinfo{author}{\bibfnamefont{S.~Y.} \bibnamefont{Li}}, \bibinfo{author}{\bibfnamefont{F.}~\bibnamefont{Lalibert{\'e}}}, \bibinfo{author}{\bibfnamefont{O.}~\bibnamefont{Cyr-Choini{\`e}re}}, \bibinfo{author}{\bibfnamefont{Y.~J.} \bibnamefont{Jo}}, \bibinfo{author}{\bibfnamefont{L.}~\bibnamefont{Balicas}}, \bibinfo{author}{\bibfnamefont{J.-Q.} \bibnamefont{Yan}}, \bibinfo{author}{\bibfnamefont{J.-S.} \bibnamefont{Zhou}}, \bibnamefont{et~al.}, \bibinfo{journal}{Nature Physics} \textbf{\bibinfo{volume}{5}}, \bibinfo{pages}{31} (\bibinfo{year}{2009}), ISSN \bibinfo{issn}{1745-2481}, \urlprefix\url{https://doi.org/10.1038/nphys1109}.

\bibitem[{\citenamefont{Legros et~al.}(2019)\citenamefont{Legros, Benhabib, Tabis, Lalibert{\'e}, Dion, Lizaire, Vignolle, Vignolles, Raffy, Li et~al.}}]{legros-2019}
\bibinfo{author}{\bibfnamefont{A.}~\bibnamefont{Legros}}, \bibinfo{author}{\bibfnamefont{S.}~\bibnamefont{Benhabib}}, \bibinfo{author}{\bibfnamefont{W.}~\bibnamefont{Tabis}}, \bibinfo{author}{\bibfnamefont{F.}~\bibnamefont{Lalibert{\'e}}}, \bibinfo{author}{\bibfnamefont{M.}~\bibnamefont{Dion}}, \bibinfo{author}{\bibfnamefont{M.}~\bibnamefont{Lizaire}}, \bibinfo{author}{\bibfnamefont{B.}~\bibnamefont{Vignolle}}, \bibinfo{author}{\bibfnamefont{D.}~\bibnamefont{Vignolles}}, \bibinfo{author}{\bibfnamefont{H.}~\bibnamefont{Raffy}}, \bibinfo{author}{\bibfnamefont{Z.~Z.} \bibnamefont{Li}}, \bibnamefont{et~al.}, \bibinfo{journal}{Nature Physics} \textbf{\bibinfo{volume}{15}}, \bibinfo{pages}{142} (\bibinfo{year}{2019}), ISSN \bibinfo{issn}{1745-2481}, \urlprefix\url{https://doi.org/10.1038/s41567-018-0334-2}.

\bibitem[{\citenamefont{Greene}(2023)}]{greene23}
\bibinfo{author}{\bibfnamefont{R.~L.} \bibnamefont{Greene}}, \bibinfo{journal}{Physica C: Superconductivity and its Applications} \textbf{\bibinfo{volume}{612}}, \bibinfo{pages}{1354319} (\bibinfo{year}{2023}), ISSN \bibinfo{issn}{0921-4534}, \urlprefix\url{https://www.sciencedirect.com/science/article/pii/S0921453423001107}.

\bibitem[{\citenamefont{Hussey and Duffy}(2022)}]{hussey-2022}
\bibinfo{author}{\bibfnamefont{N.~E.} \bibnamefont{Hussey}} \bibnamefont{and} \bibinfo{author}{\bibfnamefont{C.}~\bibnamefont{Duffy}}, \bibinfo{journal}{Sci Bull (Beijing)} \textbf{\bibinfo{volume}{67}}, \bibinfo{pages}{985} (\bibinfo{year}{2022}).

\bibitem[{\citenamefont{Yuan et~al.}(2022)\citenamefont{Yuan, Chen, Jiang, Feng, Lin, Yu, He, Zhang, Jiang, Zhang et~al.}}]{zhao22}
\bibinfo{author}{\bibfnamefont{J.}~\bibnamefont{Yuan}}, \bibinfo{author}{\bibfnamefont{Q.}~\bibnamefont{Chen}}, \bibinfo{author}{\bibfnamefont{K.}~\bibnamefont{Jiang}}, \bibinfo{author}{\bibfnamefont{Z.}~\bibnamefont{Feng}}, \bibinfo{author}{\bibfnamefont{Z.}~\bibnamefont{Lin}}, \bibinfo{author}{\bibfnamefont{H.}~\bibnamefont{Yu}}, \bibinfo{author}{\bibfnamefont{G.}~\bibnamefont{He}}, \bibinfo{author}{\bibfnamefont{J.}~\bibnamefont{Zhang}}, \bibinfo{author}{\bibfnamefont{X.}~\bibnamefont{Jiang}}, \bibinfo{author}{\bibfnamefont{X.}~\bibnamefont{Zhang}}, \bibnamefont{et~al.}, \bibinfo{journal}{Nature} \textbf{\bibinfo{volume}{602}}, \bibinfo{pages}{431} (\bibinfo{year}{2022}), ISSN \bibinfo{issn}{1476-4687}, \urlprefix\url{https://doi.org/10.1038/s41586-021-04305-5}.

\bibitem[{\citenamefont{Chang et~al.}(2024)\citenamefont{Chang, Nguyen, Remund, and Chung}}]{chung24}
\bibinfo{author}{\bibfnamefont{Y.-Y.} \bibnamefont{Chang}}, \bibinfo{author}{\bibfnamefont{K.~V.} \bibnamefont{Nguyen}}, \bibinfo{author}{\bibfnamefont{K.}~\bibnamefont{Remund}}, \bibnamefont{and} \bibinfo{author}{\bibfnamefont{C.-H.} \bibnamefont{Chung}} (\bibinfo{year}{2024}), \eprint{2406.14858}, \urlprefix\url{https://arxiv.org/abs/2406.14858}.

\bibitem[{\citenamefont{Zhang et~al.}(2024)\citenamefont{Zhang, Su, Huang, Shan, Sun, Huo, Ye, Zhang, Yang, Xu et~al.}}]{yuan24}
\bibinfo{author}{\bibfnamefont{Y.}~\bibnamefont{Zhang}}, \bibinfo{author}{\bibfnamefont{D.}~\bibnamefont{Su}}, \bibinfo{author}{\bibfnamefont{Y.}~\bibnamefont{Huang}}, \bibinfo{author}{\bibfnamefont{Z.}~\bibnamefont{Shan}}, \bibinfo{author}{\bibfnamefont{H.}~\bibnamefont{Sun}}, \bibinfo{author}{\bibfnamefont{M.}~\bibnamefont{Huo}}, \bibinfo{author}{\bibfnamefont{K.}~\bibnamefont{Ye}}, \bibinfo{author}{\bibfnamefont{J.}~\bibnamefont{Zhang}}, \bibinfo{author}{\bibfnamefont{Z.}~\bibnamefont{Yang}}, \bibinfo{author}{\bibfnamefont{Y.}~\bibnamefont{Xu}}, \bibnamefont{et~al.}, \bibinfo{journal}{Nature Physics} \textbf{\bibinfo{volume}{20}}, \bibinfo{pages}{1269} (\bibinfo{year}{2024}), ISSN \bibinfo{issn}{1745-2481}, \urlprefix\url{https://doi.org/10.1038/s41567-024-02515-y}.

\bibitem[{\citenamefont{Hartnoll and Mackenzie}(2022)}]{hartnoll-2022}
\bibinfo{author}{\bibfnamefont{S.~A.} \bibnamefont{Hartnoll}} \bibnamefont{and} \bibinfo{author}{\bibfnamefont{A.~P.} \bibnamefont{Mackenzie}}, \bibinfo{journal}{REVIEWS OF MODERN PHYSICS} \textbf{\bibinfo{volume}{94}}, \bibinfo{pages}{041002} (\bibinfo{year}{2022}), \urlprefix\url{https://doi.org/10.1103/RevModPhys.94.041002}.

\bibitem[{\citenamefont{Checkelsky et~al.}(2024)\citenamefont{Checkelsky, Bernevig, Coleman, Si, and Paschen}}]{paschen24}
\bibinfo{author}{\bibfnamefont{J.~G.} \bibnamefont{Checkelsky}}, \bibinfo{author}{\bibfnamefont{B.~A.} \bibnamefont{Bernevig}}, \bibinfo{author}{\bibfnamefont{P.}~\bibnamefont{Coleman}}, \bibinfo{author}{\bibfnamefont{Q.}~\bibnamefont{Si}}, \bibnamefont{and} \bibinfo{author}{\bibfnamefont{S.}~\bibnamefont{Paschen}}, \bibinfo{journal}{Nature Reviews Materials} \textbf{\bibinfo{volume}{9}}, \bibinfo{pages}{509} (\bibinfo{year}{2024}), ISSN \bibinfo{issn}{2058-8437}, \urlprefix\url{https://doi.org/10.1038/s41578-023-00644-z}.

\bibitem[{\citenamefont{Varma et~al.}(2002)\citenamefont{Varma, Nussinov, and {van Saarloos}}}]{varma02}
\bibinfo{author}{\bibfnamefont{C.}~\bibnamefont{Varma}}, \bibinfo{author}{\bibfnamefont{Z.}~\bibnamefont{Nussinov}}, \bibnamefont{and} \bibinfo{author}{\bibfnamefont{W.}~\bibnamefont{{van Saarloos}}}, \bibinfo{journal}{Physics Reports} \textbf{\bibinfo{volume}{361}}, \bibinfo{pages}{267} (\bibinfo{year}{2002}), ISSN \bibinfo{issn}{0370-1573}, \urlprefix\url{https://www.sciencedirect.com/science/article/pii/S0370157301000606}.

\bibitem[{\citenamefont{Wahlberg et~al.}(2021)\citenamefont{Wahlberg, Arpaia, Seibold, Rossi, Fumagalli, Trabaldo, Brookes, Braicovich, Caprara, Gran et~al.}}]{science21}
\bibinfo{author}{\bibfnamefont{E.}~\bibnamefont{Wahlberg}}, \bibinfo{author}{\bibfnamefont{R.}~\bibnamefont{Arpaia}}, \bibinfo{author}{\bibfnamefont{G.}~\bibnamefont{Seibold}}, \bibinfo{author}{\bibfnamefont{M.}~\bibnamefont{Rossi}}, \bibinfo{author}{\bibfnamefont{R.}~\bibnamefont{Fumagalli}}, \bibinfo{author}{\bibfnamefont{E.}~\bibnamefont{Trabaldo}}, \bibinfo{author}{\bibfnamefont{N.~B.} \bibnamefont{Brookes}}, \bibinfo{author}{\bibfnamefont{L.}~\bibnamefont{Braicovich}}, \bibinfo{author}{\bibfnamefont{S.}~\bibnamefont{Caprara}}, \bibinfo{author}{\bibfnamefont{U.}~\bibnamefont{Gran}}, \bibnamefont{et~al.}, \bibinfo{journal}{Science} \textbf{\bibinfo{volume}{373}}, \bibinfo{pages}{1506} (\bibinfo{year}{2021}), \eprint{https://www.science.org/doi/pdf/10.1126/science.abc8372}, \urlprefix\url{https://www.science.org/doi/abs/10.1126/science.abc8372}.

\bibitem[{\citenamefont{Phillips et~al.}(2022)\citenamefont{Phillips, Hussey, and Abbamonte}}]{phillips22}
\bibinfo{author}{\bibfnamefont{P.~W.} \bibnamefont{Phillips}}, \bibinfo{author}{\bibfnamefont{N.~E.} \bibnamefont{Hussey}}, \bibnamefont{and} \bibinfo{author}{\bibfnamefont{P.}~\bibnamefont{Abbamonte}}, \bibinfo{journal}{Science} \textbf{\bibinfo{volume}{377}}, \bibinfo{pages}{eabh4273} (\bibinfo{year}{2022}), \eprint{https://www.science.org/doi/pdf/10.1126/science.abh4273}, \urlprefix\url{https://www.science.org/doi/abs/10.1126/science.abh4273}.

\bibitem[{\citenamefont{Ciuchi and Fratini}(2023)}]{fratini23}
\bibinfo{author}{\bibfnamefont{S.}~\bibnamefont{Ciuchi}} \bibnamefont{and} \bibinfo{author}{\bibfnamefont{S.}~\bibnamefont{Fratini}}, \bibinfo{journal}{Phys. Rev. B} \textbf{\bibinfo{volume}{108}}, \bibinfo{pages}{235173} (\bibinfo{year}{2023}), \urlprefix\url{https://link.aps.org/doi/10.1103/PhysRevB.108.235173}.

\bibitem[{\citenamefont{Tomita et~al.}(2015)\citenamefont{Tomita, Kuga, Uwatoko, Coleman, and Nakatsuji}}]{tomita15}
\bibinfo{author}{\bibfnamefont{T.}~\bibnamefont{Tomita}}, \bibinfo{author}{\bibfnamefont{K.}~\bibnamefont{Kuga}}, \bibinfo{author}{\bibfnamefont{Y.}~\bibnamefont{Uwatoko}}, \bibinfo{author}{\bibfnamefont{P.}~\bibnamefont{Coleman}}, \bibnamefont{and} \bibinfo{author}{\bibfnamefont{S.}~\bibnamefont{Nakatsuji}}, \bibinfo{journal}{Science} \textbf{\bibinfo{volume}{349}}, \bibinfo{pages}{506} (\bibinfo{year}{2015}), \eprint{https://www.science.org/doi/pdf/10.1126/science.1262054}, \urlprefix\url{https://www.science.org/doi/abs/10.1126/science.1262054}.

\bibitem[{\citenamefont{Mousatov et~al.}(2020)\citenamefont{Mousatov, Berg, and Hartnoll}}]{hartnoll20}
\bibinfo{author}{\bibfnamefont{C.~H.} \bibnamefont{Mousatov}}, \bibinfo{author}{\bibfnamefont{E.}~\bibnamefont{Berg}}, \bibnamefont{and} \bibinfo{author}{\bibfnamefont{S.~A.} \bibnamefont{Hartnoll}}, \bibinfo{journal}{Proceedings of the National Academy of Sciences} \textbf{\bibinfo{volume}{117}}, \bibinfo{pages}{2852} (\bibinfo{year}{2020}), \eprint{https://www.pnas.org/doi/pdf/10.1073/pnas.1915224117}, \urlprefix\url{https://www.pnas.org/doi/abs/10.1073/pnas.1915224117}.

\bibitem[{\citenamefont{Shen et~al.}(2020)\citenamefont{Shen, Zhang, Komijani, Nicklas, Borth, Wang, Chen, Nie, Li, Lu et~al.}}]{yuan20}
\bibinfo{author}{\bibfnamefont{B.}~\bibnamefont{Shen}}, \bibinfo{author}{\bibfnamefont{Y.}~\bibnamefont{Zhang}}, \bibinfo{author}{\bibfnamefont{Y.}~\bibnamefont{Komijani}}, \bibinfo{author}{\bibfnamefont{M.}~\bibnamefont{Nicklas}}, \bibinfo{author}{\bibfnamefont{R.}~\bibnamefont{Borth}}, \bibinfo{author}{\bibfnamefont{A.}~\bibnamefont{Wang}}, \bibinfo{author}{\bibfnamefont{Y.}~\bibnamefont{Chen}}, \bibinfo{author}{\bibfnamefont{Z.}~\bibnamefont{Nie}}, \bibinfo{author}{\bibfnamefont{R.}~\bibnamefont{Li}}, \bibinfo{author}{\bibfnamefont{X.}~\bibnamefont{Lu}}, \bibnamefont{et~al.}, \bibinfo{journal}{Nature} \textbf{\bibinfo{volume}{579}}, \bibinfo{pages}{51} (\bibinfo{year}{2020}), ISSN \bibinfo{issn}{1476-4687}, \urlprefix\url{https://doi.org/10.1038/s41586-020-2052-z}.

\bibitem[{\citenamefont{Taupin and Paschen}(2022)}]{paschen22}
\bibinfo{author}{\bibfnamefont{M.}~\bibnamefont{Taupin}} \bibnamefont{and} \bibinfo{author}{\bibfnamefont{S.}~\bibnamefont{Paschen}}, \bibinfo{journal}{Crystals} \textbf{\bibinfo{volume}{12}} (\bibinfo{year}{2022}), ISSN \bibinfo{issn}{2073-4352}, \urlprefix\url{https://www.mdpi.com/2073-4352/12/2/251}.

\bibitem[{\citenamefont{Stewart}(2001)}]{stewart01}
\bibinfo{author}{\bibfnamefont{G.~R.} \bibnamefont{Stewart}}, \bibinfo{journal}{Rev. Mod. Phys.} \textbf{\bibinfo{volume}{73}}, \bibinfo{pages}{797} (\bibinfo{year}{2001}), \urlprefix\url{https://link.aps.org/doi/10.1103/RevModPhys.73.797}.

\bibitem[{\citenamefont{Rosch et~al.}(2007)\citenamefont{Rosch, Vojta, and W\"olfle}}]{vonloehneysen-2007}
\bibinfo{author}{\bibfnamefont{A.}~\bibnamefont{Rosch}}, \bibinfo{author}{\bibfnamefont{M.}~\bibnamefont{Vojta}}, \bibnamefont{and} \bibinfo{author}{\bibfnamefont{P.}~\bibnamefont{W\"olfle}}, \bibinfo{journal}{Rev. Mod. Phys.} \textbf{\bibinfo{volume}{79}}, \bibinfo{pages}{1015} (\bibinfo{year}{2007}).

\bibitem[{\citenamefont{Bruin et~al.}(2013)\citenamefont{Bruin, Sakai, Perry, and Mackenzie}}]{kenzie13}
\bibinfo{author}{\bibfnamefont{J.~A.~N.} \bibnamefont{Bruin}}, \bibinfo{author}{\bibfnamefont{H.}~\bibnamefont{Sakai}}, \bibinfo{author}{\bibfnamefont{R.~S.} \bibnamefont{Perry}}, \bibnamefont{and} \bibinfo{author}{\bibfnamefont{A.~P.} \bibnamefont{Mackenzie}}, \bibinfo{journal}{Science} \textbf{\bibinfo{volume}{339}}, \bibinfo{pages}{804} (\bibinfo{year}{2013}).

\bibitem[{\citenamefont{Fernandes et~al.}(2022)\citenamefont{Fernandes, Coldea, Ding, Fisher, Hirschfeld, and Kotliar}}]{fernandes22}
\bibinfo{author}{\bibfnamefont{R.~M.} \bibnamefont{Fernandes}}, \bibinfo{author}{\bibfnamefont{A.~I.} \bibnamefont{Coldea}}, \bibinfo{author}{\bibfnamefont{H.}~\bibnamefont{Ding}}, \bibinfo{author}{\bibfnamefont{I.~R.} \bibnamefont{Fisher}}, \bibinfo{author}{\bibfnamefont{P.~J.} \bibnamefont{Hirschfeld}}, \bibnamefont{and} \bibinfo{author}{\bibfnamefont{G.}~\bibnamefont{Kotliar}}, \bibinfo{journal}{Nature} \textbf{\bibinfo{volume}{601}}, \bibinfo{pages}{35} (\bibinfo{year}{2022}), ISSN \bibinfo{issn}{1476-4687}, \urlprefix\url{https://doi.org/10.1038/s41586-021-04073-2}.

\bibitem[{\citenamefont{Hayes et~al.}(2021)\citenamefont{Hayes, Maksimovic, Lopez, Chan, Ramshaw, McDonald, and Analytis}}]{hayes-2016}
\bibinfo{author}{\bibfnamefont{I.~M.} \bibnamefont{Hayes}}, \bibinfo{author}{\bibfnamefont{N.}~\bibnamefont{Maksimovic}}, \bibinfo{author}{\bibfnamefont{G.~N.} \bibnamefont{Lopez}}, \bibinfo{author}{\bibfnamefont{M.~K.} \bibnamefont{Chan}}, \bibinfo{author}{\bibfnamefont{B.~J.} \bibnamefont{Ramshaw}}, \bibinfo{author}{\bibfnamefont{R.~D.} \bibnamefont{McDonald}}, \bibnamefont{and} \bibinfo{author}{\bibfnamefont{J.~G.} \bibnamefont{Analytis}}, \bibinfo{journal}{Nature Physics} \textbf{\bibinfo{volume}{17}}, \bibinfo{pages}{58} (\bibinfo{year}{2021}), ISSN \bibinfo{issn}{1745-2481}, \urlprefix\url{https://doi.org/10.1038/s41567-020-0982-x}.

\bibitem[{\citenamefont{Ball}(2020)}]{ball20}
\bibinfo{author}{\bibfnamefont{P.}~\bibnamefont{Ball}}, \bibinfo{journal}{Nature Materials} \textbf{\bibinfo{volume}{19}}, \bibinfo{pages}{368} (\bibinfo{year}{2020}), ISSN \bibinfo{issn}{1476-4660}, \urlprefix\url{https://doi.org/10.1038/s41563-020-0653-4}.

\bibitem[{\citenamefont{Lyu et~al.}(2021)\citenamefont{Lyu, Tuchfeld, Verma, Tian, Watanabe, Taniguchi, Lau, Randeria, and Bockrath}}]{randeria21}
\bibinfo{author}{\bibfnamefont{R.}~\bibnamefont{Lyu}}, \bibinfo{author}{\bibfnamefont{Z.}~\bibnamefont{Tuchfeld}}, \bibinfo{author}{\bibfnamefont{N.}~\bibnamefont{Verma}}, \bibinfo{author}{\bibfnamefont{H.}~\bibnamefont{Tian}}, \bibinfo{author}{\bibfnamefont{K.}~\bibnamefont{Watanabe}}, \bibinfo{author}{\bibfnamefont{T.}~\bibnamefont{Taniguchi}}, \bibinfo{author}{\bibfnamefont{C.~N.} \bibnamefont{Lau}}, \bibinfo{author}{\bibfnamefont{M.}~\bibnamefont{Randeria}}, \bibnamefont{and} \bibinfo{author}{\bibfnamefont{M.}~\bibnamefont{Bockrath}}, \bibinfo{journal}{Phys. Rev. B} \textbf{\bibinfo{volume}{103}}, \bibinfo{pages}{245424} (\bibinfo{year}{2021}), \urlprefix\url{https://link.aps.org/doi/10.1103/PhysRevB.103.245424}.

\bibitem[{\citenamefont{Patel et~al.}(2018)\citenamefont{Patel, McGreevy, Arovas, and Sachdev}}]{patel-2018}
\bibinfo{author}{\bibfnamefont{A.~A.} \bibnamefont{Patel}}, \bibinfo{author}{\bibfnamefont{J.}~\bibnamefont{McGreevy}}, \bibinfo{author}{\bibfnamefont{D.~P.} \bibnamefont{Arovas}}, \bibnamefont{and} \bibinfo{author}{\bibfnamefont{S.}~\bibnamefont{Sachdev}}, \bibinfo{journal}{Phys. Rev. X} \textbf{\bibinfo{volume}{8}}, \bibinfo{pages}{021049} (\bibinfo{year}{2018}), \urlprefix\url{https://link.aps.org/doi/10.1103/PhysRevX.8.021049}.

\bibitem[{\citenamefont{Sachdev}(2023)}]{sachdev23_1}
\bibinfo{author}{\bibfnamefont{S.}~\bibnamefont{Sachdev}}, in \emph{\bibinfo{booktitle}{Oxford Research Encyclopedia of Physics}}, edited by \bibinfo{editor}{\bibfnamefont{B.}~\bibnamefont{Foster}} (\bibinfo{publisher}{Oxford University Press}, \bibinfo{year}{2023}), ISBN \bibinfo{isbn}{9780197851753}, \urlprefix\url{https://doi.org/10.1093/acrefore/9780190871994.013.48}.

\bibitem[{\citenamefont{Patel et~al.}(2023)\citenamefont{Patel, Guo, Esterlis, and Sachdev}}]{sachdev23_2}
\bibinfo{author}{\bibfnamefont{A.~A.} \bibnamefont{Patel}}, \bibinfo{author}{\bibfnamefont{H.}~\bibnamefont{Guo}}, \bibinfo{author}{\bibfnamefont{I.}~\bibnamefont{Esterlis}}, \bibnamefont{and} \bibinfo{author}{\bibfnamefont{S.}~\bibnamefont{Sachdev}}, \bibinfo{journal}{Science} \textbf{\bibinfo{volume}{381}}, \bibinfo{pages}{790} (\bibinfo{year}{2023}), \eprint{https://www.science.org/doi/pdf/10.1126/science.abq6011}, \urlprefix\url{https://www.science.org/doi/abs/10.1126/science.abq6011}.

\bibitem[{\citenamefont{Tulipman et~al.}(2024)\citenamefont{Tulipman, Bashan, Schmalian, and Berg}}]{schmalian-2024}
\bibinfo{author}{\bibfnamefont{E.}~\bibnamefont{Tulipman}}, \bibinfo{author}{\bibfnamefont{N.}~\bibnamefont{Bashan}}, \bibinfo{author}{\bibfnamefont{J.}~\bibnamefont{Schmalian}}, \bibnamefont{and} \bibinfo{author}{\bibfnamefont{E.}~\bibnamefont{Berg}}, \bibinfo{journal}{Phys. Rev. B} \textbf{\bibinfo{volume}{110}}, \bibinfo{pages}{155118} (\bibinfo{year}{2024}), \urlprefix\url{https://link.aps.org/doi/10.1103/PhysRevB.110.155118}.

\bibitem[{\citenamefont{Varma}(2020)}]{varma-2020}
\bibinfo{author}{\bibfnamefont{C.~M.} \bibnamefont{Varma}}, \bibinfo{journal}{Rev. Mod. Phys.} \textbf{\bibinfo{volume}{92}}, \bibinfo{pages}{031001} (\bibinfo{year}{2020}), \urlprefix\url{https://link.aps.org/doi/10.1103/RevModPhys.92.031001}.

\bibitem[{\citenamefont{Zhu et~al.}(2015)\citenamefont{Zhu, Chen, and Varma}}]{varma-2015}
\bibinfo{author}{\bibfnamefont{L.}~\bibnamefont{Zhu}}, \bibinfo{author}{\bibfnamefont{Y.}~\bibnamefont{Chen}}, \bibnamefont{and} \bibinfo{author}{\bibfnamefont{C.~M.} \bibnamefont{Varma}}, \bibinfo{journal}{Phys. Rev. B} \textbf{\bibinfo{volume}{91}}, \bibinfo{pages}{205129} (\bibinfo{year}{2015}), \urlprefix\url{https://link.aps.org/doi/10.1103/PhysRevB.91.205129}.

\bibitem[{\citenamefont{Maebashi and Varma}(2025)}]{varma-2025}
\bibinfo{author}{\bibfnamefont{H.}~\bibnamefont{Maebashi}} \bibnamefont{and} \bibinfo{author}{\bibfnamefont{C.~M.} \bibnamefont{Varma}}, \bibinfo{journal}{Phys. Rev. B} \textbf{\bibinfo{volume}{112}}, \bibinfo{pages}{245156} (\bibinfo{year}{2025}), \urlprefix\url{https://link.aps.org/doi/10.1103/cxjv-1wn6}.

\bibitem[{\citenamefont{Varma}(1989)}]{varmaMFL}
\bibinfo{author}{\bibfnamefont{C.}~\bibnamefont{Varma}}, \bibinfo{journal}{International Journal of Modern Physics B} \textbf{\bibinfo{volume}{03}}, \bibinfo{pages}{2083} (\bibinfo{year}{1989}), \eprint{https://doi.org/10.1142/S0217979289001342}, \urlprefix\url{https://doi.org/10.1142/S0217979289001342}.

\bibitem[{\citenamefont{Varma et~al.}(1989)\citenamefont{Varma, Littlewood, Schmitt-Rink, Abrahams, and Ruckenstein}}]{varma89}
\bibinfo{author}{\bibfnamefont{C.~M.} \bibnamefont{Varma}}, \bibinfo{author}{\bibfnamefont{P.~B.} \bibnamefont{Littlewood}}, \bibinfo{author}{\bibfnamefont{S.}~\bibnamefont{Schmitt-Rink}}, \bibinfo{author}{\bibfnamefont{E.}~\bibnamefont{Abrahams}}, \bibnamefont{and} \bibinfo{author}{\bibfnamefont{A.~E.} \bibnamefont{Ruckenstein}}, \bibinfo{journal}{Phys. Rev. Lett.} \textbf{\bibinfo{volume}{63}}, \bibinfo{pages}{1996} (\bibinfo{year}{1989}), \urlprefix\url{https://link.aps.org/doi/10.1103/PhysRevLett.63.1996}.

\bibitem[{\citenamefont{Arpaia et~al.}(2019)\citenamefont{Arpaia, Caprara, Fumagalli, Vecchi, Peng, Andersson, Betto, Luca, Brookes, Lombardi et~al.}}]{arpaia-2019}
\bibinfo{author}{\bibfnamefont{R.}~\bibnamefont{Arpaia}}, \bibinfo{author}{\bibfnamefont{S.}~\bibnamefont{Caprara}}, \bibinfo{author}{\bibfnamefont{R.}~\bibnamefont{Fumagalli}}, \bibinfo{author}{\bibfnamefont{G.~D.} \bibnamefont{Vecchi}}, \bibinfo{author}{\bibfnamefont{Y.~Y.} \bibnamefont{Peng}}, \bibinfo{author}{\bibfnamefont{E.}~\bibnamefont{Andersson}}, \bibinfo{author}{\bibfnamefont{D.}~\bibnamefont{Betto}}, \bibinfo{author}{\bibfnamefont{G.~M.~D.} \bibnamefont{Luca}}, \bibinfo{author}{\bibfnamefont{N.~B.} \bibnamefont{Brookes}}, \bibinfo{author}{\bibfnamefont{F.}~\bibnamefont{Lombardi}}, \bibnamefont{et~al.}, \bibinfo{journal}{Science} \textbf{\bibinfo{volume}{365}}, \bibinfo{pages}{906} (\bibinfo{year}{2019}), \eprint{https://www.science.org/doi/pdf/10.1126/science.aav1315}, \urlprefix\url{https://www.science.org/doi/abs/10.1126/science.aav1315}.

\bibitem[{\citenamefont{Arpaia et~al.}(2023)\citenamefont{Arpaia, Martinelli, Sala, Caprara, Nag, Brookes, Camisa, Li, Gao, Zhou et~al.}}]{arpaia-2023}
\bibinfo{author}{\bibfnamefont{R.}~\bibnamefont{Arpaia}}, \bibinfo{author}{\bibfnamefont{L.}~\bibnamefont{Martinelli}}, \bibinfo{author}{\bibfnamefont{M.~M.} \bibnamefont{Sala}}, \bibinfo{author}{\bibfnamefont{S.}~\bibnamefont{Caprara}}, \bibinfo{author}{\bibfnamefont{A.}~\bibnamefont{Nag}}, \bibinfo{author}{\bibfnamefont{N.~B.} \bibnamefont{Brookes}}, \bibinfo{author}{\bibfnamefont{P.}~\bibnamefont{Camisa}}, \bibinfo{author}{\bibfnamefont{Q.}~\bibnamefont{Li}}, \bibinfo{author}{\bibfnamefont{Q.}~\bibnamefont{Gao}}, \bibinfo{author}{\bibfnamefont{X.}~\bibnamefont{Zhou}}, \bibnamefont{et~al.}, \bibinfo{journal}{Nature Communications} \textbf{\bibinfo{volume}{14}}, \bibinfo{pages}{7198} (\bibinfo{year}{2023}), ISSN \bibinfo{issn}{2041-1723}, \urlprefix\url{https://doi.org/10.1038/s41467-023-42961-5}.

\bibitem[{\citenamefont{Seibold et~al.}(2021)\citenamefont{Seibold, Arpaia, Peng, Fumagalli, Braicovich, Di~Castro, Grilli, Ghiringhelli, and Caprara}}]{seibold-2021}
\bibinfo{author}{\bibfnamefont{G.}~\bibnamefont{Seibold}}, \bibinfo{author}{\bibfnamefont{R.}~\bibnamefont{Arpaia}}, \bibinfo{author}{\bibfnamefont{Y.~Y.} \bibnamefont{Peng}}, \bibinfo{author}{\bibfnamefont{R.}~\bibnamefont{Fumagalli}}, \bibinfo{author}{\bibfnamefont{L.}~\bibnamefont{Braicovich}}, \bibinfo{author}{\bibfnamefont{C.}~\bibnamefont{Di~Castro}}, \bibinfo{author}{\bibfnamefont{M.}~\bibnamefont{Grilli}}, \bibinfo{author}{\bibfnamefont{G.~C.} \bibnamefont{Ghiringhelli}}, \bibnamefont{and} \bibinfo{author}{\bibfnamefont{S.}~\bibnamefont{Caprara}}, \bibinfo{journal}{Communications Physics} \textbf{\bibinfo{volume}{4}}, \bibinfo{pages}{7} (\bibinfo{year}{2021}), ISSN \bibinfo{issn}{2399-3650}, \urlprefix\url{https://doi.org/10.1038/s42005-020-00505-z}.

\bibitem[{\citenamefont{Caprara et~al.}(2024)\citenamefont{Caprara, Di~Castro, Mirarchi, Seibold, and Grilli}}]{SFLoptics-2024}
\bibinfo{author}{\bibfnamefont{S.}~\bibnamefont{Caprara}}, \bibinfo{author}{\bibfnamefont{C.}~\bibnamefont{Di~Castro}}, \bibinfo{author}{\bibfnamefont{G.}~\bibnamefont{Mirarchi}}, \bibinfo{author}{\bibfnamefont{G.}~\bibnamefont{Seibold}}, \bibnamefont{and} \bibinfo{author}{\bibfnamefont{M.}~\bibnamefont{Grilli}}, \bibinfo{journal}{Materials} \textbf{\bibinfo{volume}{17}} (\bibinfo{year}{2024}), ISSN \bibinfo{issn}{1996-1944}, \urlprefix\url{https://www.mdpi.com/1996-1944/17/23/5849}.

\bibitem[{\citenamefont{Caprara et~al.}(2022)\citenamefont{Caprara, Castro, Mirarchi, Seibold, and Grilli}}]{caprara-2022}
\bibinfo{author}{\bibfnamefont{S.}~\bibnamefont{Caprara}}, \bibinfo{author}{\bibfnamefont{C.~D.} \bibnamefont{Castro}}, \bibinfo{author}{\bibfnamefont{G.}~\bibnamefont{Mirarchi}}, \bibinfo{author}{\bibfnamefont{G.}~\bibnamefont{Seibold}}, \bibnamefont{and} \bibinfo{author}{\bibfnamefont{M.}~\bibnamefont{Grilli}}, \bibinfo{journal}{Communications Physics} \textbf{\bibinfo{volume}{5}}, \bibinfo{pages}{10} (\bibinfo{year}{2022}), ISSN \bibinfo{issn}{2399-3650}, \urlprefix\url{https://doi.org/10.1038/s42005-021-00786-y}.

\bibitem[{\citenamefont{Grilli et~al.}(2023)\citenamefont{Grilli, Di~Castro, Mirarchi, Seibold, and Caprara}}]{grilli-2023}
\bibinfo{author}{\bibfnamefont{M.}~\bibnamefont{Grilli}}, \bibinfo{author}{\bibfnamefont{C.}~\bibnamefont{Di~Castro}}, \bibinfo{author}{\bibfnamefont{G.}~\bibnamefont{Mirarchi}}, \bibinfo{author}{\bibfnamefont{G.}~\bibnamefont{Seibold}}, \bibnamefont{and} \bibinfo{author}{\bibfnamefont{S.}~\bibnamefont{Caprara}}, \bibinfo{journal}{Symmetry} \textbf{\bibinfo{volume}{15}} (\bibinfo{year}{2023}), ISSN \bibinfo{issn}{2073-8994}, \urlprefix\url{https://www.mdpi.com/2073-8994/15/3/569}.

\bibitem[{\citenamefont{Mirarchi et~al.}(2024)\citenamefont{Mirarchi, Grilli, Seibold, and Caprara}}]{mirarchi-2024}
\bibinfo{author}{\bibfnamefont{G.}~\bibnamefont{Mirarchi}}, \bibinfo{author}{\bibfnamefont{M.}~\bibnamefont{Grilli}}, \bibinfo{author}{\bibfnamefont{G.}~\bibnamefont{Seibold}}, \bibnamefont{and} \bibinfo{author}{\bibfnamefont{S.}~\bibnamefont{Caprara}}, \bibinfo{journal}{Condensed Matter} \textbf{\bibinfo{volume}{9}} (\bibinfo{year}{2024}), ISSN \bibinfo{issn}{2410-3896}, \urlprefix\url{https://www.mdpi.com/2410-3896/9/1/14}.

\bibitem[{\citenamefont{Millis et~al.}(1990)\citenamefont{Millis, Monien, and Pines}}]{millis-1990}
\bibinfo{author}{\bibfnamefont{A.~J.} \bibnamefont{Millis}}, \bibinfo{author}{\bibfnamefont{H.}~\bibnamefont{Monien}}, \bibnamefont{and} \bibinfo{author}{\bibfnamefont{D.}~\bibnamefont{Pines}}, \bibinfo{journal}{Phys. Rev. B} \textbf{\bibinfo{volume}{42}}, \bibinfo{pages}{197} (\bibinfo{year}{1990}), \urlprefix\url{https://doi.org/10.1103/PhysRevB.42.167}.

\bibitem[{\citenamefont{Abanov et~al.}(2003)\citenamefont{Abanov, Chubukov, and Schmalian}}]{abanov-2003}
\bibinfo{author}{\bibfnamefont{A.}~\bibnamefont{Abanov}}, \bibinfo{author}{\bibfnamefont{A.~V.} \bibnamefont{Chubukov}}, \bibnamefont{and} \bibinfo{author}{\bibfnamefont{J.}~\bibnamefont{Schmalian}}, \bibinfo{journal}{Advances in Physics} \textbf{\bibinfo{volume}{52}}, \bibinfo{pages}{119} (\bibinfo{year}{2003}), \urlprefix\url{http://dx.doi.org/10.1080/0001873021000057123}.

\bibitem[{\citenamefont{Castellani et~al.}(1995)\citenamefont{Castellani, Di~Castro, and Grilli}}]{castellani-1995}
\bibinfo{author}{\bibfnamefont{C.}~\bibnamefont{Castellani}}, \bibinfo{author}{\bibfnamefont{C.}~\bibnamefont{Di~Castro}}, \bibnamefont{and} \bibinfo{author}{\bibfnamefont{M.}~\bibnamefont{Grilli}}, \bibinfo{journal}{Phys. Rev. Lett.} \textbf{\bibinfo{volume}{75}}, \bibinfo{pages}{4650} (\bibinfo{year}{1995}), \urlprefix\url{https://link.aps.org/doi/10.1103/PhysRevLett.75.4650}.

\bibitem[{\citenamefont{Castellani et~al.}(1996)\citenamefont{Castellani, Di~Castro, and Grilli}}]{reviewQCP1}
\bibinfo{author}{\bibfnamefont{C.}~\bibnamefont{Castellani}}, \bibinfo{author}{\bibfnamefont{C.}~\bibnamefont{Di~Castro}}, \bibnamefont{and} \bibinfo{author}{\bibfnamefont{M.}~\bibnamefont{Grilli}}, \bibinfo{journal}{Zeitschrift f{\"u}r Physik B Condensed Matter} \textbf{\bibinfo{volume}{103}}, \bibinfo{pages}{137} (\bibinfo{year}{1996}), ISSN \bibinfo{issn}{1431-584X}, \urlprefix\url{https://doi.org/10.1007/s002570050347}.

\bibitem[{\citenamefont{Castellani et~al.}(1998)\citenamefont{Castellani, Castro, and Grilli}}]{reviewQCP2}
\bibinfo{author}{\bibfnamefont{C.}~\bibnamefont{Castellani}}, \bibinfo{author}{\bibfnamefont{C.~D.} \bibnamefont{Castro}}, \bibnamefont{and} \bibinfo{author}{\bibfnamefont{M.}~\bibnamefont{Grilli}}, \bibinfo{journal}{Journal of Physics and Chemistry of Solids} \textbf{\bibinfo{volume}{59}}, \bibinfo{pages}{1694} (\bibinfo{year}{1998}), ISSN \bibinfo{issn}{0022-3697}, \urlprefix\url{https://www.sciencedirect.com/science/article/pii/S0022369798000857}.

\bibitem[{\citenamefont{Seibold et~al.}(2000)\citenamefont{Seibold, Becca, Bucci, Castellani, Di~Castro, and Grilli}}]{seibold-2000}
\bibinfo{author}{\bibfnamefont{G.}~\bibnamefont{Seibold}}, \bibinfo{author}{\bibfnamefont{F.}~\bibnamefont{Becca}}, \bibinfo{author}{\bibfnamefont{F.}~\bibnamefont{Bucci}}, \bibinfo{author}{\bibfnamefont{C.}~\bibnamefont{Castellani}}, \bibinfo{author}{\bibfnamefont{C.}~\bibnamefont{Di~Castro}}, \bibnamefont{and} \bibinfo{author}{\bibfnamefont{M.}~\bibnamefont{Grilli}}, \bibinfo{journal}{The European Physical Journal B - Condensed Matter and Complex Systems} \textbf{\bibinfo{volume}{13}}, \bibinfo{pages}{87} (\bibinfo{year}{2000}), ISSN \bibinfo{issn}{1434-6036}, \urlprefix\url{https://doi.org/10.1007/s100510050013}.

\bibitem[{\citenamefont{Andergassen et~al.}(2001)\citenamefont{Andergassen, Caprara, Di~Castro, and Grilli}}]{andergassen-2001}
\bibinfo{author}{\bibfnamefont{S.}~\bibnamefont{Andergassen}}, \bibinfo{author}{\bibfnamefont{S.}~\bibnamefont{Caprara}}, \bibinfo{author}{\bibfnamefont{C.}~\bibnamefont{Di~Castro}}, \bibnamefont{and} \bibinfo{author}{\bibfnamefont{M.}~\bibnamefont{Grilli}}, \bibinfo{journal}{Phys. Rev. Lett.} \textbf{\bibinfo{volume}{87}}, \bibinfo{pages}{056401} (\bibinfo{year}{2001}), \urlprefix\url{https://link.aps.org/doi/10.1103/PhysRevLett.87.056401}.

\bibitem[{\citenamefont{Caprara et~al.}(2017)\citenamefont{Caprara, Castro, Seibold, and Grilli1}}]{caprara-2017}
\bibinfo{author}{\bibfnamefont{S.}~\bibnamefont{Caprara}}, \bibinfo{author}{\bibfnamefont{C.~D.} \bibnamefont{Castro}}, \bibinfo{author}{\bibfnamefont{G.}~\bibnamefont{Seibold}}, \bibnamefont{and} \bibinfo{author}{\bibfnamefont{M.}~\bibnamefont{Grilli1}}, \bibinfo{journal}{Physical Review B} \textbf{\bibinfo{volume}{95}}, \bibinfo{pages}{224511} (\bibinfo{year}{2017}), \urlprefix\url{https://doi.org/10.1103/PhysRevB.95.224511}.

\bibitem[{\citenamefont{Tranquada et~al.}(1995)\citenamefont{Tranquada, Sternlieb, Axe, Nakamura, and Uchida}}]{tranquada-1995}
\bibinfo{author}{\bibfnamefont{J.~M.} \bibnamefont{Tranquada}}, \bibinfo{author}{\bibfnamefont{B.~J.} \bibnamefont{Sternlieb}}, \bibinfo{author}{\bibfnamefont{J.~D.} \bibnamefont{Axe}}, \bibinfo{author}{\bibfnamefont{Y.}~\bibnamefont{Nakamura}}, \bibnamefont{and} \bibinfo{author}{\bibfnamefont{S.}~\bibnamefont{Uchida}}, \bibinfo{journal}{Nature} \textbf{\bibinfo{volume}{375}}, \bibinfo{pages}{561} (\bibinfo{year}{1995}).

\bibitem[{\citenamefont{Kivelson et~al.}(2003)\citenamefont{Kivelson, Bindloss, Fradkin, Oganesyan, Tranquada, Kapitulnik, and Howald}}]{kivelson-review}
\bibinfo{author}{\bibfnamefont{S.~A.} \bibnamefont{Kivelson}}, \bibinfo{author}{\bibfnamefont{I.~P.} \bibnamefont{Bindloss}}, \bibinfo{author}{\bibfnamefont{E.}~\bibnamefont{Fradkin}}, \bibinfo{author}{\bibfnamefont{V.}~\bibnamefont{Oganesyan}}, \bibinfo{author}{\bibfnamefont{J.~M.} \bibnamefont{Tranquada}}, \bibinfo{author}{\bibfnamefont{A.}~\bibnamefont{Kapitulnik}}, \bibnamefont{and} \bibinfo{author}{\bibfnamefont{C.}~\bibnamefont{Howald}}, \bibinfo{journal}{Rev. Mod. Phys.} \textbf{\bibinfo{volume}{75}}, \bibinfo{pages}{1201} (\bibinfo{year}{2003}), \urlprefix\url{https://link.aps.org/doi/10.1103/RevModPhys.75.1201}.

\bibitem[{\citenamefont{Keimer et~al.}(2015)\citenamefont{Keimer, Kivelson, Norman, Uchida, and Zaanen}}]{keimer-2015}
\bibinfo{author}{\bibfnamefont{B.}~\bibnamefont{Keimer}}, \bibinfo{author}{\bibfnamefont{S.~A.} \bibnamefont{Kivelson}}, \bibinfo{author}{\bibfnamefont{M.~R.} \bibnamefont{Norman}}, \bibinfo{author}{\bibfnamefont{S.}~\bibnamefont{Uchida}}, \bibnamefont{and} \bibinfo{author}{\bibfnamefont{J.}~\bibnamefont{Zaanen}}, \bibinfo{journal}{Nature (London)} \textbf{\bibinfo{volume}{518}}, \bibinfo{pages}{179} (\bibinfo{year}{2015}), \urlprefix\url{www.nature.com/doifinder/10.1038/nature14165}.

\bibitem[{\citenamefont{Doiron-Leyraud et~al.}(2007)\citenamefont{Doiron-Leyraud, Proust, LeBoeuf, Levallois, Bonnemaison, Liang, Bonn, Hardy, and Taillefer}}]{taillefer-2007-1}
\bibinfo{author}{\bibfnamefont{N.}~\bibnamefont{Doiron-Leyraud}}, \bibinfo{author}{\bibfnamefont{C.}~\bibnamefont{Proust}}, \bibinfo{author}{\bibfnamefont{D.}~\bibnamefont{LeBoeuf}}, \bibinfo{author}{\bibfnamefont{J.}~\bibnamefont{Levallois}}, \bibinfo{author}{\bibfnamefont{J.-B.} \bibnamefont{Bonnemaison}}, \bibinfo{author}{\bibfnamefont{R.}~\bibnamefont{Liang}}, \bibinfo{author}{\bibfnamefont{D.~A.} \bibnamefont{Bonn}}, \bibinfo{author}{\bibfnamefont{W.~N.} \bibnamefont{Hardy}}, \bibnamefont{and} \bibinfo{author}{\bibfnamefont{L.}~\bibnamefont{Taillefer}}, \bibinfo{journal}{Nature (London)} \textbf{\bibinfo{volume}{447}}, \bibinfo{pages}{565} (\bibinfo{year}{2007}), \urlprefix\url{https://doi.org/10.1038/nature05872}.

\bibitem[{\citenamefont{Badoux et~al.}(2016)\citenamefont{Badoux, Tabis, Lalibert\'e, Grissonnanche, Vignolle, Vignolles, B\'eard, Bonn, Hardy, Liang et~al.}}]{badoux-2016}
\bibinfo{author}{\bibfnamefont{S.}~\bibnamefont{Badoux}}, \bibinfo{author}{\bibfnamefont{W.}~\bibnamefont{Tabis}}, \bibinfo{author}{\bibfnamefont{F.}~\bibnamefont{Lalibert\'e}}, \bibinfo{author}{\bibfnamefont{G.}~\bibnamefont{Grissonnanche}}, \bibinfo{author}{\bibfnamefont{B.}~\bibnamefont{Vignolle}}, \bibinfo{author}{\bibfnamefont{D.}~\bibnamefont{Vignolles}}, \bibinfo{author}{\bibfnamefont{J.}~\bibnamefont{B\'eard}}, \bibinfo{author}{\bibfnamefont{D.~A.} \bibnamefont{Bonn}}, \bibinfo{author}{\bibfnamefont{W.~N.} \bibnamefont{Hardy}}, \bibinfo{author}{\bibfnamefont{R.}~\bibnamefont{Liang}}, \bibnamefont{et~al.}, \bibinfo{journal}{Nature (London)} \textbf{\bibinfo{volume}{531}}, \bibinfo{pages}{210} (\bibinfo{year}{2016}), \urlprefix\url{https://doi.org/10.1038/nature16983}.

\bibitem[{\citenamefont{Perali et~al.}(1996)\citenamefont{Perali, Castellani, Di~Castro, and Grilli}}]{perali-1996}
\bibinfo{author}{\bibfnamefont{A.}~\bibnamefont{Perali}}, \bibinfo{author}{\bibfnamefont{C.}~\bibnamefont{Castellani}}, \bibinfo{author}{\bibfnamefont{C.}~\bibnamefont{Di~Castro}}, \bibnamefont{and} \bibinfo{author}{\bibfnamefont{M.}~\bibnamefont{Grilli}}, \bibinfo{journal}{Phys. Rev. B} \textbf{\bibinfo{volume}{54}}, \bibinfo{pages}{16216} (\bibinfo{year}{1996}), \urlprefix\url{https://link.aps.org/doi/10.1103/PhysRevB.54.16216}.

\bibitem[{\citenamefont{Hlubina and Rice}(1995)}]{hlubina}
\bibinfo{author}{\bibfnamefont{R.}~\bibnamefont{Hlubina}} \bibnamefont{and} \bibinfo{author}{\bibfnamefont{T.~M.} \bibnamefont{Rice}}, \bibinfo{journal}{Phys. Rev. B} \textbf{\bibinfo{volume}{51}}, \bibinfo{pages}{9253} (\bibinfo{year}{1995}), \urlprefix\url{https://link.aps.org/doi/10.1103/PhysRevB.51.9253}.

\bibitem[{\citenamefont{Ghiringhelli et~al.}(2012)\citenamefont{Ghiringhelli, Tacon, Minola, Blanco-Canosa, Mazzoli, Brookes, Luca, Frano, Hawthorn, He et~al.}}]{ghiringhelli-2012}
\bibinfo{author}{\bibfnamefont{G.}~\bibnamefont{Ghiringhelli}}, \bibinfo{author}{\bibfnamefont{M.~L.} \bibnamefont{Tacon}}, \bibinfo{author}{\bibfnamefont{M.}~\bibnamefont{Minola}}, \bibinfo{author}{\bibfnamefont{S.}~\bibnamefont{Blanco-Canosa}}, \bibinfo{author}{\bibfnamefont{C.}~\bibnamefont{Mazzoli}}, \bibinfo{author}{\bibfnamefont{N.~B.} \bibnamefont{Brookes}}, \bibinfo{author}{\bibfnamefont{G.~M.~D.} \bibnamefont{Luca}}, \bibinfo{author}{\bibfnamefont{A.}~\bibnamefont{Frano}}, \bibinfo{author}{\bibfnamefont{D.~G.} \bibnamefont{Hawthorn}}, \bibinfo{author}{\bibfnamefont{F.}~\bibnamefont{He}}, \bibnamefont{et~al.}, \bibinfo{journal}{Science} \textbf{\bibinfo{volume}{337}}, \bibinfo{pages}{821} (\bibinfo{year}{2012}), \eprint{https://www.science.org/doi/pdf/10.1126/science.1223532}, \urlprefix\url{https://www.science.org/doi/abs/10.1126/science.1223532}.

\bibitem[{\citenamefont{Chang et~al.}(2012)\citenamefont{Chang, Blackburn, Holmes, Christensen, Larsen, Mesot, Liang, Bonn, Hardy, Watenphul et~al.}}]{chang-2012}
\bibinfo{author}{\bibfnamefont{J.}~\bibnamefont{Chang}}, \bibinfo{author}{\bibfnamefont{E.}~\bibnamefont{Blackburn}}, \bibinfo{author}{\bibfnamefont{A.~T.} \bibnamefont{Holmes}}, \bibinfo{author}{\bibfnamefont{N.~B.} \bibnamefont{Christensen}}, \bibinfo{author}{\bibfnamefont{J.}~\bibnamefont{Larsen}}, \bibinfo{author}{\bibfnamefont{J.}~\bibnamefont{Mesot}}, \bibinfo{author}{\bibfnamefont{R.}~\bibnamefont{Liang}}, \bibinfo{author}{\bibfnamefont{D.~A.} \bibnamefont{Bonn}}, \bibinfo{author}{\bibfnamefont{W.~N.} \bibnamefont{Hardy}}, \bibinfo{author}{\bibfnamefont{A.}~\bibnamefont{Watenphul}}, \bibnamefont{et~al.}, \bibinfo{journal}{Nature Physics} \textbf{\bibinfo{volume}{8}}, \bibinfo{pages}{871} (\bibinfo{year}{2012}), ISSN \bibinfo{issn}{1745-2481}, \urlprefix\url{https://doi.org/10.1038/nphys2456}.

\bibitem[{\citenamefont{da~Silva Neto Eduardo~H et~al.}(2014)\citenamefont{da~Silva Neto Eduardo~H, Pegor, Alex, Riccardo, Enrico, Eugen, Andr\'as, Jinsheng, John, Zhijun et~al.}}]{dasilvaneto-2014}
\bibinfo{author}{\bibnamefont{da~Silva Neto Eduardo~H}}, \bibinfo{author}{\bibfnamefont{A.}~\bibnamefont{Pegor}}, \bibinfo{author}{\bibfnamefont{F.}~\bibnamefont{Alex}}, \bibinfo{author}{\bibfnamefont{C.}~\bibnamefont{Riccardo}}, \bibinfo{author}{\bibfnamefont{S.}~\bibnamefont{Enrico}}, \bibinfo{author}{\bibfnamefont{W.}~\bibnamefont{Eugen}}, \bibinfo{author}{\bibfnamefont{G.}~\bibnamefont{Andr\'as}}, \bibinfo{author}{\bibfnamefont{W.}~\bibnamefont{Jinsheng}}, \bibinfo{author}{\bibfnamefont{S.}~\bibnamefont{John}}, \bibinfo{author}{\bibfnamefont{X.}~\bibnamefont{Zhijun}}, \bibnamefont{et~al.}, \bibinfo{journal}{Science} \textbf{\bibinfo{volume}{343}}, \bibinfo{pages}{393} (\bibinfo{year}{2014}), \urlprefix\url{https://www.science.org/doi/10.1126/science.1243479}.

\bibitem[{\citenamefont{Comin and Damascelli}(2016)}]{comin-2016}
\bibinfo{author}{\bibfnamefont{R.}~\bibnamefont{Comin}} \bibnamefont{and} \bibinfo{author}{\bibfnamefont{A.}~\bibnamefont{Damascelli}}, \bibinfo{journal}{Annu. Rev. Condens. Matter Phys} \textbf{\bibinfo{volume}{7}}, \bibinfo{pages}{369} (\bibinfo{year}{2016}).

\bibitem[{\citenamefont{Arpaia and Ghiringhelli}(2021)}]{arpaia-2021}
\bibinfo{author}{\bibfnamefont{R.}~\bibnamefont{Arpaia}} \bibnamefont{and} \bibinfo{author}{\bibfnamefont{G.}~\bibnamefont{Ghiringhelli}}, \bibinfo{journal}{Journal of the Physical Society of Japan} \textbf{\bibinfo{volume}{90}}, \bibinfo{pages}{111005} (\bibinfo{year}{2021}), \urlprefix\url{https://doi.org/10.7566/JPSJ.90.111005}.

\bibitem[{\citenamefont{Guo et~al.}(2024)\citenamefont{Guo, Chen, Hoveyda-Marashi, Bettler, Chaudhuri, Kengle, Schneeloch, Zhang, Gu, Chiang et~al.}}]{abbamonte-2025}
\bibinfo{author}{\bibfnamefont{X.}~\bibnamefont{Guo}}, \bibinfo{author}{\bibfnamefont{J.}~\bibnamefont{Chen}}, \bibinfo{author}{\bibfnamefont{F.}~\bibnamefont{Hoveyda-Marashi}}, \bibinfo{author}{\bibfnamefont{S.~L.} \bibnamefont{Bettler}}, \bibinfo{author}{\bibfnamefont{D.}~\bibnamefont{Chaudhuri}}, \bibinfo{author}{\bibfnamefont{C.~S.} \bibnamefont{Kengle}}, \bibinfo{author}{\bibfnamefont{J.~A.} \bibnamefont{Schneeloch}}, \bibinfo{author}{\bibfnamefont{R.}~\bibnamefont{Zhang}}, \bibinfo{author}{\bibfnamefont{G.}~\bibnamefont{Gu}}, \bibinfo{author}{\bibfnamefont{T.-C.} \bibnamefont{Chiang}}, \bibnamefont{et~al.} (\bibinfo{year}{2024}), \eprint{2411.11164}, \urlprefix\url{https://arxiv.org/abs/2411.11164}.

\bibitem[{\citenamefont{Howald et~al.}(2003)\citenamefont{Howald, Eisaki, Kaneko, Greven, and Kapitulnik}}]{how03}
\bibinfo{author}{\bibfnamefont{C.}~\bibnamefont{Howald}}, \bibinfo{author}{\bibfnamefont{H.}~\bibnamefont{Eisaki}}, \bibinfo{author}{\bibfnamefont{N.}~\bibnamefont{Kaneko}}, \bibinfo{author}{\bibfnamefont{M.}~\bibnamefont{Greven}}, \bibnamefont{and} \bibinfo{author}{\bibfnamefont{A.}~\bibnamefont{Kapitulnik}}, \bibinfo{journal}{Phys. Rev. B} \textbf{\bibinfo{volume}{67}}, \bibinfo{pages}{014533} (\bibinfo{year}{2003}), \urlprefix\url{https://link.aps.org/doi/10.1103/PhysRevB.67.014533}.

\bibitem[{\citenamefont{Hanaguri et~al.}(2004)\citenamefont{Hanaguri, Lupien, Kohsaka, Lee, Azuma, Takano, Takagi, and Davis}}]{hana4}
\bibinfo{author}{\bibfnamefont{T.}~\bibnamefont{Hanaguri}}, \bibinfo{author}{\bibfnamefont{C.}~\bibnamefont{Lupien}}, \bibinfo{author}{\bibfnamefont{Y.}~\bibnamefont{Kohsaka}}, \bibinfo{author}{\bibfnamefont{D.-H.} \bibnamefont{Lee}}, \bibinfo{author}{\bibfnamefont{M.}~\bibnamefont{Azuma}}, \bibinfo{author}{\bibfnamefont{M.}~\bibnamefont{Takano}}, \bibinfo{author}{\bibfnamefont{H.}~\bibnamefont{Takagi}}, \bibnamefont{and} \bibinfo{author}{\bibfnamefont{J.~C.} \bibnamefont{Davis}}, \bibinfo{journal}{Nature} \textbf{\bibinfo{volume}{430}}, \bibinfo{pages}{1001} (\bibinfo{year}{2004}), ISSN \bibinfo{issn}{1476-4687}, \urlprefix\url{https://doi.org/10.1038/nature02861}.

\bibitem[{\citenamefont{Vershinin et~al.}(2004)\citenamefont{Vershinin, Misra, Ono, Abe, Ando, and Yazdani}}]{versh04}
\bibinfo{author}{\bibfnamefont{M.}~\bibnamefont{Vershinin}}, \bibinfo{author}{\bibfnamefont{S.}~\bibnamefont{Misra}}, \bibinfo{author}{\bibfnamefont{S.}~\bibnamefont{Ono}}, \bibinfo{author}{\bibfnamefont{Y.}~\bibnamefont{Abe}}, \bibinfo{author}{\bibfnamefont{Y.}~\bibnamefont{Ando}}, \bibnamefont{and} \bibinfo{author}{\bibfnamefont{A.}~\bibnamefont{Yazdani}}, \bibinfo{journal}{Science} \textbf{\bibinfo{volume}{303}}, \bibinfo{pages}{1995} (\bibinfo{year}{2004}), \eprint{https://www.science.org/doi/pdf/10.1126/science.1093384}, \urlprefix\url{https://www.science.org/doi/abs/10.1126/science.1093384}.

\bibitem[{\citenamefont{Caprara et~al.}(2007)\citenamefont{Caprara, Grilli, Di~Castro, and Enss}}]{enss-2007}
\bibinfo{author}{\bibfnamefont{S.}~\bibnamefont{Caprara}}, \bibinfo{author}{\bibfnamefont{M.}~\bibnamefont{Grilli}}, \bibinfo{author}{\bibfnamefont{C.}~\bibnamefont{Di~Castro}}, \bibnamefont{and} \bibinfo{author}{\bibfnamefont{T.}~\bibnamefont{Enss}}, \bibinfo{journal}{Phys. Rev. B} \textbf{\bibinfo{volume}{75}}, \bibinfo{pages}{140505(R)} (\bibinfo{year}{2007}), \urlprefix\url{https://link.aps.org/doi/10.1103/PhysRevB.75.140505}.

\bibitem[{\citenamefont{Michon et~al.}(2019)\citenamefont{Michon, Girod, Badoux, Ka{\v{c}}mar{\v{c}}{\'i}k, Ma, Dragomir, Dabkowska, Gaulin, Zhou, Pyon et~al.}}]{michon-2019}
\bibinfo{author}{\bibfnamefont{B.}~\bibnamefont{Michon}}, \bibinfo{author}{\bibfnamefont{C.}~\bibnamefont{Girod}}, \bibinfo{author}{\bibfnamefont{S.}~\bibnamefont{Badoux}}, \bibinfo{author}{\bibfnamefont{J.}~\bibnamefont{Ka{\v{c}}mar{\v{c}}{\'i}k}}, \bibinfo{author}{\bibfnamefont{Q.}~\bibnamefont{Ma}}, \bibinfo{author}{\bibfnamefont{M.}~\bibnamefont{Dragomir}}, \bibinfo{author}{\bibfnamefont{H.~A.} \bibnamefont{Dabkowska}}, \bibinfo{author}{\bibfnamefont{B.~D.} \bibnamefont{Gaulin}}, \bibinfo{author}{\bibfnamefont{J.-S.} \bibnamefont{Zhou}}, \bibinfo{author}{\bibfnamefont{S.}~\bibnamefont{Pyon}}, \bibnamefont{et~al.}, \bibinfo{journal}{Nature} \textbf{\bibinfo{volume}{567}}, \bibinfo{pages}{218} (\bibinfo{year}{2019}), ISSN \bibinfo{issn}{1476-4687}, \urlprefix\url{https://doi.org/10.1038/s41586-019-0932-x}.

\bibitem[{\citenamefont{Ketmaier et~al.}(2026)\citenamefont{Ketmaier, L., Kurchan, and Zamponi}}]{ketmaier-2026}
\bibinfo{author}{\bibfnamefont{M.}~\bibnamefont{Ketmaier}, \bibfnamefont{Grilli}}, \bibinfo{author}{\bibnamefont{L.}}, \bibinfo{author}{\bibfnamefont{J.}~\bibnamefont{Kurchan}}, \bibnamefont{and} \bibinfo{author}{\bibfnamefont{F.}~\bibnamefont{Zamponi}}, \bibinfo{journal}{unpublished}  (\bibinfo{year}{2026}).

\bibitem[{\citenamefont{Ament et~al.}(2011)\citenamefont{Ament, van Veenendaal, Devereaux, Hill, and van~den Brink}}]{devereaux_2011}
\bibinfo{author}{\bibfnamefont{L.~J.~P.} \bibnamefont{Ament}}, \bibinfo{author}{\bibfnamefont{M.}~\bibnamefont{van Veenendaal}}, \bibinfo{author}{\bibfnamefont{T.~P.} \bibnamefont{Devereaux}}, \bibinfo{author}{\bibfnamefont{J.~P.} \bibnamefont{Hill}}, \bibnamefont{and} \bibinfo{author}{\bibfnamefont{J.}~\bibnamefont{van~den Brink}}, \bibinfo{journal}{Rev. Mod. Phys.} \textbf{\bibinfo{volume}{83}}, \bibinfo{pages}{705} (\bibinfo{year}{2011}), \urlprefix\url{https://link.aps.org/doi/10.1103/RevModPhys.83.705}.

\bibitem[{\citenamefont{Allen}(2015)}]{allen}
\bibinfo{author}{\bibfnamefont{P.~B.} \bibnamefont{Allen}}, \bibinfo{journal}{Phys. Rev. B} \textbf{\bibinfo{volume}{92}}, \bibinfo{pages}{054305} (\bibinfo{year}{2015}), \urlprefix\url{https://link.aps.org/doi/10.1103/PhysRevB.92.054305}.

\bibitem[{\citenamefont{Marconi et~al.}(2012)\citenamefont{Marconi, Pop, and Pop}}]{marconi-2012}
\bibinfo{author}{\bibfnamefont{D.}~\bibnamefont{Marconi}}, \bibinfo{author}{\bibfnamefont{M.}~\bibnamefont{Pop}}, \bibnamefont{and} \bibinfo{author}{\bibfnamefont{A.}~\bibnamefont{Pop}}, \bibinfo{journal}{Journal of Alloys and Compounds} \textbf{\bibinfo{volume}{513}}, \bibinfo{pages}{586} (\bibinfo{year}{2012}), ISSN \bibinfo{issn}{0925-8388}, \urlprefix\url{https://www.sciencedirect.com/science/article/pii/S0925838811021001}.

\bibitem[{\citenamefont{Park et~al.}(2025)\citenamefont{Park, Lee, Gu, and Hwang}}]{park-2025}
\bibinfo{author}{\bibfnamefont{H.}~\bibnamefont{Park}}, \bibinfo{author}{\bibfnamefont{S.-S.} \bibnamefont{Lee}}, \bibinfo{author}{\bibfnamefont{G.~D.} \bibnamefont{Gu}}, \bibnamefont{and} \bibinfo{author}{\bibfnamefont{J.}~\bibnamefont{Hwang}} (\bibinfo{year}{2025}), \eprint{2509.19675}, \urlprefix\url{https://arxiv.org/abs/2509.19675}.

\bibitem[{\citenamefont{Spera and Caprara}(2024)}]{spera-2026}
\bibinfo{author}{\bibfnamefont{F.}~\bibnamefont{Spera}} \bibnamefont{and} \bibinfo{author}{\bibfnamefont{S.}~\bibnamefont{Caprara}}, \bibinfo{journal}{Sapienza University of Rome, Master Thesis}  (\bibinfo{year}{2024}).

\bibitem[{\citenamefont{Uykur et~al.}(2011)\citenamefont{Uykur, Tanaka, Masui, Miyasaka, and Tajima}}]{uykur-2011}
\bibinfo{author}{\bibfnamefont{E.}~\bibnamefont{Uykur}}, \bibinfo{author}{\bibfnamefont{K.}~\bibnamefont{Tanaka}}, \bibinfo{author}{\bibfnamefont{T.}~\bibnamefont{Masui}}, \bibinfo{author}{\bibfnamefont{S.}~\bibnamefont{Miyasaka}}, \bibnamefont{and} \bibinfo{author}{\bibfnamefont{S.}~\bibnamefont{Tajima}}, \bibinfo{journal}{Phys. Rev. B} \textbf{\bibinfo{volume}{84}}, \bibinfo{pages}{184527} (\bibinfo{year}{2011}), \urlprefix\url{https://link.aps.org/doi/10.1103/PhysRevB.84.184527}.

\bibitem[{\citenamefont{Michon et~al.}(2023)\citenamefont{Michon, Berthod, Rischau, Ataei, Chen, Komiya, Ono, Taillefer, van~der Marel, and Georges}}]{michon-2023}
\bibinfo{author}{\bibfnamefont{B.}~\bibnamefont{Michon}}, \bibinfo{author}{\bibfnamefont{C.}~\bibnamefont{Berthod}}, \bibinfo{author}{\bibfnamefont{C.~W.} \bibnamefont{Rischau}}, \bibinfo{author}{\bibfnamefont{A.}~\bibnamefont{Ataei}}, \bibinfo{author}{\bibfnamefont{L.}~\bibnamefont{Chen}}, \bibinfo{author}{\bibfnamefont{S.}~\bibnamefont{Komiya}}, \bibinfo{author}{\bibfnamefont{S.}~\bibnamefont{Ono}}, \bibinfo{author}{\bibfnamefont{L.}~\bibnamefont{Taillefer}}, \bibinfo{author}{\bibfnamefont{D.}~\bibnamefont{van~der Marel}}, \bibnamefont{and} \bibinfo{author}{\bibfnamefont{A.}~\bibnamefont{Georges}}, \bibinfo{journal}{Nature Communications} \textbf{\bibinfo{volume}{14}}, \bibinfo{pages}{3033} (\bibinfo{year}{2023}), ISSN \bibinfo{issn}{2041-1723}, \urlprefix\url{https://doi.org/10.1038/s41467-023-38762-5}.

\bibitem[{\citenamefont{Caprara et~al.}(2011)\citenamefont{Caprara, Di~Castro, Muschler, Prestel, Hackl, Lambacher, Erb, Komiya, Ando, and Grilli}}]{caprara11}
\bibinfo{author}{\bibfnamefont{S.}~\bibnamefont{Caprara}}, \bibinfo{author}{\bibfnamefont{C.}~\bibnamefont{Di~Castro}}, \bibinfo{author}{\bibfnamefont{B.}~\bibnamefont{Muschler}}, \bibinfo{author}{\bibfnamefont{W.}~\bibnamefont{Prestel}}, \bibinfo{author}{\bibfnamefont{R.}~\bibnamefont{Hackl}}, \bibinfo{author}{\bibfnamefont{M.}~\bibnamefont{Lambacher}}, \bibinfo{author}{\bibfnamefont{A.}~\bibnamefont{Erb}}, \bibinfo{author}{\bibfnamefont{S.}~\bibnamefont{Komiya}}, \bibinfo{author}{\bibfnamefont{Y.}~\bibnamefont{Ando}}, \bibnamefont{and} \bibinfo{author}{\bibfnamefont{M.}~\bibnamefont{Grilli}}, \bibinfo{journal}{Phys. Rev. B} \textbf{\bibinfo{volume}{84}}, \bibinfo{pages}{054508} (\bibinfo{year}{2011}), \urlprefix\url{https://link.aps.org/doi/10.1103/PhysRevB.84.054508}.

\bibitem[{\citenamefont{Caprara et~al.}(2005)\citenamefont{Caprara, Di~Castro, Grilli, and Suppa}}]{suppa05}
\bibinfo{author}{\bibfnamefont{S.}~\bibnamefont{Caprara}}, \bibinfo{author}{\bibfnamefont{C.}~\bibnamefont{Di~Castro}}, \bibinfo{author}{\bibfnamefont{M.}~\bibnamefont{Grilli}}, \bibnamefont{and} \bibinfo{author}{\bibfnamefont{D.}~\bibnamefont{Suppa}}, \bibinfo{journal}{Phys. Rev. Lett.} \textbf{\bibinfo{volume}{95}}, \bibinfo{pages}{117004} (\bibinfo{year}{2005}), \urlprefix\url{https://link.aps.org/doi/10.1103/PhysRevLett.95.117004}.

\bibitem[{\citenamefont{Devereaux and Hackl}(2007)}]{dev07}
\bibinfo{author}{\bibfnamefont{T.~P.} \bibnamefont{Devereaux}} \bibnamefont{and} \bibinfo{author}{\bibfnamefont{R.}~\bibnamefont{Hackl}}, \bibinfo{journal}{Rev. Mod. Phys.} \textbf{\bibinfo{volume}{79}}, \bibinfo{pages}{175} (\bibinfo{year}{2007}), \urlprefix\url{https://link.aps.org/doi/10.1103/RevModPhys.79.175}.

\bibitem[{\citenamefont{Venturini et~al.}(2002)\citenamefont{Venturini, Opel, Hackl, Berger, Forró, and Revaz}}]{venturini02}
\bibinfo{author}{\bibfnamefont{F.}~\bibnamefont{Venturini}}, \bibinfo{author}{\bibfnamefont{M.}~\bibnamefont{Opel}}, \bibinfo{author}{\bibfnamefont{R.}~\bibnamefont{Hackl}}, \bibinfo{author}{\bibfnamefont{H.}~\bibnamefont{Berger}}, \bibinfo{author}{\bibfnamefont{L.}~\bibnamefont{Forró}}, \bibnamefont{and} \bibinfo{author}{\bibfnamefont{B.}~\bibnamefont{Revaz}}, \bibinfo{journal}{Journal of Physics and Chemistry of Solids} \textbf{\bibinfo{volume}{63}}, \bibinfo{pages}{2345} (\bibinfo{year}{2002}), ISSN \bibinfo{issn}{0022-3697}, \bibinfo{note}{proceedings of the Conference on Spectroscopies in Novel Superconductors}, \urlprefix\url{https://www.sciencedirect.com/science/article/pii/S0022369702002391}.

\bibitem[{\citenamefont{Opel et~al.}(2000)\citenamefont{Opel, Nemetschek, Hoffmann, Philipp, M\"uller, Hackl, T\"utt\ifmmode~\mbox{\H{o}}\else \H{o}\fi{}, Erb, Revaz, Walker et~al.}}]{opel00}
\bibinfo{author}{\bibfnamefont{M.}~\bibnamefont{Opel}}, \bibinfo{author}{\bibfnamefont{R.}~\bibnamefont{Nemetschek}}, \bibinfo{author}{\bibfnamefont{C.}~\bibnamefont{Hoffmann}}, \bibinfo{author}{\bibfnamefont{R.}~\bibnamefont{Philipp}}, \bibinfo{author}{\bibfnamefont{P.~F.} \bibnamefont{M\"uller}}, \bibinfo{author}{\bibfnamefont{R.}~\bibnamefont{Hackl}}, \bibinfo{author}{\bibfnamefont{I.}~\bibnamefont{T\"utt\ifmmode~\mbox{\H{o}}\else \H{o}\fi{}}}, \bibinfo{author}{\bibfnamefont{A.}~\bibnamefont{Erb}}, \bibinfo{author}{\bibfnamefont{B.}~\bibnamefont{Revaz}}, \bibinfo{author}{\bibfnamefont{E.}~\bibnamefont{Walker}}, \bibnamefont{et~al.}, \bibinfo{journal}{Phys. Rev. B} \textbf{\bibinfo{volume}{61}}, \bibinfo{pages}{9752} (\bibinfo{year}{2000}), \urlprefix\url{https://link.aps.org/doi/10.1103/PhysRevB.61.9752}.

\bibitem[{\citenamefont{Feile}(1989)}]{feile89}
\bibinfo{author}{\bibfnamefont{R.}~\bibnamefont{Feile}}, \bibinfo{journal}{Physica C: Superconductivity} \textbf{\bibinfo{volume}{159}}, \bibinfo{pages}{1} (\bibinfo{year}{1989}), ISSN \bibinfo{issn}{0921-4534}, \urlprefix\url{https://www.sciencedirect.com/science/article/pii/0921453489900993}.

\bibitem[{\citenamefont{Thomsen et~al.}(1988)\citenamefont{Thomsen, Cardona, Gegenheimer, Liu, and Simon}}]{thomsen88}
\bibinfo{author}{\bibfnamefont{C.}~\bibnamefont{Thomsen}}, \bibinfo{author}{\bibfnamefont{M.}~\bibnamefont{Cardona}}, \bibinfo{author}{\bibfnamefont{B.}~\bibnamefont{Gegenheimer}}, \bibinfo{author}{\bibfnamefont{R.}~\bibnamefont{Liu}}, \bibnamefont{and} \bibinfo{author}{\bibfnamefont{A.}~\bibnamefont{Simon}}, \bibinfo{journal}{Phys. Rev. B} \textbf{\bibinfo{volume}{37}}, \bibinfo{pages}{9860(R)} (\bibinfo{year}{1988}), \urlprefix\url{https://link.aps.org/doi/10.1103/PhysRevB.37.9860}.

\bibitem[{\citenamefont{van~der Marel et~al.}(2003)\citenamefont{van~der Marel, Molegraaf, Zaanen, Nussinov, Carbone, Damascelli, Eisaki, Greven, Kes, and Li}}]{marel03}
\bibinfo{author}{\bibfnamefont{D.}~\bibnamefont{van~der Marel}}, \bibinfo{author}{\bibfnamefont{H.~J.~A.} \bibnamefont{Molegraaf}}, \bibinfo{author}{\bibfnamefont{J.}~\bibnamefont{Zaanen}}, \bibinfo{author}{\bibfnamefont{Z.}~\bibnamefont{Nussinov}}, \bibinfo{author}{\bibfnamefont{F.}~\bibnamefont{Carbone}}, \bibinfo{author}{\bibfnamefont{A.}~\bibnamefont{Damascelli}}, \bibinfo{author}{\bibfnamefont{H.}~\bibnamefont{Eisaki}}, \bibinfo{author}{\bibfnamefont{M.}~\bibnamefont{Greven}}, \bibinfo{author}{\bibfnamefont{P.~H.} \bibnamefont{Kes}}, \bibnamefont{and} \bibinfo{author}{\bibfnamefont{M.}~\bibnamefont{Li}}, \bibinfo{journal}{Nature} \textbf{\bibinfo{volume}{425}}, \bibinfo{pages}{271} (\bibinfo{year}{2003}).

\bibitem[{\citenamefont{Puchkov et~al.}(1996)\citenamefont{Puchkov, Basov, and Timusk}}]{puchkov-1999}
\bibinfo{author}{\bibfnamefont{A.~V.} \bibnamefont{Puchkov}}, \bibinfo{author}{\bibfnamefont{D.~N.} \bibnamefont{Basov}}, \bibnamefont{and} \bibinfo{author}{\bibfnamefont{T.}~\bibnamefont{Timusk}}, \bibinfo{journal}{Journal of Physics: Condensed Matter} \textbf{\bibinfo{volume}{8}}, \bibinfo{pages}{10049} (\bibinfo{year}{1996}), \urlprefix\url{https://doi.org/10.1088/0953-8984/8/48/023}.

\bibitem[{\citenamefont{Hwang et~al.}(2004)\citenamefont{Hwang, Timusk, and Gu}}]{hwang-2004}
\bibinfo{author}{\bibfnamefont{J.}~\bibnamefont{Hwang}}, \bibinfo{author}{\bibfnamefont{T.}~\bibnamefont{Timusk}}, \bibnamefont{and} \bibinfo{author}{\bibfnamefont{G.~D.} \bibnamefont{Gu}}, \bibinfo{journal}{Nature} \textbf{\bibinfo{volume}{427}}, \bibinfo{pages}{714} (\bibinfo{year}{2004}), ISSN \bibinfo{issn}{1476-4687}, \urlprefix\url{https://doi.org/10.1038/nature02347}.

\bibitem[{\citenamefont{Basov and Timusk}(2005)}]{timusk-2005}
\bibinfo{author}{\bibfnamefont{D.~N.} \bibnamefont{Basov}} \bibnamefont{and} \bibinfo{author}{\bibfnamefont{T.}~\bibnamefont{Timusk}}, \bibinfo{journal}{Rev. Mod. Phys.} \textbf{\bibinfo{volume}{77}}, \bibinfo{pages}{721} (\bibinfo{year}{2005}), \urlprefix\url{https://link.aps.org/doi/10.1103/RevModPhys.77.721}.

\bibitem[{\citenamefont{v.~L\"ohneysen et~al.}(1994)\citenamefont{v.~L\"ohneysen, Pietrus, Portisch, Schlager, Schr\"oder, Sieck, and Trappmann}}]{vonloehneysen-1994}
\bibinfo{author}{\bibfnamefont{H.}~\bibnamefont{v.~L\"ohneysen}}, \bibinfo{author}{\bibfnamefont{T.}~\bibnamefont{Pietrus}}, \bibinfo{author}{\bibfnamefont{G.}~\bibnamefont{Portisch}}, \bibinfo{author}{\bibfnamefont{H.}~\bibnamefont{Schlager}}, \bibinfo{author}{\bibfnamefont{A.}~\bibnamefont{Schr\"oder}}, \bibinfo{author}{\bibfnamefont{M.}~\bibnamefont{Sieck}}, \bibnamefont{and} \bibinfo{author}{\bibfnamefont{T.}~\bibnamefont{Trappmann}}, \bibinfo{journal}{Phys. Rev. Lett} \textbf{\bibinfo{volume}{72}}, \bibinfo{pages}{3262} (\bibinfo{year}{1994}), \urlprefix\url{https://doi.org/10.1103/PhysRevLett.72.3262}.

\bibitem[{\citenamefont{Zhang et~al.}(2023)\citenamefont{Zhang, Berg, and Chubukov}}]{chubukov-2023}
\bibinfo{author}{\bibfnamefont{S.-S.} \bibnamefont{Zhang}}, \bibinfo{author}{\bibfnamefont{E.}~\bibnamefont{Berg}}, \bibnamefont{and} \bibinfo{author}{\bibfnamefont{A.~V.} \bibnamefont{Chubukov}}, \bibinfo{journal}{Phys. Rev. B} \textbf{\bibinfo{volume}{107}}, \bibinfo{pages}{144507} (\bibinfo{year}{2023}), \urlprefix\url{https://link.aps.org/doi/10.1103/PhysRevB.107.144507}.

\bibitem[{\citenamefont{Ziman}(1960)}]{ziman}
\bibinfo{author}{\bibfnamefont{J.~M.} \bibnamefont{Ziman}} (\bibinfo{year}{1960}).

\bibitem[{\citenamefont{Mirarchi et~al.}(2026)\citenamefont{Mirarchi, Seibold, Grilli, and Caprara}}]{mirarchi-2026}
\bibinfo{author}{\bibfnamefont{G.}~\bibnamefont{Mirarchi}}, \bibinfo{author}{\bibfnamefont{G.}~\bibnamefont{Seibold}}, \bibinfo{author}{\bibfnamefont{M.}~\bibnamefont{Grilli}}, \bibnamefont{and} \bibinfo{author}{\bibfnamefont{S.}~\bibnamefont{Caprara}}, \bibinfo{journal}{arXiv:2607.02449}  (\bibinfo{year}{2026}).

\bibitem[{\citenamefont{Becca et~al.}(1996)\citenamefont{Becca, Tarquini, Grilli, and Di~Castro}}]{becca-1996}
\bibinfo{author}{\bibfnamefont{F.}~\bibnamefont{Becca}}, \bibinfo{author}{\bibfnamefont{M.}~\bibnamefont{Tarquini}}, \bibinfo{author}{\bibfnamefont{M.}~\bibnamefont{Grilli}}, \bibnamefont{and} \bibinfo{author}{\bibfnamefont{C.}~\bibnamefont{Di~Castro}}, \bibinfo{journal}{Phys. Rev. B} \textbf{\bibinfo{volume}{54}}, \bibinfo{pages}{12443} (\bibinfo{year}{1996}), \urlprefix\url{https://link.aps.org/doi/10.1103/PhysRevB.54.12443}.

\bibitem[{\citenamefont{Fedele et~al.}(2026)\citenamefont{Fedele, Merzoni, Sala, Rosa, Brookes, Lombardi, Caprara, Ghiringhelli, and Arpaia}}]{fedele-2026}
\bibinfo{author}{\bibfnamefont{M.}~\bibnamefont{Fedele}}, \bibinfo{author}{\bibfnamefont{G.}~\bibnamefont{Merzoni}}, \bibinfo{author}{\bibfnamefont{M.~M.} \bibnamefont{Sala}}, \bibinfo{author}{\bibfnamefont{F.}~\bibnamefont{Rosa}}, \bibinfo{author}{\bibfnamefont{N.~B.} \bibnamefont{Brookes}}, \bibinfo{author}{\bibfnamefont{F.}~\bibnamefont{Lombardi}}, \bibinfo{author}{\bibfnamefont{S.}~\bibnamefont{Caprara}}, \bibinfo{author}{\bibfnamefont{G.}~\bibnamefont{Ghiringhelli}}, \bibnamefont{and} \bibinfo{author}{\bibfnamefont{R.}~\bibnamefont{Arpaia}} (\bibinfo{year}{2026}), \eprint{2602.18112}, \urlprefix\url{https://arxiv.org/abs/2602.18112}.

\bibitem[{\citenamefont{Michon et~al.}(2018)\citenamefont{Michon, Ataei, Bourgeois-Hope, Collignon, Li, Badoux, Gourgout, Lalibert\'e, Zhou, Doiron-Leyraud et~al.}}]{michon-2018}
\bibinfo{author}{\bibfnamefont{B.}~\bibnamefont{Michon}}, \bibinfo{author}{\bibfnamefont{A.}~\bibnamefont{Ataei}}, \bibinfo{author}{\bibfnamefont{P.}~\bibnamefont{Bourgeois-Hope}}, \bibinfo{author}{\bibfnamefont{C.}~\bibnamefont{Collignon}}, \bibinfo{author}{\bibfnamefont{S.~Y.} \bibnamefont{Li}}, \bibinfo{author}{\bibfnamefont{S.}~\bibnamefont{Badoux}}, \bibinfo{author}{\bibfnamefont{A.}~\bibnamefont{Gourgout}}, \bibinfo{author}{\bibfnamefont{F.}~\bibnamefont{Lalibert\'e}}, \bibinfo{author}{\bibfnamefont{J.-S.} \bibnamefont{Zhou}}, \bibinfo{author}{\bibfnamefont{N.}~\bibnamefont{Doiron-Leyraud}}, \bibnamefont{et~al.}, \bibinfo{journal}{Phys. Rev. X} \textbf{\bibinfo{volume}{8}}, \bibinfo{pages}{041010} (\bibinfo{year}{2018}), \urlprefix\url{https://link.aps.org/doi/10.1103/PhysRevX.8.041010}.

\bibitem[{\citenamefont{Georges and Mravlje}(2021)}]{georges-2021}
\bibinfo{author}{\bibfnamefont{A.}~\bibnamefont{Georges}} \bibnamefont{and} \bibinfo{author}{\bibfnamefont{J.}~\bibnamefont{Mravlje}}, \bibinfo{journal}{Phys. Rev. Res.} \textbf{\bibinfo{volume}{3}}, \bibinfo{pages}{043132} (\bibinfo{year}{2021}), \urlprefix\url{https://doi.org/10.1103/PhysRevResearch.3.043132}.

\bibitem[{not()}]{note-fit}
\bibinfo{note}{The Nd-LSCO sample used to measure $S/T$ in Ref. \onlinecite{gourgout-2022} is the same or similar to the one used to measure the specific heat coefficient in Ref. \onlinecite{michon-2019}. So one would expect that the fitting of $\gamma(T)$ provides the same parameters in both cases. This is not the case for several reasons. First of all we fit the Seebeck in a crossover regime, where the presence of parameters like $\eta$ do not allow for an exact correspondence. Moreover one should remember that the $\gamma(T,p)$ expression fitting the specific heat was (arbitrarily) assumed to incorporate both the fermionic and bosonic contribution. This also makes the comparison difficult and the $\gamma(T)$ expressions from the two experiments should be similar but cannot coincide.}

\bibitem[{\citenamefont{Gourgout et~al.}(2022)\citenamefont{Gourgout, Grissonnanche, Lalibert\'e, Ataei, Chen, Verret, Zhou, Mravlje, Georges, Doiron-Leyraud et~al.}}]{gourgout-2022}
\bibinfo{author}{\bibfnamefont{A.}~\bibnamefont{Gourgout}}, \bibinfo{author}{\bibfnamefont{G.}~\bibnamefont{Grissonnanche}}, \bibinfo{author}{\bibfnamefont{F.}~\bibnamefont{Lalibert\'e}}, \bibinfo{author}{\bibfnamefont{A.}~\bibnamefont{Ataei}}, \bibinfo{author}{\bibfnamefont{L.}~\bibnamefont{Chen}}, \bibinfo{author}{\bibfnamefont{S.}~\bibnamefont{Verret}}, \bibinfo{author}{\bibfnamefont{J.-S.} \bibnamefont{Zhou}}, \bibinfo{author}{\bibfnamefont{J.}~\bibnamefont{Mravlje}}, \bibinfo{author}{\bibfnamefont{A.}~\bibnamefont{Georges}}, \bibinfo{author}{\bibfnamefont{N.}~\bibnamefont{Doiron-Leyraud}}, \bibnamefont{et~al.}, \bibinfo{journal}{Phys. Rev. X} \textbf{\bibinfo{volume}{12}}, \bibinfo{pages}{011037} (\bibinfo{year}{2022}), \urlprefix\url{https://link.aps.org/doi/10.1103/PhysRevX.12.011037}.

\bibitem[{\citenamefont{Boebinger et~al.}(1996)\citenamefont{Boebinger, Ando, Passner, Kimura, Okuya, Shimoyama, Kishio, Tamasaku, Ichikawa, and Uchida}}]{boebinger-1996}
\bibinfo{author}{\bibfnamefont{G.~S.} \bibnamefont{Boebinger}}, \bibinfo{author}{\bibfnamefont{Y.}~\bibnamefont{Ando}}, \bibinfo{author}{\bibfnamefont{A.}~\bibnamefont{Passner}}, \bibinfo{author}{\bibfnamefont{T.}~\bibnamefont{Kimura}}, \bibinfo{author}{\bibfnamefont{M.}~\bibnamefont{Okuya}}, \bibinfo{author}{\bibfnamefont{J.}~\bibnamefont{Shimoyama}}, \bibinfo{author}{\bibfnamefont{K.}~\bibnamefont{Kishio}}, \bibinfo{author}{\bibfnamefont{K.}~\bibnamefont{Tamasaku}}, \bibinfo{author}{\bibfnamefont{N.}~\bibnamefont{Ichikawa}}, \bibnamefont{and} \bibinfo{author}{\bibfnamefont{S.}~\bibnamefont{Uchida}}, \bibinfo{journal}{Phys. Rev. Lett.} \textbf{\bibinfo{volume}{77}}, \bibinfo{pages}{5417} (\bibinfo{year}{1996}), \urlprefix\url{https://doi.org/10.1103/PhysRevLett.77.5417}.

\bibitem[{\citenamefont{Ayres et~al.}(2021)\citenamefont{Ayres, Berben, {\v{C}}ulo, Hsu, van Heumen, Huang, Zaanen, Kondo, Takeuchi, Cooper et~al.}}]{ayres-2021}
\bibinfo{author}{\bibfnamefont{J.}~\bibnamefont{Ayres}}, \bibinfo{author}{\bibfnamefont{M.}~\bibnamefont{Berben}}, \bibinfo{author}{\bibfnamefont{M.}~\bibnamefont{{\v{C}}ulo}}, \bibinfo{author}{\bibfnamefont{Y.-T.} \bibnamefont{Hsu}}, \bibinfo{author}{\bibfnamefont{E.}~\bibnamefont{van Heumen}}, \bibinfo{author}{\bibfnamefont{Y.}~\bibnamefont{Huang}}, \bibinfo{author}{\bibfnamefont{J.}~\bibnamefont{Zaanen}}, \bibinfo{author}{\bibfnamefont{T.}~\bibnamefont{Kondo}}, \bibinfo{author}{\bibfnamefont{T.}~\bibnamefont{Takeuchi}}, \bibinfo{author}{\bibfnamefont{J.~R.} \bibnamefont{Cooper}}, \bibnamefont{et~al.}, \bibinfo{journal}{Nature} \textbf{\bibinfo{volume}{595}}, \bibinfo{pages}{661} (\bibinfo{year}{2021}), ISSN \bibinfo{issn}{1476-4687}, \urlprefix\url{https://doi.org/10.1038/s41586-021-03622-z}.

\bibitem[{\citenamefont{Ashcroft and Mermin}(1976)}]{ashcroft}
\bibinfo{author}{\bibfnamefont{N.~W.} \bibnamefont{Ashcroft}} \bibnamefont{and} \bibinfo{author}{\bibfnamefont{N.~D.} \bibnamefont{Mermin}}, \emph{\bibinfo{title}{{S}olid {S}tate {P}hysics}} (\bibinfo{publisher}{Holt-Saunders}, \bibinfo{year}{1976}).

\bibitem[{\citenamefont{Luo and Miley}(2002)}]{luo02}
\bibinfo{author}{\bibfnamefont{N.}~\bibnamefont{Luo}} \bibnamefont{and} \bibinfo{author}{\bibfnamefont{G.}~\bibnamefont{Miley}}, \bibinfo{journal}{Physica C: Superconductivity} \textbf{\bibinfo{volume}{371}}, \bibinfo{pages}{259} (\bibinfo{year}{2002}), ISSN \bibinfo{issn}{0921-4534}, \urlprefix\url{https://www.sciencedirect.com/science/article/pii/S0921453401011017}.

\bibitem[{\citenamefont{Harris et~al.}(1995)\citenamefont{Harris, Yan, Matl, Ong, Anderson, Kimura, and Kitazawa}}]{harris}
\bibinfo{author}{\bibfnamefont{J.~M.} \bibnamefont{Harris}}, \bibinfo{author}{\bibfnamefont{Y.~F.} \bibnamefont{Yan}}, \bibinfo{author}{\bibfnamefont{P.}~\bibnamefont{Matl}}, \bibinfo{author}{\bibfnamefont{N.~P.} \bibnamefont{Ong}}, \bibinfo{author}{\bibfnamefont{P.~W.} \bibnamefont{Anderson}}, \bibinfo{author}{\bibfnamefont{T.}~\bibnamefont{Kimura}}, \bibnamefont{and} \bibinfo{author}{\bibfnamefont{K.}~\bibnamefont{Kitazawa}}, \bibinfo{journal}{Phys. Rev. Lett.} \textbf{\bibinfo{volume}{75}}, \bibinfo{pages}{1391} (\bibinfo{year}{1995}), \urlprefix\url{https://link.aps.org/doi/10.1103/PhysRevLett.75.1391}.

\bibitem[{\citenamefont{Hussey}(2003)}]{hussey03}
\bibinfo{author}{\bibfnamefont{N.~E.} \bibnamefont{Hussey}}, \bibinfo{journal}{The European Physical Journal B - Condensed Matter and Complex Systems} \textbf{\bibinfo{volume}{31}}, \bibinfo{pages}{495} (\bibinfo{year}{2003}), ISSN \bibinfo{issn}{1434-6036}, \urlprefix\url{https://doi.org/10.1140/epjb/e2003-00059-9}.

\bibitem[{\citenamefont{Hussey}(2008)}]{hussey08}
\bibinfo{author}{\bibfnamefont{N.~E.} \bibnamefont{Hussey}}, \bibinfo{journal}{Journal of Physics: Condensed Matter} \textbf{\bibinfo{volume}{20}}, \bibinfo{pages}{123201} (\bibinfo{year}{2008}), \urlprefix\url{https://doi.org/10.1088/0953-8984/20/12/123201}.

\bibitem[{\citenamefont{Giraldo-Gallo et~al.}(2018)\citenamefont{Giraldo-Gallo, Galvis, Stegen, Modic, Balakirev, Betts, Lian, Moir, Riggs, Wu et~al.}}]{giraldo-gallo-2018}
\bibinfo{author}{\bibfnamefont{P.}~\bibnamefont{Giraldo-Gallo}}, \bibinfo{author}{\bibfnamefont{J.~A.} \bibnamefont{Galvis}}, \bibinfo{author}{\bibfnamefont{Z.}~\bibnamefont{Stegen}}, \bibinfo{author}{\bibfnamefont{K.~A.} \bibnamefont{Modic}}, \bibinfo{author}{\bibfnamefont{F.~F.} \bibnamefont{Balakirev}}, \bibinfo{author}{\bibfnamefont{J.~B.} \bibnamefont{Betts}}, \bibinfo{author}{\bibfnamefont{X.}~\bibnamefont{Lian}}, \bibinfo{author}{\bibfnamefont{C.}~\bibnamefont{Moir}}, \bibinfo{author}{\bibfnamefont{S.~C.} \bibnamefont{Riggs}}, \bibinfo{author}{\bibfnamefont{J.}~\bibnamefont{Wu}}, \bibnamefont{et~al.}, \bibinfo{journal}{Science} \textbf{\bibinfo{volume}{361}}, \bibinfo{pages}{479} (\bibinfo{year}{2018}), \eprint{https://www.science.org/doi/pdf/10.1126/science.aan3178}, \urlprefix\url{https://www.science.org/doi/abs/10.1126/science.aan3178}.

\bibitem[{\citenamefont{Ataei et~al.}(2022)\citenamefont{Ataei, Gourgout, Grissonnanche, Chen, Baglo, Boulanger, Lalibert{\'e}, Badoux, Doiron-Leyraud, Oliviero et~al.}}]{ataei-2022}
\bibinfo{author}{\bibfnamefont{A.}~\bibnamefont{Ataei}}, \bibinfo{author}{\bibfnamefont{A.}~\bibnamefont{Gourgout}}, \bibinfo{author}{\bibfnamefont{G.}~\bibnamefont{Grissonnanche}}, \bibinfo{author}{\bibfnamefont{L.}~\bibnamefont{Chen}}, \bibinfo{author}{\bibfnamefont{J.}~\bibnamefont{Baglo}}, \bibinfo{author}{\bibfnamefont{M.-E.} \bibnamefont{Boulanger}}, \bibinfo{author}{\bibfnamefont{F.}~\bibnamefont{Lalibert{\'e}}}, \bibinfo{author}{\bibfnamefont{S.}~\bibnamefont{Badoux}}, \bibinfo{author}{\bibfnamefont{N.}~\bibnamefont{Doiron-Leyraud}}, \bibinfo{author}{\bibfnamefont{V.}~\bibnamefont{Oliviero}}, \bibnamefont{et~al.}, \bibinfo{journal}{Nature Physics} \textbf{\bibinfo{volume}{18}}, \bibinfo{pages}{1420} (\bibinfo{year}{2022}), ISSN \bibinfo{issn}{1745-2481}, \urlprefix\url{https://doi.org/10.1038/s41567-022-01763-0}.

\bibitem[{\citenamefont{Chambers}(1952)}]{chambers}
\bibinfo{author}{\bibfnamefont{R.~G.} \bibnamefont{Chambers}}, \bibinfo{journal}{Proceedings of the Physical Society. Section A} \textbf{\bibinfo{volume}{65}}, \bibinfo{pages}{458} (\bibinfo{year}{1952}), \urlprefix\url{https://doi.org/10.1088/0370-1298/65/6/114}.

\bibitem[{\citenamefont{Hinlopen et~al.}(2022)\citenamefont{Hinlopen, Hinlopen, Ayres, and Hussey}}]{hussey22}
\bibinfo{author}{\bibfnamefont{R.~D.~H.} \bibnamefont{Hinlopen}}, \bibinfo{author}{\bibfnamefont{F.~A.} \bibnamefont{Hinlopen}}, \bibinfo{author}{\bibfnamefont{J.}~\bibnamefont{Ayres}}, \bibnamefont{and} \bibinfo{author}{\bibfnamefont{N.~E.} \bibnamefont{Hussey}}, \bibinfo{journal}{Phys. Rev. Res.} \textbf{\bibinfo{volume}{4}}, \bibinfo{pages}{033195} (\bibinfo{year}{2022}), \urlprefix\url{https://link.aps.org/doi/10.1103/PhysRevResearch.4.033195}.

\bibitem[{\citenamefont{Mirarchi and Caprara}(2024)}]{caprara24}
\bibinfo{author}{\bibfnamefont{G.}~\bibnamefont{Mirarchi}} \bibnamefont{and} \bibinfo{author}{\bibfnamefont{S.}~\bibnamefont{Caprara}}, \bibinfo{journal}{Condensed Matter} \textbf{\bibinfo{volume}{9}} (\bibinfo{year}{2024}), ISSN \bibinfo{issn}{2410-3896}, \urlprefix\url{https://www.mdpi.com/2410-3896/9/4/52}.

\bibitem[{\citenamefont{Lorenzana and Seibold}(2002)}]{sei02}
\bibinfo{author}{\bibfnamefont{J.}~\bibnamefont{Lorenzana}} \bibnamefont{and} \bibinfo{author}{\bibfnamefont{G.}~\bibnamefont{Seibold}}, \bibinfo{journal}{Phys. Rev. Lett.} \textbf{\bibinfo{volume}{89}}, \bibinfo{pages}{136401} (\bibinfo{year}{2002}), \urlprefix\url{https://link.aps.org/doi/10.1103/PhysRevLett.89.136401}.

\bibitem[{\citenamefont{Abbamonte et~al.}(2005)\citenamefont{Abbamonte, Rusydi, Smadici, Gu, Sawatzky, and Feng}}]{abba05}
\bibinfo{author}{\bibfnamefont{P.}~\bibnamefont{Abbamonte}}, \bibinfo{author}{\bibfnamefont{A.}~\bibnamefont{Rusydi}}, \bibinfo{author}{\bibfnamefont{S.}~\bibnamefont{Smadici}}, \bibinfo{author}{\bibfnamefont{G.~D.} \bibnamefont{Gu}}, \bibinfo{author}{\bibfnamefont{G.~A.} \bibnamefont{Sawatzky}}, \bibnamefont{and} \bibinfo{author}{\bibfnamefont{D.~L.} \bibnamefont{Feng}}, \bibinfo{journal}{Nature Physics} \textbf{\bibinfo{volume}{1}}, \bibinfo{pages}{155} (\bibinfo{year}{2005}), ISSN \bibinfo{issn}{1745-2481}, \urlprefix\url{https://doi.org/10.1038/nphys178}.

\bibitem[{\citenamefont{Radaelli et~al.}(2026)\citenamefont{Radaelli, Patel, Zhu, Lipscombe, Stewart, Sachdev, and Hayden}}]{hayden-2026}
\bibinfo{author}{\bibfnamefont{J.}~\bibnamefont{Radaelli}}, \bibinfo{author}{\bibfnamefont{A.~A.} \bibnamefont{Patel}}, \bibinfo{author}{\bibfnamefont{M.}~\bibnamefont{Zhu}}, \bibinfo{author}{\bibfnamefont{O.~J.} \bibnamefont{Lipscombe}}, \bibinfo{author}{\bibfnamefont{J.~R.} \bibnamefont{Stewart}}, \bibinfo{author}{\bibfnamefont{S.}~\bibnamefont{Sachdev}}, \bibnamefont{and} \bibinfo{author}{\bibfnamefont{S.~M.} \bibnamefont{Hayden}}, \bibinfo{journal}{Nature Communications} \textbf{\bibinfo{volume}{17}}, \bibinfo{pages}{4564} (\bibinfo{year}{2026}), \urlprefix\url{https://doi.org/10.1038/s41467-026-71319-w}.

\bibitem[{\citenamefont{Ma et~al.}(2021)\citenamefont{Ma, Rule, Cronkwright, Dragomir, Mitchell, Smith, Chi, Kolesnikov, Stone, and Gaulin}}]{gaulin21}
\bibinfo{author}{\bibfnamefont{Q.}~\bibnamefont{Ma}}, \bibinfo{author}{\bibfnamefont{K.~C.} \bibnamefont{Rule}}, \bibinfo{author}{\bibfnamefont{Z.~W.} \bibnamefont{Cronkwright}}, \bibinfo{author}{\bibfnamefont{M.}~\bibnamefont{Dragomir}}, \bibinfo{author}{\bibfnamefont{G.}~\bibnamefont{Mitchell}}, \bibinfo{author}{\bibfnamefont{E.~M.} \bibnamefont{Smith}}, \bibinfo{author}{\bibfnamefont{S.}~\bibnamefont{Chi}}, \bibinfo{author}{\bibfnamefont{A.~I.} \bibnamefont{Kolesnikov}}, \bibinfo{author}{\bibfnamefont{M.~B.} \bibnamefont{Stone}}, \bibnamefont{and} \bibinfo{author}{\bibfnamefont{B.~D.} \bibnamefont{Gaulin}}, \bibinfo{journal}{Phys. Rev. Res.} \textbf{\bibinfo{volume}{3}}, \bibinfo{pages}{023151} (\bibinfo{year}{2021}), \urlprefix\url{https://link.aps.org/doi/10.1103/PhysRevResearch.3.023151}.

\bibitem[{\citenamefont{Shi et~al.}(2025)\citenamefont{Shi, Zhang, Jiang, Liang, Zhang, Xu, Zhao, Chen, Cheng, Wang et~al.}}]{shi25}
\bibinfo{author}{\bibfnamefont{Q.}~\bibnamefont{Shi}}, \bibinfo{author}{\bibfnamefont{J.}~\bibnamefont{Zhang}}, \bibinfo{author}{\bibfnamefont{X.}~\bibnamefont{Jiang}}, \bibinfo{author}{\bibfnamefont{X.}~\bibnamefont{Liang}}, \bibinfo{author}{\bibfnamefont{R.}~\bibnamefont{Zhang}}, \bibinfo{author}{\bibfnamefont{J.}~\bibnamefont{Xu}}, \bibinfo{author}{\bibfnamefont{Z.}~\bibnamefont{Zhao}}, \bibinfo{author}{\bibfnamefont{J.}~\bibnamefont{Chen}}, \bibinfo{author}{\bibfnamefont{W.}~\bibnamefont{Cheng}}, \bibinfo{author}{\bibfnamefont{X.}~\bibnamefont{Wang}}, \bibnamefont{et~al.}, \bibinfo{journal}{Phys. Rev. B} \textbf{\bibinfo{volume}{112}}, \bibinfo{pages}{184510} (\bibinfo{year}{2025}), \urlprefix\url{https://link.aps.org/doi/10.1103/zlqq-zf1b}.

\bibitem[{\citenamefont{Comin et~al.}(2015)\citenamefont{Comin, Sutarto, da~Silva~Neto, Chauviere, Liang, Hardy, Bonn, He, Sawatzky, and Damascelli}}]{comin-2015}
\bibinfo{author}{\bibfnamefont{R.}~\bibnamefont{Comin}}, \bibinfo{author}{\bibfnamefont{R.}~\bibnamefont{Sutarto}}, \bibinfo{author}{\bibfnamefont{E.~H.} \bibnamefont{da~Silva~Neto}}, \bibinfo{author}{\bibfnamefont{L.}~\bibnamefont{Chauviere}}, \bibinfo{author}{\bibfnamefont{R.}~\bibnamefont{Liang}}, \bibinfo{author}{\bibfnamefont{W.~N.} \bibnamefont{Hardy}}, \bibinfo{author}{\bibfnamefont{D.~A.} \bibnamefont{Bonn}}, \bibinfo{author}{\bibfnamefont{F.}~\bibnamefont{He}}, \bibinfo{author}{\bibfnamefont{G.~A.} \bibnamefont{Sawatzky}}, \bibnamefont{and} \bibinfo{author}{\bibfnamefont{A.}~\bibnamefont{Damascelli}}, \bibinfo{journal}{Science} \textbf{\bibinfo{volume}{347}}, \bibinfo{pages}{1335} (\bibinfo{year}{2015}), \eprint{https://www.science.org/doi/pdf/10.1126/science.1258399}, \urlprefix\url{https://www.science.org/doi/abs/10.1126/science.1258399}.

\bibitem[{\citenamefont{Hashimoto et~al.}(2010)\citenamefont{Hashimoto, He, Tanaka, Testaud, Meevasana, Moore, Lu, Yao, Yoshida, Eisaki et~al.}}]{hashimoto-2010}
\bibinfo{author}{\bibfnamefont{M.}~\bibnamefont{Hashimoto}}, \bibinfo{author}{\bibfnamefont{R.-H.} \bibnamefont{He}}, \bibinfo{author}{\bibfnamefont{K.}~\bibnamefont{Tanaka}}, \bibinfo{author}{\bibfnamefont{J.-P.} \bibnamefont{Testaud}}, \bibinfo{author}{\bibfnamefont{W.}~\bibnamefont{Meevasana}}, \bibinfo{author}{\bibfnamefont{R.~G.} \bibnamefont{Moore}}, \bibinfo{author}{\bibfnamefont{D.}~\bibnamefont{Lu}}, \bibinfo{author}{\bibfnamefont{H.}~\bibnamefont{Yao}}, \bibinfo{author}{\bibfnamefont{Y.}~\bibnamefont{Yoshida}}, \bibinfo{author}{\bibfnamefont{H.}~\bibnamefont{Eisaki}}, \bibnamefont{et~al.}, \bibinfo{journal}{Nature Physics} \textbf{\bibinfo{volume}{6}}, \bibinfo{pages}{414} (\bibinfo{year}{2010}), ISSN \bibinfo{issn}{1745-2481}, \urlprefix\url{https://doi.org/10.1038/nphys1632}.

\bibitem[{\citenamefont{Tallon and Loram}(2001)}]{tallon-2001}
\bibinfo{author}{\bibfnamefont{J.~L.} \bibnamefont{Tallon}} \bibnamefont{and} \bibinfo{author}{\bibfnamefont{J.~W.} \bibnamefont{Loram}}, \bibinfo{journal}{Physica C} \textbf{\bibinfo{volume}{349}}, \bibinfo{pages}{53} (\bibinfo{year}{2001}).

\bibitem[{\citenamefont{Chen et~al.}(2023)\citenamefont{Chen, Lowder, Bakali, Andrews, Schrenk, Waas, Svagera, Eguchi, Prochaska, Wang et~al.}}]{natelson-2023}
\bibinfo{author}{\bibfnamefont{L.}~\bibnamefont{Chen}}, \bibinfo{author}{\bibfnamefont{D.~T.} \bibnamefont{Lowder}}, \bibinfo{author}{\bibfnamefont{E.}~\bibnamefont{Bakali}}, \bibinfo{author}{\bibfnamefont{A.~M.} \bibnamefont{Andrews}}, \bibinfo{author}{\bibfnamefont{W.}~\bibnamefont{Schrenk}}, \bibinfo{author}{\bibfnamefont{M.}~\bibnamefont{Waas}}, \bibinfo{author}{\bibfnamefont{R.}~\bibnamefont{Svagera}}, \bibinfo{author}{\bibfnamefont{G.}~\bibnamefont{Eguchi}}, \bibinfo{author}{\bibfnamefont{L.}~\bibnamefont{Prochaska}}, \bibinfo{author}{\bibfnamefont{Y.}~\bibnamefont{Wang}}, \bibnamefont{et~al.}, \bibinfo{journal}{Science} \textbf{\bibinfo{volume}{382}}, \bibinfo{pages}{907} (\bibinfo{year}{2023}), \eprint{https://www.science.org/doi/pdf/10.1126/science.abq6100}, \urlprefix\url{https://www.science.org/doi/abs/10.1126/science.abq6100}.

\bibitem[{\citenamefont{Valentinis et~al.}(2026)\citenamefont{Valentinis, Schmalian, Sachdev, and Patel}}]{valentinis-2026}
\bibinfo{author}{\bibfnamefont{D.}~\bibnamefont{Valentinis}}, \bibinfo{author}{\bibfnamefont{J.}~\bibnamefont{Schmalian}}, \bibinfo{author}{\bibfnamefont{S.}~\bibnamefont{Sachdev}}, \bibnamefont{and} \bibinfo{author}{\bibfnamefont{A.}~\bibnamefont{Patel}}, \bibinfo{journal}{Phys. Rev. Research} \textbf{\bibinfo{volume}{8}}, \bibinfo{pages}{013299} (\bibinfo{year}{2026}), \urlprefix\url{https://doi.org/10.1103/32ts-qh8d}.

\end{thebibliography}
 
\end{document}